\documentclass[12pt,a4paper]{article} 
\pdfoutput=1
\usepackage{style}

\usepackage{color,graphicx,amsmath,amssymb,lipsum}

\usepackage[normalem]{ulem}

\usepackage[utf8]{inputenc}
\usepackage[T1]{fontenc}
\usepackage{mathtools}
\usepackage{bm}

\usepackage{multirow}

\usepackage{ulem}
\usepackage{diagbox}

\usepackage[dvipsnames]{xcolor}
\newcommand{\rdrag}{r_{\rm drag}}
\newcommand{\kms}{{\rm \,\,km\,s^{-1}Mpc^{-1}}}

\newcommand{\eV}{\,\text{eV}}

\newcommand\Mpc{\text{Mpc}}
\newcommand\zeff{z_\text{eff}}

\renewcommand\thesection{\arabic{section}}
\renewcommand\thesubsection{\arabic{section}.\arabic{subsection}}
\newcommand{\be}{\begin{equation}}
\newcommand{\ee}{\end{equation}}
\newcommand{\beqa}{\begin{eqnarray}}
\newcommand{\eeqa}{\end{eqnarray}}
\newcommand{\bseq}{\begin{subequations}}
\newcommand{\eseq}{\end{subequations}}

\renewcommand\l{\lambda}
\renewcommand\r{\rho}

\renewcommand{\tanh}{\mathop{\rm th}\nolimits}

\renewcommand{\ln}{\mathop{\rm ln}\nolimits}

\def\l{\left(}
\def\r{\right)}
\def\Neff{N_{\rm eff}}
\def\Mnu{\sum m_\nu}
\def\DE{\rm DE}

\def\tr{\rm tr}
\def\bestf{{\rm best\text{-}fit}}

\def\lmax{\ell_{\rm max}}
\def\Ndof{N_{\rm dof}}

\def\beqn#1{\begin{equation}\label{#1}}
\def\eeqn{\end{equation}}

\def\beqa#1{\begin{eqnarray}\label{#1}}
\def\eeqa{\end{eqnarray}}

\def\Z2{$\mathcal{Z_2}$}

\newcommand {\ignore}[1]{}

\title{Exploring $\bm{\Lambda}$CDM extensions with SPT-3G and Planck data: \\
4\,$\sigma$ evidence for neutrino masses and \\
implications of extended dark energy models for cosmological tensions
}

\affiliation[a]{Department of Physics \& Astronomy, McMaster University,\\ 
1280 Main Street West, Hamilton, ON L8S 4M1, Canada}
\affiliation[b]{Institute for Nuclear Research of the Russian Academy of Sciences, \\ 
60th October Anniversary Prospect, 7a, 117312
Moscow, Russia}
\affiliation[c]{Moscow Institute of Physics and Technology,\\
	Institutsky lane 9, Dolgoprudny, Moscow region, 141700, Russia}
\affiliation[d]{Department of Particle Physics and Cosmology, Physics Faculty, M.V. Lomonosov Moscow State University, \\ Vorobjevy Gory, 119991 Moscow, Russia}

\author[a,b]{Anton Chudaykin\footnote{\texttt{chudayka@mcmaster.ca}},}
\author[b,c]{Dmitry Gorbunov\footnote{\texttt{gorby@ms2.inr.ac.ru}},}

\author[b,d]{Nikita Nedelko\footnote{\texttt{nikita.nedelko1999@gmail.com}}}

\begin{document}

\abstract{
We present new cosmological constraints in a set of motivated extensions of the $\Lambda$CDM model using the polarization and gravitational lensing measurements from the South Pole Telescope and the Planck CMB temperature observations at large angular scales. 
In all cosmological scenarios, this CMB data brings the clustering measurements into agreement with the low-redshift probes of large-scale structure.
Combining the SPT-3G, SPTpol and Planck large-scale temperature data with the latest full-shape BOSS and BAO measurements, information from the weak lensing and photometric galaxy clustering, and Pantheon supernova set we find a $4\sigma$ evidence for nonzero neutrino mass, $\sum m_\nu=0.22\pm0.06\eV$.
Breaking the CMB degeneracies between $\sum m_\nu$ and the cosmological parameters by the BOSS data is a major contribution to our neutrino mass measurement.
The future CMB data would allow for investigating this measurement.
Then we explore the possibility of dynamical dark energy with two model-independent approaches: one introduces a phantom crossing in dark energy equation of state, another provides with a sharp transition in the dark energy evolution. 
For the combination of all data considered, the both models predict $H_0\simeq68\kms$ being in a $\sim3\sigma$ tension with the SH0ES constraint.
However, when the local Type Ia supernovae are calibrated by Cepheids, the late Universe scenarios suggest significantly higher values of $H_0$ consistent with SH0ES.
Our work draws attention to the supernova absolute magnitude calibration as one of the issues on the way to reconcile the $H_0$ tension.

}

\begin{flushright}
	INR-TH-2022-006
\end{flushright}

\maketitle
\flushbottom

\section{Introduction}
\label{sec:intro}

Modern cosmology demonstrated a significant progress in the last decade. The most outstanding results came from the cosmic microwave background (CMB) which remains the most precise cosmological probe to date. The Planck measurements of CMB anisotropies have provided a fantastic confirmation of the standard $\Lambda$ Cold Dark Matter ($\Lambda$CDM) cosmological model, which parameters have been determined with unprecedented accuracy.
However, the increase of the experimental sensitivity has led to several statistically significant tensions between the early-time CMB measurements and other low-redshift cosmological probes.

The most significant tension refers to the difference between the values of the Hubble constant ($H_0$) directly measured in the late Universe and extracted from the CMB assuming the $\Lambda$CDM cosmology~\cite{Abdalla:2022yfr}. Local distance ladder approach utilizing photometry of 75 Milky Way Cepheids and Gaia EDR3 parallaxes yields $H_0=73.2\pm1.3\kms$~\cite{Riess:2020fzl}, which exhibits a 4.2$\sigma$ discrepancy with the number extracted from the Planck CMB data under $\Lambda$CDM, $H_0=67.36\pm0.54\kms$~\cite{Planck:2018vyg}. 
The latest SH0ES measurement, $H_0=73.04 \pm 1.04\kms$,~\cite{Riess:2021jrx} tightens the tension with the CMB estimate to $5\sigma$.
This discrepancy is conventionally treated as the Hubble tension, or even the Hubble crisis.
Other direct low-redshift probes have inferred the values of $H_0$ consistent with SH0ES, however the uncertainties associated to these measurements are considerably larger~\cite{Abdalla:2022yfr}.
The Type Ia supernovae calibrated by the Tip of the Red Giant Branch yield a somewhat lower value, $H_0=69.6 \pm 1.9 \kms$~\cite{Freedman:2020dne}.
The measurement of time delays in strongly lensed quasar systems leads to $H_0=73.3^{+1.7}_{-1.8}\kms$~\cite{Wong:2019kwg} which is independent of the cosmic distance ladder. Relaxing the assumptions on the mass density profile of the lensing galaxies, the TDCOSMO collaboration obtains $H_0=74.5^{+5.6}_{-6.1}\kms$, and $H_0=67.4^{+4.1}_{-3.2}\kms$ by combining the time-delay lenses with non time-delay lenses from the SLACS sample~\cite{Birrer:2020tax}.

In addition to the long-standing $H_0$ disagreement, the low-redshift measurements predict a systematically lower clustering amplitude compared to that measured by Planck from CMB~\cite{DiValentino:2020vvd}. 
This tension has been supported by results from Dark Energy Survey (DES), $S_8=0.776\pm0.017$~\cite{DES:2021wwk}, and Kilo-Degree Survey (KiDS), $S_8=0.759_{-0.021}^{+0.024}$~\cite{KiDS:2020suj}, where the $S_8=\sigma_8\sqrt{\Omega_m/0.3}$ parameter modulates the amplitude of the weak lensing measurements.~\footnote{When this paper was in the final preparation stage, the Hyper Suprime-Cam (HSC) Year 3 results have been announced~\cite{Dalal:2023olq}. They reported $S_8=0.776_{-0.033}^{+0.032}$ which is in excellent agreement with the other cosmic shear measurements.}
Being combined DES-Y3 and KiDS-1000 measurements are in tension with the Planck baseline result at the 3.3$\sigma$ level which is $S_8=0.832 \pm 0.013$~\cite{Planck:2018vyg}.
Full-shape analysis of galaxy power spectra and bispectrum~\cite{Philcox:2021kcw} along with traditional measurements of redshift-space distortions~\cite{Nunes:2021ipq} also bring consistently low values of $S_8$. 

While the $H_0$- and $S_8$-tensions can hint at cracks in the standard cosmological paradigm and the necessity for new physics, these discrepancies can still be in part the result of systematic errors in the experiments.

Intriguingly, there are a couple of curious features in the Planck data that lead to moderate tensions in parameter consistency tests.
The most significant feature refers to an oscillatory residual of the temperature (TT) power spectrum in the range $1000\lesssim\ell\lesssim2000$ that mimics the extra smoothing of acoustic CMB peaks generated by gravitational lensing~\cite{Aghanim:2016sns}.~\footnote{Although the oscillatory pattern looks similar to gravitational lensing at high multipoles, an implausibly large change in the foreground model can give a difference in the predicted spectra with a similar oscillatory component, see the related discussion in~\cite{Aghanim:2016sns}.}
The amount of lensing determined from the smoothing of the acoustic peaks in the CMB spectra is $2.8\sigma$ too high when compared with the $\Lambda$CDM expectation based on the "unlensed" temperature and polarization power spectra~\cite{Motloch:2019gux}.
Even within $\Lambda$CDM, the Planck internal features drive a moderate tension between the low-multipoles ($\ell<800$) and high-multipoles ($\ell>800$) constraints.~\footnote{Part of the difference between the low- and high-multipole ranges is caused by the dip in the Planck TT power spectrum in the range $20\lesssim\ell\lesssim30$~\cite{Aghanim:2016sns}.
}
In particular, the Planck TT $\ell>800$ data favours higher fluctuation amplitude $A_s$ and matter density $\Omega_mh^2$ as compared to the lower multipole range by about $3\sigma$~\cite{Aghanim:2016sns}.
Even though the significance of any individual shift is reduced in the multi-dimentional parameter space,
this disagreement drives the conspicuous differences in $\sigma_8$ and $H_0$ posteriors, which play more significant role in comparison with low-redshift cosmological probes.
Moreover, in some extensions of the base-$\Lambda$CDM model the overly enhanced smoothing of the CMB acoustic peaks could strongly affect the parameter constraints.
For instance, the neutrino mass lowers the predicted lensing power compared to $\Lambda$CDM that leads to surprisingly tight limit, $\Mnu<0.26\eV$ at $95\%$ confidence level (CL)~\cite{Planck:2018vyg}. 
If one marginalizes over the lensing information contained in the smoothing of the peaks of the CMB power spectra, 
the Planck constraint degrades to $\Mnu<0.87\eV$ at $95\%$ CL~\cite{Motloch:2019gux}.
In the cosmological model with extra relativistic degrees of freedom in the plasma, parameterized by an effective number of neutrinos $\Neff$, the arbitrary gravitational lensing opens up a new degeneracy direction between $H_0$ and $\Neff$ parameters thereby introducing an interesting avenue to reduce the $H_0$ tension~\cite{Motloch:2019gux}. 
Alternative CMB measurements especially on small angular scales can provide an important consistency check of the Planck results.

The small-scale CMB anisotropies can be probed by ground-based telescopes with exquisite precision.
The most accurate measurements of the CMB temperature and polarization power spectra have been taken by the South Pole Telescope (SPT-3G)~\cite{SPT-3G:2021eoc} and the Atacama Cosmology Telescope (ACT Data Release 4, ACT-DR4)~\cite{ACT:2020gnv}. Interestingly, these observations show no deviation from the standard lensing effect predicted for the base $\Lambda$CDM model.
Since the ground-based experiments have a higher sensitivity to small scales, it is highly beneficial to combine the full-sky and ground-based CMB measurements in the cosmological analysis.
Indeed, Ref.~\cite{Chudaykin:2020acu} showed that the Planck large-scale temperature data, the SPTpol polarization and lensing measurements combined within $\Lambda$CDM predict a substantially lower value of $S_8$ being consistent with the direct probes in the late Universe.
This result suggests that the $S_8$ tension can be driven by the extra smoothing of acoustic peaks in the Planck data that pulls the late-time amplitude to higher values.
This CMB setup also alleviates the Hubble tension down to $2.5\sigma$ statistical significance.
The same methodology has been applied in the Early Dark Energy (EDE) scenario to explore the cosmological tensions~\cite{Chudaykin:2020igl}.
Generally, the combined data approach yields robust measurements of cosmological parameters with only modestly larger error bars compared to the baseline Planck analysis, see Refs.~\cite{Chudaykin:2020acu,Chudaykin:2020igl}.

While the cosmological tensions can be partially explained by the internal features in the Planck data, they may also constitute hints towards new physics in the early or/and late Universe, see the recent review~\cite{Abdalla:2022yfr}.
The class of late-time scenarios which invokes modifications in the dark energy sector has been extensively investigated in the literature~\cite{Yang:2018qmz, DiValentino:2019dzu, Vagnozzi:2019ezj, Keeley:2019esp, Yang:2021flj, Roy:2022fif,Sharma:2022ifr}. 
These models assume variations in the dark energy equation of state parameter $w_{\DE}$, as well as the dark energy density $\rho_{\DE}$.
Such cosmological scenarios typically solve the Hubble tension within $2\sigma$ at the price of a phantom-like dark energy $w_{\DE}<-1$.
At the same time, model-independent studies based on the late Universe reconstruction point towards possible phantom crossing in the dark energy equation of state, see e.g.~\cite{Zhao:2012aw,Zhao:2017cud,Wang:2018fng,Dutta:2018vmq,Capozziello:2018jya}.
Moreover, the generic analytical approach~\cite{Heisenberg:2022lob} showed that solving both the $H_0$ and $S_8$ tensions necessarily requires the $w_{\DE}(z)$ to cross the value $w_{\DE}=-1$~\cite{Alestas:2021xes}.
It is important to investigate the possibility of dynamical dark energy with phantom crossing to alleviate the cosmological tensions when using the alternative CMB measurements.

In {\it this} work, we revisit the combined data analysis~\cite{Chudaykin:2020acu} by considering the latest SPT-3G polarization measurements. 
To be specific, we utilize the SPT-3G TE and EE power spectra, the SPTpol lensing reconstruction and the Planck TT $\ell<1000$ data.
First, we validate a statistical agreement amongst the different CMB measurements in the base-$\Lambda$CDM model. 
Then, we explore two physically well-motivated extensions: $\Lambda$CDM with massive active neutrinos ($\rm \Lambda$CDM+$\sum m_\nu$) and $\Lambda$CDM with extra relativistic degrees of freedom ($\rm \Lambda$CDM+$\Neff$). The main goal of this study is to obtain the alternative parameter constraints not affected by the Planck lensing-like anomaly. 
In passing, we explore the potential of $\rm \Lambda$CDM+$\sum m_\nu$ and $\rm \Lambda$CDM+$\Neff$ models to alleviate one or both cosmological tensions. Finally, we confront our results to that in the baseline Planck analysis.

We further explore the possibility of dynamical dark energy with two model-independent approaches.
The first scenario dubbed Phantom Dark Energy (PDE)~\cite{DiValentino:2020naf} parameterizes the dark energy density $\rho_{\DE}(z)$ through a Taylor series expansion truncated at certain order. 
There is no assumption about the physical entity of dark energy apart of that it has a phantom crossing during the course of its evolution.
This model was argued to be capable of alleviating the tension between the early and late Universe determinations of $H_0$~\cite{DiValentino:2020naf}.
At the same time, when the combination of all data is considered, the PDE scenario can not solve the $S_8$ tension which is largely driven by the Planck high-$\ell$ TT data. 
The second appealing scenario is the Transitional Dark Energy (TDE) originally suggested in~\cite{Keeley:2019esp}. 
This is a four parameter dynamical dark energy model based on a model-independent reconstruction of the effective dark energy equation of state, $w_{\DE}^{\rm eff}$,  
defined by $\rho_{\DE}(z)=\rho_{\DE}(0)(1+z)^{3(1+w_{\DE}^{\rm eff})}$~\cite{Jassal:2006gf}. 
Then Ref.~\cite{Keeley:2019esp} argues that a sharp transition in $w_{\DE}^{\rm eff}$ at $1<z<2$ could simultaneously explain the $H_0$ and $S_8$ tensions. 
We access the possibility of the PDE and TDE scenarios to alleviate the cosmological tensions using the alternative CMB data along with large-scale structure and supernova measurements.

This research improves the previous analyses~\cite{Chudaykin:2020acu,Chudaykin:2020igl} in the following directions. First, we utilize the latest CMB polarization measurements collected by the SPT-3G instrument~\cite{SPT-3G:2021eoc} which substantially improves upon the previous SPTpol results~\cite{Henning:2017nuy}. 
Second, we perform a full-shape analysis of the BOSS DR12 galaxy data including information from the power spectrum multipoles~\cite{Ivanov:2019pdj}, the real-space power spectrum~\cite{Ivanov:2021haa}, the reconstructed power spectrum~\cite{Philcox:2020vvt} and the bispectrum monopole~\cite{Philcox:2021kcw}. 
In addition, we consider multiple BAO measurements based on catalogs of emission-line galaxies, quasars, Ly$\alpha$ absorption and cross-correlation between the last two
that allows us to trace the cosmological evolution back to earlier times. 
Third, we use the Pantheon supernova data which helps to constrain the background cosmology in late-time modifying scenarios.
Fourth, we utilize the entire distance ladder which replaces the standard Gaussian constraint on $H_0$.

The outline of this paper is as follows. In Section \ref{sec:param} we describe our methodology and introduce all data sets used in the analysis. In Section \ref{sec:res} we brief a reader on our main results. In Section \ref{sec:valid} we validate our CMB setup. In Section \ref{sec:LCDM} we present cosmological constraints in the $\Lambda$CDM scenario. In Section \ref{sec:nuLCDM} we fit the parameters of $\rm \Lambda$CDM+$\sum m_\nu$ and $\rm \Lambda$CDM+$\Neff$ models to cosmological data and compare our results with those in the Planck analysis. In Section \ref{sec:PDE} we examine the PDE scenario against up-to-date cosmological data. In Section \ref{sec:TDE} we explore the implication of the TDE model for the cosmological tensions. We conclude in Section \ref{sec:concl}.

Five appendices contain supplementary materials. 
Appendix \ref{app:chi2} presents a complete breakdown of the best-fit $\chi^2_{\min}$ values per experiment for all models. 
In Appendix \ref{app:split} we assess the consistency between our CMB data set and the Planck TT $\ell>1000$ power spectrum. We also examine the sensitivity of our CMB-based parameter constraints to the choice of a Planck TT data cutoff.
Appendix \ref{app:PDEcom} presents the parameter constraints in the full Planck data analysis inside the PDE framework. 
In Appendix \ref{app:H0} we illustrate the difference between the entire distance ladder approach and the traditional Gaussian constraint on $H_0$ in the PDE model.
In Appendix \ref{app:Prior} we examine the sensitivity of parameter constraints to the choice of the TDE priors.

\section{Method and data}
\label{sec:param}

In this Section we describe our analysis procedure and data sets. 

\subsection{Method}
\label{sec:meth}

We obtain cosmological parameter constraints using the modified Einstein--Boltzmann code \texttt{CLASS-PT}~\cite{Chudaykin:2020aoj}, interfaced with the \texttt{Montepython} Monte Carlo sampler~\cite{Audren:2012wb,Brinckmann:2018cvx}. We perform the Markov Chain Monte Carlo (MCMC) analysis, sampling from the posterior distributions using the Metropolis-Hastings algorithm~\cite{Lewis:2002ah,Lewis:2013hha}.
The plots and marginalized constraints are generated with the latest version~\footnote{\href{https://getdist.readthedocs.io/en/latest/}{
\textcolor{blue}{https://getdist.readthedocs.io/en/latest/}}
} of the \texttt{getdist} package~\cite{Lewis:2019xzd}.  

In the $\Lambda$CDM model we vary the following set of cosmological parameters ($\omega_{cdm}$, $\omega_b$, $H_0$, $\ln(10^{10}A_s)$, $n_s$, $\tau$), where 
$H_0$ is the Hubble constant, which value can be recast as $H_0\equiv h\times 100$\,km\,s$^{-1}$\,Mpc$^{-1}$. Then,   
$\omega_{cdm}\equiv \Omega_{cdm} h^2$, $\omega_b\equiv\Omega_b h^2$ with $\Omega_{cdm}$ and $\Omega_b$ standing for the relative contribution of cold dark matter and baryons to the present energy density of the Universe. $A_{\rm s}$ and $n_s$ are the amplitude and the tilt of the primordial spectrum of scalar perturbations, $\tau$ denotes the reionization optical depth. In $\Lambda$CDM we assume the normal neutrino hierarchy with the total active neutrino mass $\sum m_\nu=0.06\eV$ and fix $\Neff$ to the default value $3.046$. Additionally, we run $\sum m_\nu$ in $\rm \Lambda$CDM+$\sum m_\nu$ and $\Neff$ in $\rm \Lambda$CDM+$\Neff$ models, respectively. In $\rm \Lambda$CDM+$\sum m_\nu$ model we approximate the neutrino sector with three degenerate massive states to boost the evaluation of the Einstein-Boltzmann code.  
In the PDE and TDE models we extend the dark energy sector accordingly along the lines of Secs.\,\ref{sec:PDE} and \ref{sec:TDE}.

Throughout our analysis the Hubble parameter $H_0$ is measured in units of km\,s$^{-1}$\,Mpc$^{-1}$, the sum of neutrino masses $\sum m_\nu$ is in units of eV, the present size of the horizon at the drag epoch $\rdrag$ is in Mpc, the angular diameter distance $D_A\equiv 1/(1+z)\int_0^z dz'/H(z')$ is in units inversed of the Hubble parameter,  km$^{-1}$\,s\,Mpc.

\subsection{Data}
\label{sec:data}

Hereafter we describe all data sets involved in this analysis.

{\bf $\bf PlanckTT\text{-}low\ell$:} We use the Planck \texttt{Plik} likelihood for the temperature (TT) power spectrum truncated at multipoles $30\leq\ell<1000$. We combined it with the \texttt{Commander} TT data in the angular multipole range $2\leq\ell<30$~\cite{Planck:2018vyg}.

{\bf SPT-3G:} We utilize the SPT-3G measurements of the E-mode (EE) polarization power spectrum and the temperature-E (TE) cross-power spectrum undertaken during a four-month period of 2018~\cite{SPT-3G:2021eoc}. 

This data includes the six EE and TE cross-frequency power spectra over the angular multipole range $300\leq\ell<3000$.
Following the original analysis~\cite{SPT-3G:2021eoc}, we include modeling of polarized Galactic dust for TE and EE spectra and Poisson-distributed point sources in the EE power spectrum. The CMB theoretical spectra are modified in order to account for the effects of instrumental calibration, aberration, super-sample lensing and survey geometry.~\footnote{We made the SPT-3G likelihood for the Montepython environment publicly available at \href{https://github.com/ksardase/SPT3G-montepython}{https://github.com/ksardase/SPT3G-montepython}}

{\bf Lens:} We use the measurement of the lensing potential power spectrum, $C_\ell^{\phi\phi}$, in the multipole range $100<\ell<2000$ from the SPTpol survey~\cite{Wu:2019hek}. 
The lensing potential is reconstructed from a minimum-variance quadratic estimator that combines both the temperature and polarization CMB maps. 
We incorporate the effects of the survey geometry and correct the $C_\ell^{\phi\phi}$ for a difference between the fiducial cosmology assumed in the lensing reconstruction and the cosmology of the SPTpol patch
following the procedure described in~\cite{Wu:2019hek}.~\footnote{The SPTpol likelihood used in this analysis is publicly available at \href{ https://github.com/ksardase/SPTPol-montepython}{https://github.com/ksardase/SPTPol-montepython}}

We use a recent measurement of the reionization optical depth from Ref.~\cite{DeBelsunce:2021xcp}. 
We impose a Gaussian constraint, 
\begin{equation}
\label{tau}
\tau=0.0581\pm0.0055\,,
\end{equation}
determined from the Planck \texttt{SRoll2} polarization (EE) maps using the likelihood approximation scheme \texttt{momento}.~\footnote{Note that the Planck 2018 legacy release High Frequency Instrument (HFI) polarization maps are based the \texttt{SRoll1} map-making algorithm. The improved map-making algorithm \texttt{SRoll2} significantly reduces large-scale polarization systematics compared to the \texttt{SRoll1} processing~\cite{DeBelsunce:2021xcp}. 
This results in the $40\%$ tighter constraint on $\tau$ \eqref{tau} compared to the Planck legacy release~\cite{Planck:2018vyg}.
}
We include the measurement \eqref{tau} in all data analyses. 
We do not mention it in data set names for brevity.

We combine all the above CMB measurements into one data set {\bf Base}.

To provide an additional test, we replace the Lens likelihood with the Planck lensing reconstruction from~\cite{Planck:2018vyg}. We refer to this combination as $\bf Base'$.

{\bf Planck 2018:} For the standard CMB analysis we use the official Planck TT, TE, EE+lensing and low-$\ell$ TT likelihoods~\cite{Planck:2018vyg}.
Note that we do not include the large-scale polarization data from Planck, choosing instead to constrain the optical depth $\tau$ via the Gaussian prior \eqref{tau}, as described above.
It allows us to perform a direct comparison.

{\bf LSS:} We perform a full-shape analysis of the large-scale power spectrum and bispectrum of the BOSS DR12 galaxy data. The galaxies were observed in the North and South Galactic Caps (NGC and SGC, respectively). We divide each sample into the two non-overlapping redshift slices with effective redshifts $\zeff=0.38$ and $\zeff=0.61$, giving a total of four data chunks. We apply window-free approach~\cite{Philcox:2020vbm,Philcox:2021ukg} which allows one to measure the unwindowed power spectrum and bispectrum directly from the observational data. We analyze the following data:~\footnote{The previous full-shape BOSS analyses were affected by an error in the public BOSS power spectra due to invalid approximation in the power spectrum normalization, for details see Ref.~\cite{Beutler:2021eqq}. In the window-free approach we do not require to model the mask, so our analysis is not affected by this problem.
}
\begin{itemize}
    \item {\it Redshift-Space Power Spectrum}: We use the pre-reconstracted power spectrum monopole, quadrupole and hexadecopole in the mode range $k\in[0.01,0.2]\,h\Mpc^{-1}$ as presented in Ref.\,\cite{Chudaykin:2020ghx}.
    \item {\it Real-Space Power Spectrum}: We use the analog to real space power spectrum for $k\in[0.2,0.4]\,h\Mpc^{-1}$ introduced in Ref.~\cite{Ivanov:2021haa}. It allows us to avoid limitations related to fingers-of-God modeling and access significantly smaller scales.
    \item {\it BAO}: We include the BAO measurements extracted from the post-reconstructed power spectra using a joint covariance matrix, as discussed in Ref.\,\cite{Philcox:2020vvt}.
    \item {\it Bispectrum}: We include the bispectrum monopole in the range $k\in[0.01,0.08]\,h\Mpc^{-1}$ with step $\Delta k=0.01\,h\Mpc^{-1}$ following~\cite{Philcox:2021kcw}. In total, it generates $62$ bispectrum bins. 
\end{itemize}
To model the above statistics, we utilize the effective field theory (EFT) of large scale structure as implemented in the \texttt{CLASS-PT} code~\cite{Chudaykin:2020aoj}. For consistency, we compute the power spectrum (bispectrum) up to one-loop (tree-level) order in the cosmological perturbation theory.
Our analysis features a full treatment of all necessary components: nonlinear corrections, galaxy bias, ultraviolet counterterms (to consistently account for short-scale physics), infrared resummation (to treat long-wavelength displacements) and stochastic bias. We marginalize the posteriors over all relevant nuisance parameters for each data chunk separately along the lines of Ref.~\cite{Philcox:2021kcw}. Detailed information about the standard EFT theoretical model and nuisance parameters can be found in Refs.~\cite{Ivanov:2019pdj,Chudaykin:2020aoj}.

We complement the BOSS DR12 measurements described above with the following BAO data:
\begin{itemize}
    \item 6dFGS at $\zeff=0.106$~\cite{Ross:2014qpa}
    \item SDSS DR7 MGS at $\zeff=0.15$~\cite{Beutler:2011hx}
    \item eBOSS quasar sample at $\zeff=1.48$~\cite{Neveux:2020voa}
    \item Auto-correlation of Ly$\alpha$ absorption and its cross correlation with quasars at $\zeff=2.33$ from the final eBOSS data release~\cite{duMasdesBourboux:2020pck}
    \item eBOSS emission line galaxy sample at $\zeff=0.845$~\cite{deMattia:2020fkb}~\footnote{We do not include the full-shape measurements of emission line galaxies because their impact on the eventual parameter constraints is rather limited as shown in~\cite{Ivanov:2021zmi}.}.
\end{itemize}

{\bf $\rm \bf S_8$:} We consider the DES-Y3 photometric galaxy clustering, galaxy-galaxy lensing, and cosmic shear measurements~\cite{DES:2021wwk}, in addition to weak gravitational lensing measurements from KiDS-1000~\cite{KiDS:2020suj} and HSC~\cite{HSC:2018mrq}. We combine these results in the form of a Gaussian prior,
\begin{equation}
\label{S8}
S_8=0.772\pm0.013\,.
\end{equation}
We treat this $S_8$ measurement separately from the other LSS data since it provides with the consistency test of individual likelihoods before combining them into a single set.

{\bf $\rm \bf SH0ES$:} 
We include the distance measurements of Type Ia supernovae in the Hubble flow calibrated with local geometric anchors via the Cepheid period luminosity relation.
We utilize the local distance ladder approach as implemented in the \texttt{distanceladder} package~\footnote{\href{https://github.com/kylargreene/distanceladder}{
\textcolor{blue}{https://github.com/kylargreene/distanceladder}}}~\cite{Greene:2021shv}.
To match the SH0ES methodology, we set up the upper redshift cut at $z=0.15$ for supernova sample.
The \texttt{distanceladder} using Cepheid calibration yields the absolute magnitude of Type Ia supernova~\cite{Greene:2021shv},
\begin{equation}
    \label{Msn}
M_B=-19.226\pm0.039\,,
\end{equation}
which closely reproduces the SH0ES result~\cite{Camarena:2021jlr}.
Assuming $\Lambda$CDM cosmology, the Cepheid calibration recovers an accurate mean values of $H_0$ compared to the SH0ES result~\cite{Riess:2020fzl},~\footnote{The \texttt{distanceladder} likelihood finds $H_0=73.14\pm1.39\kms$~\cite{Greene:2021shv}. The slight difference with the SH0ES value \eqref{H0} steams from the \texttt{distanceladder} only having access to the LMC and NGC4258 as anchors while the SH0ES analysis~\cite{Riess:2020fzl} uses the LMC, NGC4258, and Milky Way Cepheids. We neglect this difference in our analysis and refer to \eqref{H0} when assessing consistency between data sets.}
\begin{equation}
    \label{H0}
H_0=73.2\pm1.3\kms\,.
\end{equation}
The difference between the entire distance ladder approach and the traditional Gaussian prior on $H_0$ is highlighted in Appendix \ref{app:Prior}.

{\bf \bf SN:} Alternatively, we use the luminosity distance data of 1048 type Ia supernovae from the Pantheon catalog~\cite{Pan-STARRS1:2017jku}.

\section{Summary of our Main results}
\label{sec:res}

Let us briefly summarize our main results before going into the technical details. We fit the model parameters to the cosmological data considering five different cosmological scenarios: $\Lambda$CDM, $\rm \Lambda$CDM+$\sum m_\nu$, $\rm \Lambda$CDM+$\Neff$, PDE and TDE.   

\begin{figure}[!htb]
    \begin{center}
        \includegraphics[width=0.6\columnwidth]{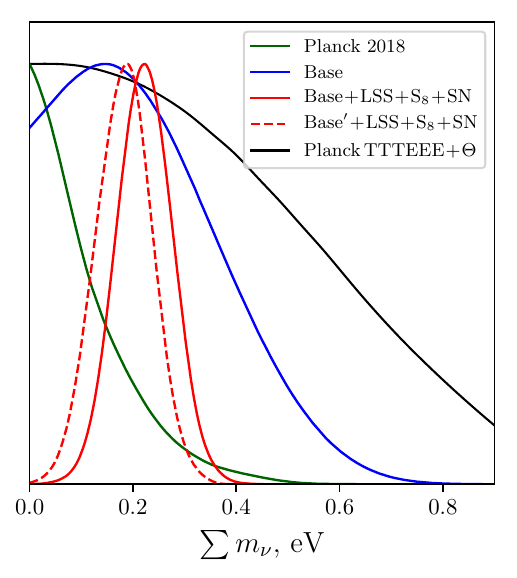}
        \caption {Marginalized 1d posterior distributions of $\sum m_\nu$ for the Planck 2018 (green), Base (blue), $\rm Base+LSS+S_8+SN$ (red), $\rm Base'+LSS+S_8+SN$ (dashed red) and Planck $\rm TTTEEE+\Theta$ (black) analyses. The $\rm Base'$ includes the Planck lensing reconstruction from Ref.\,\cite{Planck:2018vyg}. Planck $\rm TTTEEE+\Theta$ refers to the result after marginalizing over lensing information in the CMB maps from Ref.~\cite{Motloch:2019gux}.}
        \label{fig:31}
    \end{center}
\end{figure}

Figure\,\ref{fig:31} shows our main results in the $\rm \Lambda$CDM+$\sum m_\nu$ model.

The Base data yields a substantially weaker constraint on $\sum m_\nu$ compared to the full Planck analysis.
The high-$\ell$ temperature spectrum in the Planck 2018 data favours more lensing than allowed in $\Lambda$CDM that strengthens the constraint on the total neutrino mass~\cite{Planck:2018vyg}.
The $\rm Base+LSS+S_8+SN$ data suggests a $3.9\sigma$ preference of nonzero neutrino masses, $\sum m_\nu=0.22\pm0.06$\,eV. Using the Planck measurement of the lensing-potential power spectrum we infer a consistent estimate $\Mnu=0.18\pm0.06$\,eV. 
The LSS data plays a crucial role in our neutrino mass measurements by breaking the CMB degeneracies between $\Mnu$ and the other cosmological parameters.
We also display the Planck limit after marginalizing over the lensing information in the CMB power spectra~\cite{Motloch:2019gux}. 
This illustrates that our measurements agree with model-independent Planck lensing constraints.

Our neutrino mass measurements agree with the results of Ref.~\cite{DiValentino:2021imh} which analyzes the SPT-3G and ACT-DR4 data when combined with WMAP. Specifically, the SPT-3G+WMAP+BAO data mildly suggests a neutrino mass with $\Mnu=0.22_{-0.14}^{+0.056}\eV$. Our analysis improves the accuracy of this measurement mainly due to the full-shape BOSS analysis which has not been considered in~\cite{DiValentino:2021imh}.

\begin{figure}
\centering
  \vspace*{0.6cm}
   \includegraphics[width=.49\linewidth]{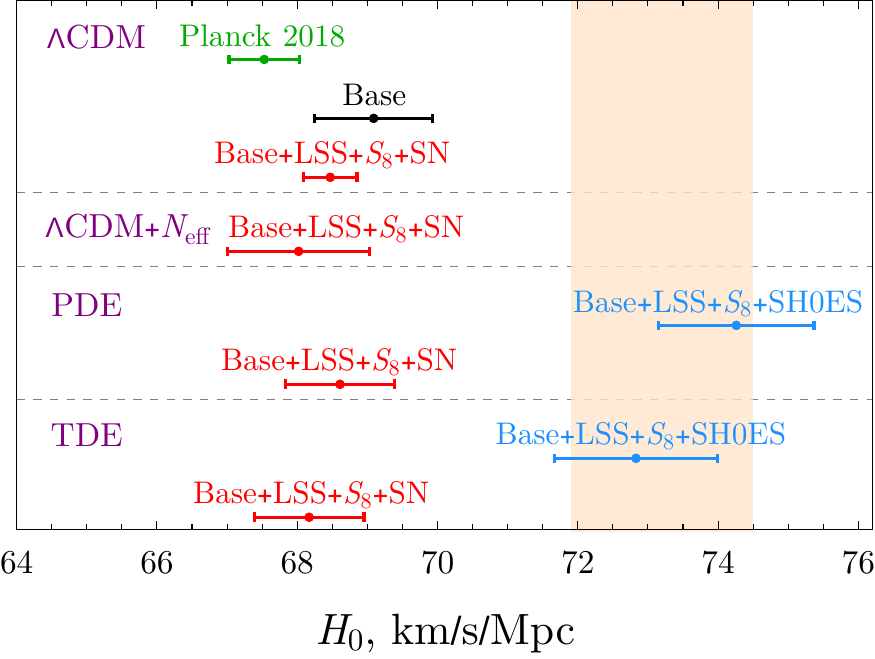}
   \includegraphics[width=.49\linewidth]{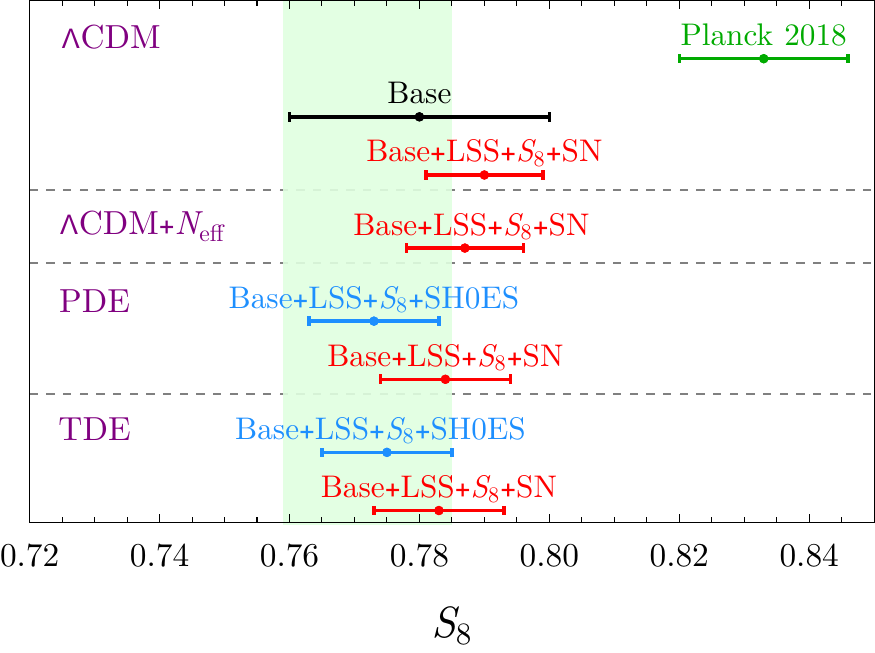}
  \label{fig:sub2}
\caption{Measurements (mean value with $1\sigma$ error bar) of the Hubble constant $H_0$ ({\it left panel}) and the late-time amplitude $S_8\equiv\sigma_8\sqrt{\Omega_m/0.3}$ ({\it right panel}) in the $\Lambda$CDM, $\Lambda$CDM+$\Neff$, PDE and TDE models. The orange band represents the direct measurement of $H_0$ \eqref{H0} reported by SH0ES, whereas the green band shows a combined constraint on $S_8$ \eqref{S8} coming from the photometric surveys DES-Y3, KiDS-1000 and HSC (both are given at $68\%$ CL).}
\label{fig:71}
\end{figure}

Fig. \ref{fig:71} summarizes the $H_0$ and $S_8$ constraints in different models. 
In all scenarios our analysis yields systematically lower values of $S_8$ being in good agreement with the low-redshift cosmological probes \eqref{S8}.
Note that the Planck 2018 data exhibits the $S_8$ tension at the $3.3\sigma$ significance level.
In $\Lambda$CDM the Base analysis predicts a moderately higher value of $H_0$ alleviating the Hubble tension to a $2.7\sigma$ level.
The $\rm Base+LSS+S_8+SN$ data shrinks the error bars on $H_0$ and $S_8$ in half.
The $\Lambda$CDM+$\Neff$ model partially alleviates the Hubble tension at the cost of inflating the error on $H_0$. 
The late-time scenarios (PDE and TDE), which drastically modify the dark energy sector, opens an avenue towards combining with the SH0ES data.
In the both models, the $\rm Base+LSS+S_8+SH0ES$ data yields significantly higher values of $H_0$ consistent with SH0ES. 
However, the $\rm Base+LSS+S_8+SN$ combination suggests a systematically lower $H_0$ being in a moderate ($\sim3\sigma$) tension with the SH0ES constraint \eqref{H0}.
The difference in the $H_0$ recovery reflects the tension between the SN calibration produced by CMB+BAO and the local astrophysical calibration by Cepheids.

We conclude that the $H_0$ tension can not be resolved by a non-trivial dynamics in the dark energy sector when all data are take into account. Our results reinforce the previous analyses~\cite{Lemos:2018smw,Poulin:2018zxs,Roy:2022fif,Dinda:2021ffa,Keeley:2022ojz} which show through the late Universe reconstruction that CMB, BAO and SN data do not allow for high $H_0$ values.

\section{CMB setup}
\label{sec:valid}

In this section we validate our CMB setup.

Our main CMB combination dubbed Base includes the Planck TT power spectrum in the multipole range $2\leq\ell<1000$, the TE and EE spectra over $300\leq\ell<3000$ from the SPT-3G data, and the power spectrum of the lensing potential at $100<\ell<2000$ measured from the SPTpol survey.~\footnote{We ignore the correlation between 2- and 4-point functions as it has been shown to be negligible at current sensitivities~\cite{Schmittfull:2013uea,Peloton:2016kbw}.} This upgrades the CMB setup used in the previous analysis~\cite{Chudaykin:2020acu} by featuring the latest SPT-3G polarization measurements.

\begin{figure}[!htb]
    \begin{center}
        \includegraphics[width=0.85\columnwidth]{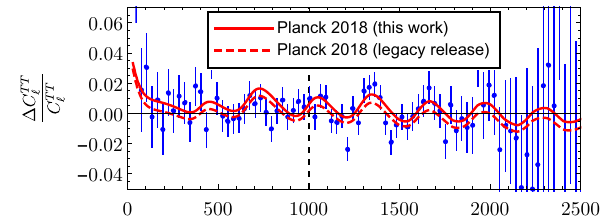}
        \includegraphics[width=0.85\columnwidth]{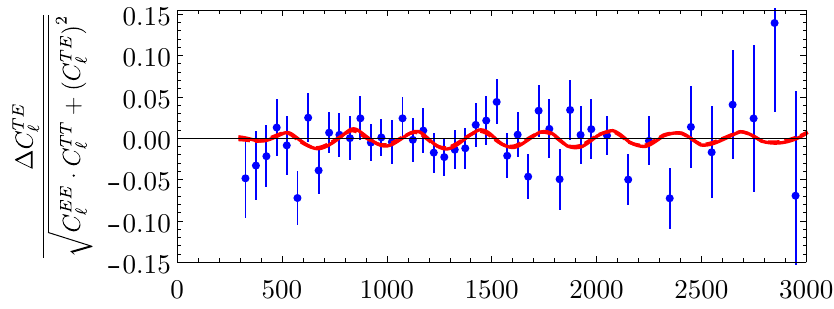}
        \includegraphics[width=0.85\columnwidth]{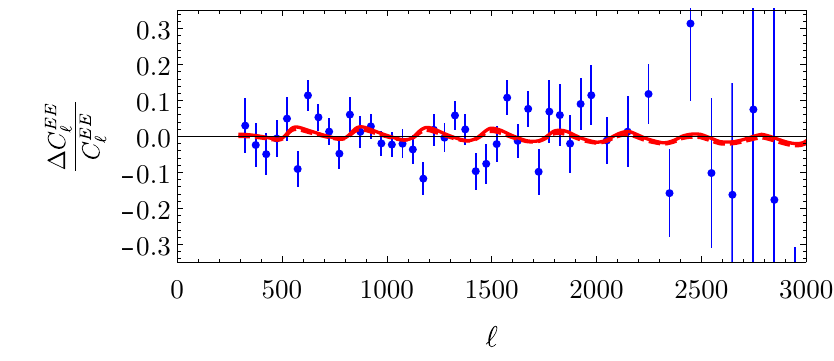}
        \vspace{-0.2cm}
        \caption {CMB residuals of the Planck TT (top panel), SPT-3G TE (middle panel) and EE (bottom panel) data with respect to the reference $\Lambda$CDM best-fit model of the Base likelihood (blue points). The red line corresponds to the difference between the $\Lambda$CDM best-fit prediction to the full Planck 2018 likelihood and the reference $\Lambda$CDM model (this work). 
        The dashed red line is the same for the official Planck best-fit model~\cite{Planck:2018vyg} (legacy release).
        The dashed black line flags the maximum multipole $\ell=1000$ used when fitting the reference $\Lambda$CDM model. 
        }
        \label{fig:res}
    \end{center}
\end{figure}

First, we test the consistency of our CMB setup at the level of the spectra. We fit the Base data within $\Lambda$CDM by varying all cosmological and nuisance parameters along the lines of Sec.~\ref{sec:meth}. Fig.~\ref{fig:res} shows the Planck TT, SPT-3G TE and EE residuals with respect to the reference $\Lambda$CDM best-fit model of the Base data.
To improve readability, we show the Planck TT power spectrum in the bands of width $\Delta \ell\approx31$ from the \texttt{Plik\_lite} likelihood~\cite{Planck:2018vyg}. 
As far as the SPT-3G data is concerned, we display the minimum-variance TE and EE bandpowers with the error bars corresponding to the diagonal elements of the bandpower covariance matrix~\footnote{Note that the SPT-3G bandpower covariance matrix do not include beam and calibration uncertainties~\cite{SPT-3G:2021eoc}.}.
We show the CMB residuals in units of $\sigma_{\rm CV}$, the cosmic variance error per multipole moment, defined as
\be
\sigma_{\rm CV} =
\begin{cases}
\sqrt{\frac{2}{2\ell+1}} C_\ell^{TT}, & {\rm TT} ,\\
\sqrt{\frac{1}{2\ell+1}}  \sqrt{C_\ell^{TT} C_\ell^{EE} + (C_\ell^{TE})^2}, & {\rm TE} ,\\
\sqrt{\frac{2}{2\ell+1}}  C_\ell^{EE}, & {\rm EE} . \\
\end{cases}
\ee

We found that our reference $\Lambda$CDM model matches both the Planck TT data in the range $30\leq\ell<1000$ and the SPT-3G TE and EE measurements (across the entire multipole range) within the statistical uncertainty. 
We detect the oscillatory residuals in the temperature power spectrum at $\ell>1000$ which can not be captured by our best-fit prediction. The associated difference is attributed to an extra peak-smoothing effect observed in the Planck high-$\ell$ TT data.
The residuals are not obviously anomalous being always within a $1.5\sigma$ statistical uncertainty, however they represent an oscillatory pattern across the broad range of angular scales which can impact the parameter constraints, for detail see~\cite{Addison:2015wyg,Aghanim:2016sns}.
When fitting the entire Planck 2018 spectra (red line), the best-fit model restores an agreement with the Planck high-$\ell$ TT data.
This is achieved at the cost of shifting cosmological parameters, mainly $A_s$ and $\omega_{cdm}$, which are pulled higher by around $2\sigma$~\cite{Planck:2018vyg}.
At the same time, the Planck 2018 prediction slightly deteriorates the fit to the PlanckTT-low$\ell$ data compared to the reference $\Lambda$CDM model.
So, the oscillatory residual in the Planck TT data has moderate impact on cosmological parameters even within the $\Lambda$CDM model.~\footnote{When the CMB $\ell>1000$ likelihood is combined with the Planck lensing reconstruction, the difference between the low- and high-multipole Planck constraints reduces but not disappears, see Fig.~22 of~\cite{Planck:2018vyg}.}
In extended cosmologies, the Planck internal features can introduce even larger shifts in the parameter constraints.

It is important to elucidate the difference between our Planck 2018 analysis and the Planck legacy release. To that end, in Fig. \ref{fig:res} we show the residuals of the official Planck best-fit model~\cite{Planck:2018vyg} with respect to the reference $\Lambda$CDM prediction (dashed red line). Our results demonstrate good agreement between the two Planck predictions. The Planck 2018 analysis implies a $\sim10\%$ higher $\tau$ compared to that in the legacy release.
This leads to a $1\sigma$ upward shift in $A_s$ which increases lensing smoothing and, therefore, provides a better fit to the Planck TT data at $\ell>1000$.
The Planck 2018 model also features a $0.6\sigma$ higher value of $A_se^{-2\tau}$ that causes a positive shift in the $C_\ell^{\rm TT}$ at large scales.
While the two Planck analyses yield the consistent CMB spectra, we choose to use the Planck 2018 data to be in line with the $\tau$ measurement \eqref{tau} used in the Base combination.

In order to assess consistency of our CMB setup, we consider a $\chi^2$ test for each individual likelihood.
Tab. \ref{tab:chi2_1} presents the $\chi^2_{\min}$ values for the best-fit $\Lambda$CDM models to the Planck 2018 and Base data as well as the associated degrees of freedom $\Ndof$.~\footnote{Since the constraints on nuisance parameters for the Planck and SPT-3G data are dominated by their priors, we accounted for the 5 free $\Lambda$CDM parameters.}
\begin{table}[t!]
	\renewcommand{\arraystretch}{1.1}
	\centering
	\begin{tabular} {| c || c |c | c|}
		\hline
		$\Lambda$CDM & Planck 2018 & Base & $\Ndof$ \\
		\hline
		\hline
		$\rm SPT$-$\rm 3G$ 
        & $530.36$  
        & $522.31$  
        & $523$\\
		$\rm Planck\, TT, \ell<30$ 
        & $23.22$ 
        & $21.15$   
        & $28$\\
		$\rm Planck\, TT, 30\leq\ell<1000$ 
        & $410.45$ 
        & $406.05$  
        & $444$\\
		$\rm Lens$ 
        & $7.93$ 
        & $5.57$  
        & $10$\\
		$\tau$-prior & $0.31$ 
        & $0.01$ 
        & $1$\\
		\hline
		Total $\chi^2_{\min}$ 
        & $972.27$ 
        & $955.09$  
        & $1006$ \\
		\hline
	\end{tabular}
	\caption {$\chi^2_{\min}$ values for the $\Lambda$CDM best-fit models to the Planck 2018 (second column) and Base (third column) data. The $\tau$-prior is set by \eqref{tau}. $\Ndof$ gives the number of degrees of freedom equal to the difference between the number of data points and the number of model parameters adjusted to produce the best-fit theory curve.}
	\label{tab:chi2_1}
\end{table}
The Base data approach improves the $\chi^2$ statistic for all CMB likelihoods with respect to the Planck 2018 analysis. The most significant contribution originates from the SPT-3G bandpowers which give $\Delta\chi^2_{\rm SPT\text{-}3G}=-8.05$. The Base analysis also improves the fit to the PlanckTT-low$\ell$ data and the CMB lensing but the corresponding improvement is modest given a number of the degrees of freedom $\Ndof$. In total, the cumulative $\chi^2_{\min}$ in the Base data approach improves by $\Delta\chi^2_{\rm tot}=-17.18$ relative to the Planck 2018 analysis. Our results demonstrate that the Base combination is mutually consistent and can be used in cosmological analyses.

We found that the Base data and the Planck TT $\ell>1000$ power spectrum are in a mild $2.4\sigma$ tension when analyzing the shifts in the full parameter space (see Appendix \ref{app:split}).~\footnote{The Planck TT $\ell<1000$ and $\ell>1000$ data are broadly consistent at the level of $1.6-1.8\sigma$~\cite{Addison:2015wyg,Aghanim:2016sns} which justifies the combination of these measurements in one data set.} Note that the individual cosmological parameters, $\omega_{cdm}$ and $H_0$, which play more significant role in comparison with low-redshift cosmological probes,~\footnote{In $\Lambda$CDM $\omega_{cdm}$ determines the broadband shape of the galaxy power spectrum measured by the BOSS collaboration and has impact on the weak-lensing parameter $S_8$.} differ by $3\sigma$. As discussed before, this disagreement is mainly caused by the overly enhanced smoothing of the CMB acoustic peaks that pulls $\sigma_8$ and $\omega_{cdm}$ to higher values. 
For this reason, we do not combine the Base and the Planck TT $\ell>1000$ into one data set.

Our $\rm PlanckTT\text{-}low\ell$ likelihood can be viewed as an emulation of the WMAP measurements. 
Indeed, the WMAP-9 and Planck TT data agree very closely at the level of the CMB power spectrum across $\ell<1000$ (see Fig. 48 of Ref.~\cite{Planck:2015bpv}). 
As the WMAP temperature maps reach the signal-to-noise ratio of unity by $\lmax\simeq600$~\cite{Aghanim:2016sns}, the Planck TT $\ell<600$ data serves as a proxy of the WMAP measurements. 
In Appendix \ref{app:split} we examine the sensitivity of our parameter constraints to the choice of a Planck TT data cutoff, $\lmax^{\rm TT}$, and find nearly indistinguishable results for $\lmax^{\rm TT}=600$ and $\lmax^{\rm TT}=1000$. 
Thus, the $\rm PlanckTT\text{-}low\ell$ data used in this work can be seen as a proxy for WMAP.

\section{$\Lambda$CDM model}
\label{sec:LCDM}

In this section we present the parameter measurements in the $\Lambda$CDM model. 
First, we scrutinize the cosmological inference from the Base data set. Second, we present the parameter constraints using the large-scale structure and supernova data. 

\subsection{Base data}
\label{sec:par}

\begin{figure}[!htb]
    \begin{center}
        \includegraphics[width=1.0\columnwidth]{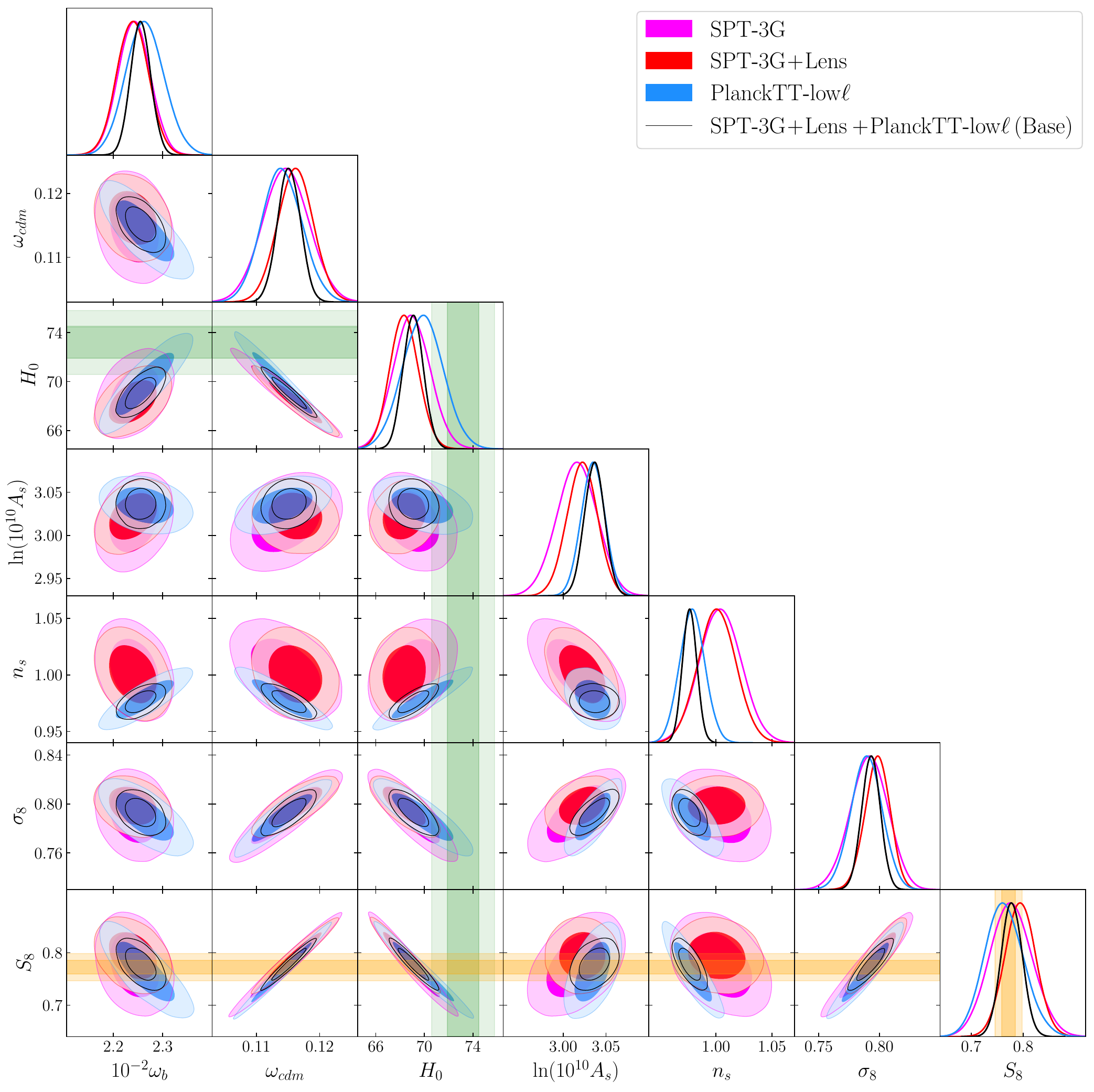}
        \caption {Marginalized 2d posterior distributions of the cosmological parameters in the $\Lambda$CDM model for $\rm SPT$-$\rm 3G$ (magenta), $\rm SPT$-$\rm 3G\!+\!Lens$ (red), $\rm PlanckTT\text{-}low\ell$ (blue), combined $\rm SPT$-$\rm 3G\!+\!Lens\!+\!PlanckTT\text{-}low\ell$ (black) data sets. The Gaussian prior on $\tau$ \eqref{tau} is always adopted. The yellow bands represent $1\sigma$ and $2\sigma$ constraints on $S_8$ \eqref{S8} coming from the photometric surveys (DES-Y3, KiDS, HSC), whereas the green bands refer to the $H_0$ measurement\,\eqref{H0} reported by the SH0ES collaboration.}
        \label{fig:1}
    \end{center}
\end{figure}

To assess the information gain coming from individual experiments we explore the parameter constraints from the SPT and Planck data separately. 
Fig. \ref{fig:1} shows the two-dimensional (2d) posterior distributions for various data set combinations.
The corresponding one-dimensional (1d) marginalized parameter constraints are tabulated in Tab.\,\ref{tab:1}. 
\begin{table}[!t]
    \renewcommand{\arraystretch}{1.2}
    \small
    \centering
    \begin{tabular} {| c || c |c | c | c |}
  \hline
		& \multicolumn{4}{c|}{$\Lambda$CDM}\\
		\hline
        \hline
        Parameter & $\rm SPT$-$\rm 3G$ & $\rm SPT$-$\rm 3G\!+\!Lens$ & $\rm \!PlanckTT\text{-}low\ell\!$ & $\rm Base$ \\
        \hline
        $100\,\omega_b$ & $2.243\pm0.033$ &
        $2.239\pm0.033$ &
        $2.264\pm0.039$ &
        $2.255\pm0.020$  \\ 
        $10\,\omega_{cdm}$ & $1.147\pm0.036$ &
        $1.162\pm0.029$ &
        $1.141\pm0.032$ &
        $1.151\pm0.018$ \\ 
        $H_0$ & $68.98\pm1.51$ &
        $68.36\pm1.20$ &
        $69.87\pm1.68$ &
        $69.09\pm0.84$ \\ 
        $\tau$ & $0.058\pm0.006$ &
        $0.058\pm0.006$ &
        $0.058\pm0.006$ &
        $0.058\pm0.005$ \\ 
        $\ln(10^{10} A_s)$ & $3.016\pm0.023$ &
        $3.022\pm0.018$ &
        $3.035\pm0.014$ &
        $3.036\pm0.012$ \\ 
        $n_s$ & $1.004\pm0.019$ &
        $1.001\pm0.017$ &
        $0.979\pm0.011$ &
        $0.977\pm0.006$ \\ 
        \hline   
        $\rdrag$ & $148.47\pm0.98$ &
        $148.10\pm0.76$ &
        $148.38\pm0.59$ &
        $148.18\pm0.43$ \\ 
        $\Omega_m$ & $0.290\pm0.020$ &
        $0.298\pm0.016$ &
        $0.282\pm0.019$ &
        $0.290\pm0.010$ \\ 
        $\sigma_8$ & $0.791\pm0.016$ &
        $0.798\pm0.011$ &
        $0.789\pm0.013$ &
        $0.793\pm0.008$ \\ 
        $S_8$ & $0.778\pm0.041$ &
        $0.796\pm0.030$ &
        $0.766\pm0.038$ &
        $0.780\pm0.020$ \\
        \hline
    \end{tabular}
    \caption {Marginalized 1d constraints on cosmological parameters in the standard $\Lambda$CDM
          model for four data sets. Recall that the Base data set includes $\rm \!SPT$-$\rm 3G\!+\!PlanckTT\text{-}low\ell\!+\!Lens$.}
    \label{tab:1}
\end{table}

Let us start with the $\rm SPT\text{-}3G$ data. Our parameter estimates agree with those from the SPT-3G official release~\cite{SPT-3G:2021eoc} at the precision level of $0.1\sigma$ in terms of the statistical error.~\footnote{Small difference can be explained by a different Gaussian constraint on $\tau$ used in Ref.~\cite{SPT-3G:2021eoc}.} These measurements significantly improve upon the previous results from the SPTpol survey~\cite{Henning:2017nuy}. The parameter constraints are also competitive with those from other current ground-based experiments~\cite{ACT:2020gnv}.

Next, we combine the $\rm SPT\text{-}3G$ data with the Lens measurement. Adding information on the lensing potential power spectrum significantly shrinks the error bars on cosmological parameters. In particular, the $H_0$ and $\sigma_8$ measurements improve by $20\%$ and $30\%$, respectively, upon including the Lens data. Overall, the parameters constraints are fully compatible with those from the $\rm SPT\text{-}3G$ analysis in agreement with~\cite{Bianchini:2019vxp}. 

As a next step, we examine the cosmological inference from the $\rm PlanckTT\text{-}low\ell$ data. 
We found that the parameter constraints are highly competitive with those from the $\rm SPT\text{-}3G\!+\!Lens$ analysis.
In particular, the $\rm SPT\text{-}3G\!+\!Lens$ analysis imposes tighter constraints on $\omega_b$, $\omega_c$, $H_0$ and $\sigma_8$ parameters whereas the $\rm PlanckTT\text{-}low\ell$ data provides more stringent bounds on $\ln(10^{10} A_s)$ and $n_s$. Thus, the two data sets naturally complement each other, and combining them at the likelihood level will yield a large information gain.

We combine the Planck and SPT measurements into one data set (Base). Our findings reinforce that the parameter constraints significantly improve upon those inferred from the $\rm SPT\text{-}3G\!+\!Lens$ and $\rm PlanckTT\text{-}low\ell$ data separately. In particular, the error bars on $H_0$ and $S_8$ shrink by $50\%$ compared to that in the $\rm PlanckTT\text{-}low\ell$ analysis, namely 
\begin{equation}
S_8=0.780\pm0.020   \qquad\qquad   H_0=69.09\pm0.84\kms
\end{equation}
Our constraint on $S_8$ is perfectly consistent with the direct measurements \eqref{S8}. In turn, the statistical difference between the CMB-based estimate of $H_0$ and the local measurement of this parameter \eqref{H0} reported by the SH0ES collaboration decreases from $4.2\sigma$ to $2.7\sigma$ level. Thus, the Hubble tension reduces compared to that if one would use the full Planck likelihood but still remains statistically implausible. We will examine the remaining tension in extended cosmologies in the following sections.

It is instructive to compare our results with those of the previous work~\cite{Chudaykin:2020acu} that uses the $\rm PlanckTT\text{-}low\ell$ data along with the SPTpol polarization and lensing measurements. Our analysis predicts $1\sigma$ higher values of $\sigma_8$ and $S_8$ compared to the previous research. This effect is attributed to the latest SPT-3G data which favours higher values of the late-time fluctuation amplitude~\cite{SPT-3G:2021eoc}. Overall, our analysis improves cosmological constraints by $10-20\%$ over that in Ref.~\cite{Chudaykin:2020acu}.

\subsection{Full data}
\label{sec:final}

\begin{figure}[!htb]
	\begin{center}
		\includegraphics[width=1.0\columnwidth]{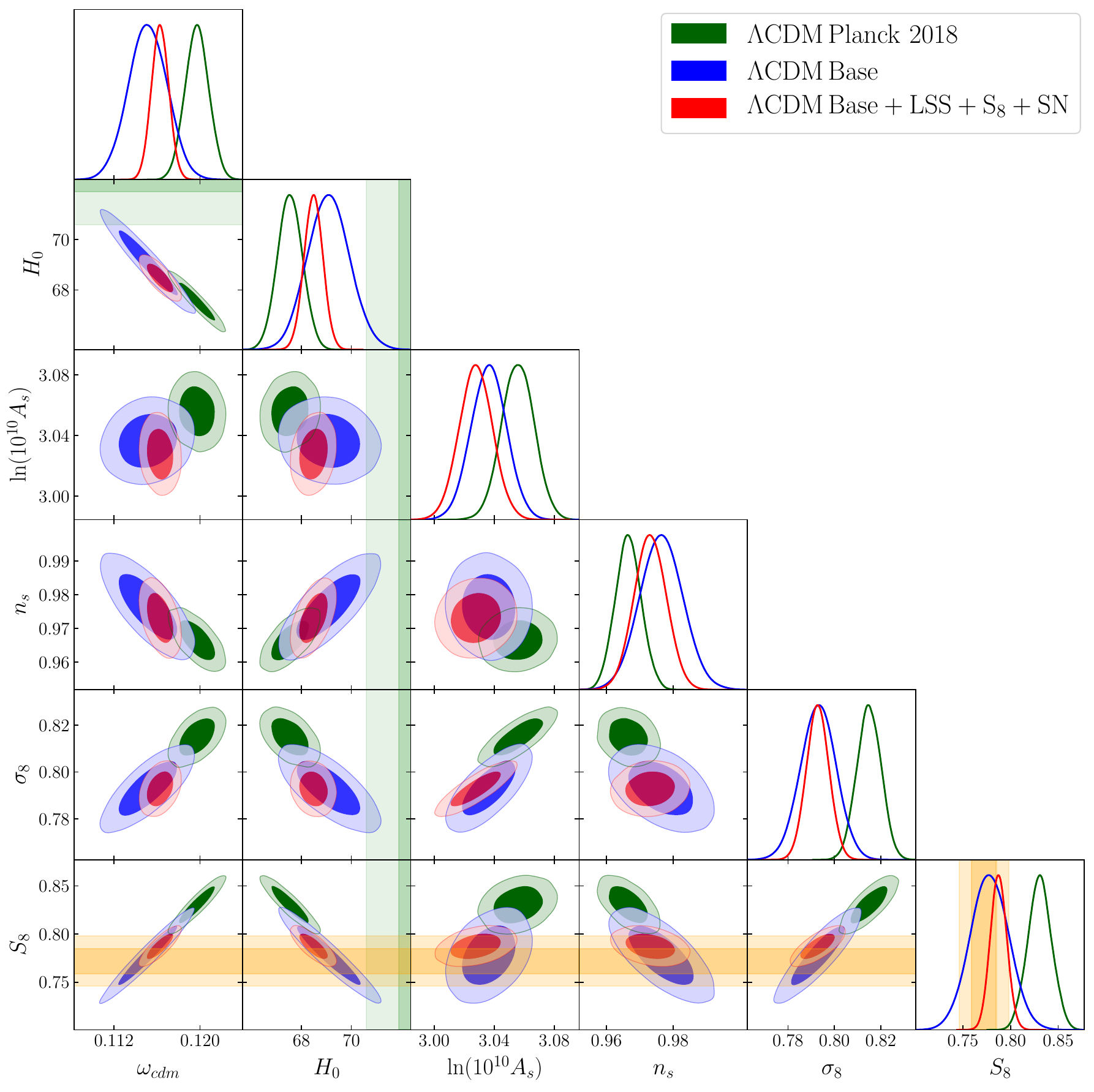}
		\caption {Marginalized 2d posterior distributions of the cosmological parameters in the $\Lambda$CDM model for Planck 2018 (green), Base (blue) and $\rm Base+LSS+S_8+SN$ (red) data sets. The Gaussian prior on $\tau$ \eqref{tau} is always set. The yellow bands represent $1\sigma$ and $2\sigma$ constraints on $S_8$ \eqref{S8} coming from the photometric surveys (DES-Y3, KiDS, HSC), whereas the green bands refer to the $H_0$ measurement \eqref{H0} reported by the SH0ES collaboration.}
		\label{fig:2}
	\end{center}
\end{figure}

Let us compare our CMB-based parameter constraints with those in the full Planck analysis. 
The 1d marginalized constraints on cosmological parameters are listed in Tab.\,\ref{tab:2}.
\begin{table}
	\renewcommand{\arraystretch}{1.2}
    \small
	\centering
	\begin{tabular} {|c|| c |c | c | c | c |}
		\hline
		& \multicolumn{5}{c|}{$\Lambda$CDM}\\
		\hline
		\hline
		\multirow{2}{*}{\!\!\! Parameter\!} & \multirow{2}{*}{$\rm Planck\,2018$} & \multirow{2}{*}{$\rm Base$} & \multirow{2}{*}{$\rm Base\!+\!LSS$} & $\rm \!Base\!+\!LSS\!$
		& $\rm \!Base\!+\!LSS\!$\\
		& & & & $\rm +S_8$ & $\rm +S_8\!+\!SN$ \\
		\hline
		$100\,\omega_b$ & $2.241\pm0.015$ &
		$2.255\pm0.020$ &
		$2.240\pm0.018$ &
		$2.247\pm0.018$ & $2.245\pm0.018$  \\ 
		$10\,\omega_{cdm}$ & $1.197\pm0.011$ &
		$1.151\pm0.018$ &
		$1.174\pm0.010$ &
		$1.163\pm0.008$ &  $1.163\pm0.008$ \\ 
		$H_0$ & $67.53\pm0.50$ &
		$69.09\pm0.84$ &
		$68.01\pm0.46$ &
		$68.49\pm0.38$ & $68.47\pm0.38$ \\ 
		$\tau$ & $0.060\pm0.005$ &
		$0.058\pm0.005$ &
		$0.055\pm0.005$ &
		$0.053\pm0.005$ & $0.053\pm0.005$ \\ 
		$\!{\rm ln}(10^{10} A_s)\!$ & $3.055\pm0.011$ &
		$3.036\pm0.012$ &
		$3.034\pm0.012$ &
		$3.028\pm0.011$ &	$3.027\pm0.011$  \\ 
		$n_s$ & $0.967\pm0.004$ &
		$0.977\pm0.006$ &
		$0.971\pm0.005$ &
		$0.973\pm0.005$ & $0.973\pm0.005$ \\ 
		\hline   
		$\rdrag$ & $147.12\pm0.25$ &
		$148.18\pm0.43$ &
		$147.75\pm0.31$ &
		$147.98\pm0.28$ & $147.97\pm0.28$ \\ 
		$\Omega_m$ & $0.313\pm0.007$ &
		$0.290\pm0.010$ &
		$0.304\pm0.006$ &
		$0.297\pm0.005$ & $0.298\pm0.005$ \\ 
		$\sigma_8$ & $0.815\pm0.005$ &
		$0.793\pm0.008$ &
		$0.799\pm0.006$ &
		$0.793\pm0.005$ & $0.793\pm0.005$  \\ 
		$S_8$ & $0.833\pm0.013$ &
		$0.780\pm0.020$ &
		$0.803\pm0.012$ &
		$0.789\pm0.009$ & $0.790\pm0.009$ \\ \hline
	\end{tabular}
	\caption {Parameter constraints in the standard $\Lambda$CDM
		model with $1\sigma$ errors. The Gaussian prior on $\tau$\,\eqref{tau} is adopted. The Base data set includes $\rm \!PlanckTT\text{-}low\ell\!+\!SPT$-$\rm 3G\!+\!Lens$.}
	\label{tab:2}
\end{table}
The resulting 2d posterior distributions for different data sets are shown in Fig.\,\ref{fig:2}. 

We found that the full Planck approach and the Base data lead to considerably different parameter constraints. 
In detail, the shifts in the posterior means between the Base and Planck 2018 analyses are 
\be
\label{diff1}
\begin{gathered}
\Delta\omega_b=0.6\sigma, \quad\Delta\omega_{cdm}=-2.2\sigma, \quad\Delta H_0=1.6\sigma,\\
\Delta\ln(10^{10}A_s)=-1.2\sigma, \quad\Delta n_s=1.4\sigma, \quad\Delta\sigma_8=-2.3\sigma,
\end{gathered}
\ee
expressed in the units of the posterior error of the two experiments combined in quadrature.~\footnote{Strictly speaking, the Planck 2018 and Base data sets are not independent since they share the common Planck TT $\ell<1000$ likelihood and $\tau$ measurement \eqref{tau}. However, combining the posterior errors in quadrature while neglecting their cross-covariance overestimates the actual size of the parameter errors. So, our estimates of the parameter shifts are conservative.} Although the cosmological constraints in these two CMB analyses are not obviously discrepant, the Planck 2018 data drives more severe tensions with the low-redshift cosmological constraints. Specifically, 
the CMB analysis based on the full Planck likelihood demonstrates the $S_8$ tension at the level of $3.3\sigma$. 
This effect is attributed to the overly enhanced smoothing of the acoustic peaks in the Planck data that pulls the late-time fluctuation amplitude $\sigma_8$ and, hence, $S_8$ to higher values.
The Base combination features the Planck TT data over $\ell<1000$, so our analysis is insensitive to the oscillatory residual in the Planck TT spectrum.~\footnote{The amount of lensing determined from the smoothing of the acoustic peaks in the SPT-3G power spectra is fully consistent with the $\Lambda$CDM expectation~\cite{SPT-3G:2021eoc}.}
The $H_0$ constraint inferred from the Planck 2018 data is also in a $4.2\sigma$ tension with the SH0ES measurement. A significantly lower value of $H_0$ in the full Planck analysis can be explained by an anti-correlation between $\sigma_8$ and $H_0$ parameters present in the CMB data as shown in Fig. \ref{fig:2}~\cite{Aghanim:2016sns}.

Next, we perform a joint analysis of the Base CMB data and the LSS likelihood (without $\rm S_8$). 
The accuracy of cosmological constraints drastically improves upon including the LSS information. In particular, the LSS data brings a twice more accurate measurement of $\omega_{cdm}$. 
This effect is attributed to the full-shape BOSS measurements which primarily constrain this parameter.
The LSS data also shrinks the error bars on $H_0$ and $S_8$ by $45\%$ and $40\%$, respectively, when compared with the Base only results. This leads to a more severe $3.8\sigma$ tension with the SH0ES constraint.
Remarkably, the $\rm Base\!+\!LSS$ data analysis is consistent with the direct probes of $S_8$ at the $1.7\sigma$ level. It justifies further account for the $\rm S_8$ data.

On the next step, we add the data on weak lensing and photometric galaxy clustering in the form of the Gaussian constraint on $S_8$ \eqref{S8}. 
We emphasize that the mean value of $S_8$ changes only by $1.1\sigma$ upon including the $\rm S_8$ information. This illustrates a good agreement between the $\rm Base+LSS$ and $S_8$ data sets. Interestingly, the mean value of $H_0$ raises up by $1\sigma$ that slightly alleviates the Hubble tension down to $3.5\sigma$ level, cf. with \eqref{H0}.

Finally, we add the supernova data. We found that the parameter constraints upon including the SN data remain essentially unchanged. 
This result can be understood as follows. 
In $\Lambda$CDM the supernova sample mainly constrains $\Omega_m$, which leads to $\Omega_m=0.298\pm0.022$~\cite{Pan-STARRS1:2017jku}. However, our $\rm Base\!+\!LSS\!+\!S_8$ data imposes a much tighter constraint on this parameter, $\Omega_m=0.297\pm0.005$, which is mainly driven by the CMB and full-shape BOSS measurements. So, the SN data has little statistical power compared to the $\rm Base\!+\!LSS\!+\!S_8$ combination. Our final constraints inferred from the $\rm Base\!+\!LSS\!+\!S_8\!+\!SN$ data read 
\begin{equation}
S_8=0.790\pm0.009\,, \qquad\qquad H_0=68.47\pm0.38\kms\,.
\end{equation}
Our results demonstrate good agreement with the direct measurements of $S_8$ (at $1.1\sigma$ level). The Hubble tension persists at the $3.5\sigma$ level.

The parameter constraints inferred from the $\rm Base\!+\!LSS\!+\!S_8\!+\!SN$ data considerably deviate from that in the full Planck analysis.

Namely, the shifts in the posteriors means between the $\rm Base\!+\!LSS\!+\!S_8\!+\!SN$ and Planck 2018 analyses read
\be
\label{diff2}
\begin{gathered}
\Delta\omega_b=0.2\sigma, \quad\Delta\omega_{cdm}=-2.5\sigma, \quad\Delta H_0=1.5\sigma,\\
\Delta{\rm ln}(10^{10} A_s)=-1.8\sigma, \quad\Delta n_s=0.9\sigma, \quad\Delta\sigma_8=-3.1\sigma,
\end{gathered}
\ee
Our approach predicts considerably smaller $\omega_{cdm}$ and $\sigma_8$ that pulls the $S_8$ value into consistency with the low-redshift probes \eqref{S8}
To a lesser extent, our CMB framework alleviates the $H_0$ tension. 
Interestingly, the shifts in $\omega_{cdm}$, ${\rm ln}(10^{10} A_s)$ and $\sigma_8$ parameters relative to the Planck 2018 values have amplified while including the $\rm LSS+S_8+SN$ data, cf. with \eqref{diff1}. 
Thus, the large-scale structure and supernova data support the cosmological inference based on the Base data.

\section{Minimal extensions of the base-$\Lambda$CDM model}
\label{sec:nuLCDM}

In this section we explore the parameter constraints in the $\rm \Lambda$CDM+$\sum m_\nu$ and $\rm \Lambda$CDM+$\Neff$ models.

\subsection{$\rm \Lambda$CDM+$\sum m_\nu$}
\label{sec:nuLCDM1}

\begin{figure}[!htb]
    \begin{center}
        \includegraphics[width=1.0\columnwidth]{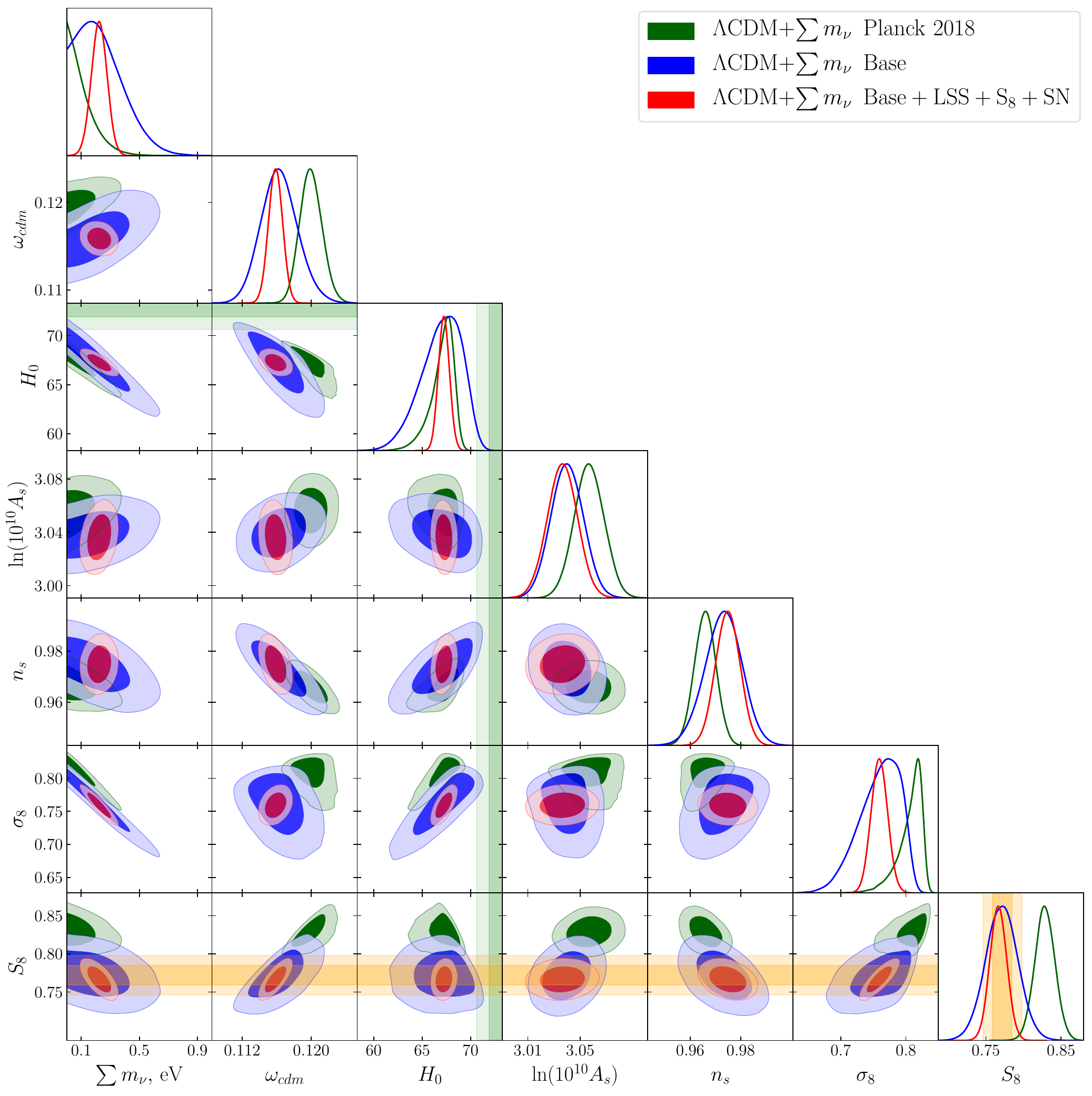}
        \caption {Marginalized 2d posterior distributions of the cosmological parameters in the $\rm \Lambda$CDM+$\sum m_\nu$ model for Planck 2018 (green), Base (blue) and $\rm Base+LSS+S_8+SN$ (red) data sets. The Gaussian prior on $\tau$\,\eqref{tau} is imposed. The yellow bands represent $1\sigma$ and $2\sigma$ constraints on $S_8$ \eqref{S8} coming from the photometric surveys (DES-Y3, KiDS, HSC), whereas the green bands refer to the $H_0$ measurement \eqref{H0} reported by the SH0ES collaboration.}
        \label{fig:3}
    \end{center}
\end{figure}

We start with the $\rm \Lambda$CDM+$\sum m_\nu$ scenario. 
Tab.\,\ref{tab:3} presents the 1d marginalized constraints on cosmological parameters in the $\rm \Lambda$CDM+$\sum m_\nu$ model. 
\begin{table}
	\renewcommand{\arraystretch}{1.2}
    \small
	\centering
	\begin{tabular} {|c|| c |c | c |c|c|}
		\hline
		& \multicolumn{5}{c|}{$\rm \Lambda$CDM+$\sum m_\nu$} \\
		\hline
		\hline
		\multirow{2}{*}{\!\!\! Parameter\!} & \multirow{2}{*}{$\rm Planck\,2018$} & \multirow{2}{*}{$\rm Base$} & \multirow{2}{*}{$\rm Base\!+\!LSS$} & $\rm \!Base\!+\!LSS\!$
		& $\rm \!Base\!+\!LSS\!$\\
		& & & & $\rm +S_8$ & $\rm +S_8\!+\!SN$ \\
		\hline
		$100\,\omega_b$ & $2.239\pm0.015$ &
		$2.246\pm0.022$ &
		$2.246\pm0.018$ &
		$2.247\pm0.018$ & $2.248\pm0.018$  \\ 
		$10\,\omega_{cdm}$ & $1.200\pm0.013$ &
		$1.163\pm0.021$ &
		$1.162\pm0.012$ &
		$1.159\pm0.008$ & $1.158\pm0.008$ \\ 
		$H_0$ & $67.03^{+1.47}_{-0.71}$ &
		$67.02^{+2.54}_{-1.61}$ &
		$67.15\pm0.59$ &
		$67.15\pm0.60$ & $67.32\pm0.57$ \\ 
		$\tau$ & $0.060\pm0.005$ &
		$0.058\pm0.005$ &
		$0.057\pm0.005$ &
		$0.057\pm0.005$ & $0.057\pm0.005$ \\ 
		$\!{\rm ln}(10^{10} A_s)\!$ & $3.057\pm0.011$ &
		$3.040\pm0.012$ &
		$3.037\pm0.012$ &
		$3.037\pm0.012$ & $3.036\pm0.012$ \\ 
		$n_s$ & $0.966\pm0.004$ &
		$0.973\pm0.007$ &
		$0.974\pm0.005$ &
		$0.975\pm0.005$ & $0.975\pm0.005$ \\ 
		$\Mnu$ & $<0.30$ &
		$<0.51$ &
		$0.22\pm0.07$ &
		$0.23\pm0.06$ & $0.22\pm0.06$ \\ 
		\hline   
		$\rdrag$ & $147.07\pm0.28$ &
		$147.93\pm0.49$ &
		$147.98\pm0.34$ &
		$148.03\pm0.28$ & $148.05\pm0.28$ \\    
		$\Omega_m$ & $0.320\pm0.016$ &
		$0.316\pm0.027$ &
		$0.313\pm0.007$ &
		$0.313\pm0.007$ & $0.310\pm0.007$ \\ 
		$\sigma_8$ & $0.806\pm0.019$ &
		$0.760\pm0.031$ &
		$0.761\pm0.018$ &
		$0.758\pm0.013$ & $0.760\pm0.013$ \\ 
		$S_8$ & $0.832\pm0.013$ &
		$0.778\pm0.021$ &
		$0.777\pm0.018$ &
		$0.774\pm0.010$ & $0.773\pm0.010$ \\
		\hline
	\end{tabular}
	\caption {Parameter constraints in the standard $\rm \Lambda$CDM+$\sum m_\nu$ model with $1\sigma$ errors. The upper limits on neutrino masses are given at $95\%$ CL. 
 Recall, the Base data set includes $\rm \!PlanckTT\text{-}low\ell\!+\!SPT$-$\rm 3G\!+\!Lens$.}
	\label{tab:3}
\end{table}
Fig.\,\ref{fig:3} displays the 2d posterior distributions for different analyses.

Let us illuminate the main differences between our approach and the full Planck analysis. 
First, the Base data predicts a $2.2\sigma$ lower value of $S_8$ when compared with the Planck 2018 result. This makes our analysis entirely consistent with the direct measurements of $S_8$, whereas the full Planck approach is in a $3.3\sigma$ tension with the $S_8$ data \eqref{S8}. 
Second, the $H_0$ measurements agree in the both analyses but the Base data leads to a two-times larger error bar on this parameter. 
Finally, we obtain the CMB-based constraint on the total neutrino mass, $\sum m_\nu<0.51\,\eV$ at $95\%$ CL. 
This limit is considerably weaker than the Planck 2018 result.~\footnote{Note that our Planck 2018 limit on the total neutrino mass is somewhat weaker than the Planck legacy release constraint, $\Mnu<0.24\,\eV$ at $95\%$ CL~\cite{Planck:2018vyg}. There are two main contributors to this difference. First, we adopt the Gaussian prior on $\tau$ \eqref{tau} which inflates the error on $\Mnu$ compared to that if one would use the large-scale polarization data.
Second, we exploit a $0.6\sigma$ higher value of $\tau$ \eqref{tau} which raises up $A_s$ and requires somewhat higher values of $\Mnu$. 
} 
The main reason is a larger smoothing of the Planck TT power spectrum peaks and troughs at $\ell>1000$ which artificially strengthens the Planck 2018 constraint on the neutrino mass~\cite{Planck:2018vyg,Motloch:2019gux}.

It is interesting to compare our Base limit on the neutrino mass with the results of the other CMB analyses which are 
insensitive to the lensing-induced smoothing of the acoustic peaks. 
First, one can marginalize over the lensing information that removes any effect of the peak smoothing in the CMB power spectra on constraints of the cosmological parameters. 
Allowing arbitrary gravitational lensing in the Planck TT,TE,EE maps the constraint on the total neutrino mass degrades to $\sum m_\nu<0.87\eV$ at $95\%$ CL~\cite{Motloch:2019gux}.~\footnote{The method applied in~\cite{Motloch:2019gux} allows one to constrain the lensing potential power spectrum in a model independent way by modeling the principal components of the gravitational lensing potential. It should be contrasted with the standard approach of introducing a phenomenological parameter $A_L$ which multiplies $C_\ell^{\phi\phi}$ at each point of the parameter space and can not be interpreted in terms of the lensing potential~\cite{Motloch:2018pjy}. }
Second, the combination of the Planck measurement of the CMB acoustic scale ($\theta_*$), the CMB lensing reconstruction power spectrum and BAO data leads to the limit $\sum m_\nu<0.60\eV$ at $95\%$ CL~\cite{Planck:2018vyg} which is almost independent of lensing effects in the CMB spectra. 
Both measurements agree with the neutrino mass constraint inferred from the Base data. The Base analysis yields the considerably tighter bound due to the SPT-3G data which independently constrains $\sum m_\nu$ through the lensing-induced smoothing of CMB acoustic peaks.

We further assess the impact of the LSS data on cosmological constraints. 
The LSS data tremendously improves (more than 3 times) the accuracy of the $H_0$ recovery. This effect is driven by the distance information encoded in the BOSS galaxy spectra and anisotropic BAO measurements at intermediate redshifts. The LSS data significantly shrinks the error bars on other cosmological parameters with an exception of ${\rm ln}(10^{10} A_s)$, which is primarily constrained by CMB, and $\tau$ governed by \eqref{tau}.
Our analysis does not feature the data on weak lensing and photometric galaxy clustering, but its result is perfectly consistent with the direct probes of $S_8$. Intriguingly, we found the $3.1\sigma$ evidence for nonzero neutrino masses, namely $\Mnu=0.22\pm0.07\eV$. 
The LSS data helps to break the CMB degeneracies between the $\Mnu$ and the cosmological parameters which significantly improves the neutrino mass constraint. 

Next, we add the $S_8$ data. 
As expected, including $S_8$ information substantially improves the bounds on $\sigma_8$ and $S_8$ parameters. 
It also tightens the $\omega_{cdm}$ constraint as this parameter largely controls the growth rate of cosmological matter perturbations. Striking, the limit on $\Mnu$ remains essentially intact. This indicates that the information on neutrino masses comes from breaking the degeneracies between the LSS and CMB rather than from the direct measurements of the late-time parameter $S_8$. All other constraints only barely change that demonstrates an excellent agreement between the $\rm Base+LSS$ and $\rm S_8$ data sets.

Finally, we include the supernova data. We found that the parameter constraints remain virtually unchanged. The reason is the same as in the $\Lambda$CDM scenario: the background evolution is tightly constrained by CMB and LSS measurements, so the gain from adding the SN data is very modest. The $\rm Base+LSS+S_8+SN$ analysis suggests the $3.9\sigma$ preference for nonzero $\Mnu$,
\be
\label{Mnu}
\Mnu=0.22\pm0.06\,\eV\,.
\ee
This measurement is consistent with both neutrino mass hierarchies. We emphasize that the information gain comes from breaking of the degeneracies between the LSS and CMB data and not from the $S_8$ constraint \eqref{S8}. In the full Planck data approach, the extra smoothing of CMB acoustic peaks strengthens the constraints on neutrino masses making higher values of $\Mnu$ implausible~\cite{Planck:2018vyg}.~\footnote{
When additionally including the Planck lensing reconstruction, together with the BAO and SN measurements, the neutrino mass constraint degrades only by $20\%$ over the Planck limit after marginalizing over lensing information~\cite{Motloch:2019gux}.
}
To validate the robustness of our result, we consider the $\rm Base'+LSS+S_8$ data which features the Planck lensing reconstruction~\cite{Planck:2018vyg}.
This analysis yields $\Mnu=0.18\pm0.06\eV$ which implies a nonzero neutrino mass at the $3.1\sigma$ level. 

Let us compare our result with the recent measurements of neutrino masses from Ref.~\cite{DiValentino:2021imh}.
First, our estimate \eqref{Mnu} is entirely consistent with the SPT-3G+WMAP+BAO constraint, $\Mnu=0.22_{-0.14}^{+0.056}\eV$,~\cite{DiValentino:2021imh}.
This agreement is not surprising because the $\rm PlanckTT\text{-}low\ell$ data used in our analysis emulates the WMAP measurement, see the related discussion in Sec. \ref{sec:valid}.
Importantly, our approach brings the more accurate measurement of $\Mnu$ owning to the full-shape BOSS analysis which has not been considered in Ref.~\cite{DiValentino:2021imh}. 
Second, the ACT-DR4+WMAP+BAO provides a weak upper limit of $\Mnu<0.19\eV$ at $68\%$ CL, which is also consistent with our constraint \eqref{Mnu}.

To evaluate the performance of $\Lambda$CDM+$\Mnu$ and $\Lambda$CDM models,
we show the difference in the best-fit $\chi^2$ values, $\Delta\chi^2_{\min}$, to different data sets in Tab. \ref{tab:imp1}.
\begin{table}[!t]
    \renewcommand{\arraystretch}{1.1}
    \centering
    \begin{tabular} {| c || c |c | c | c |}
        \hline
        \multirow{2}{*}{Parameter}  & \multirow{2}{*}{$\rm Base$} &
		\multirow{2}{*}{$\rm Base\!+\!LSS$} &
		$\rm \!Base\!+\!LSS\!$ &
		$\rm \!Base\!+\!LSS\!$ \\
        & & & $\rm +S_8$ & $\rm +S_8\!+\!SN$\\
        \hline
        \hline
        $\Delta\chi^2_{\rm min}$ & $-1.22$  &
        $-4.37$ &
        $-6.22$ &
        $-5.91$  \\ 
        $\Delta{\rm AIC}$ & $+0.78$ &
        $-2.37$ &
        $-4.22$ &
        $-3.91$ \\ 
        \hline
    \end{tabular}
    \caption {The $\Delta\chi^2_{\rm min}$ and $\Delta{\rm AIC}$ values between the best-fit $\Lambda$CDM+$\Mnu$ and $\Lambda$CDM models to different data sets.}
    \label{tab:imp1}
\end{table}
As the $\Delta\chi^2_{\min}$ is expected to follow the $\chi^2$ distribution with one degree of freedom (the number of extra parameters introduced by $\Lambda$CDM+$\sum m_\nu$), we compute the associated confidence interval at which the $\Lambda$CDM+$\sum m_\nu$ model is preferred over $\Lambda$CDM. For the Base data analysis we found an insignificant ($1.1\sigma$) improvement in the $\Lambda$CDM+$\sum m_\nu$ fit over $\Lambda$CDM. The $\rm Base+LSS+S_8+SN$ data shows a $2.4\sigma$ preference for the $\Lambda$CDM+$\sum m_\nu$ scenario. The improvement in the $\Lambda$CDM+$\sum m_\nu$ fit over $\Lambda$CDM is mainly driven by the LSS data: $\Delta\chi^2_{\rm LSS}=-4.36$/$-3.51$/$-2.61$ for the $\rm Base+LSS$/$\rm Base+LSS+S_8$/$\rm Base+LSS+S_8+SN$ analyses. This effect can be attributed to a systimatically lower value of $\sigma_8$ inferred from the BOSS DR12 data~\cite{Philcox:2021kcw,Ivanov:2019pdj}. Massive neutrinos suppress the growth of linear density field on scales smaller than neutrino free-streaming length that moves the inferred cosmological constraints into better agreement with the BOSS measurements.

To further assess the robustness of the overall preference for the $\Lambda$CDM+$\sum m_\nu$ scenario over $\Lambda$CDM, we use the Akaike Information Criteria (AIC)~\cite{1100705} defined by ${\rm AIC}=\chi^2_{\rm min}+2N_p$, where $N_p$ is the number of free parameters in the model. Then, the difference $\Delta{\rm AIC}=\Delta\chi^2_{\rm min}+2\Delta N_p$ sets a penalty proportional to the number of extra parameters introduced by a more complex model ($\Delta N_p=1$ for $\Lambda$CDM+$\sum m_\nu$). 
The Base data shows a preference in favor of the standard $\Lambda$CDM model. In contrast, for the $\rm Base+LSS+S_8+SN$ analysis we found $\Delta{\rm AIC}=-3.91$, which corresponds to a positive preference for the $\Lambda$CDM+$\sum m_\nu$ scenario over $\Lambda$CDM. Our result is stable against removing $\rm S_8$ or SN data sets. This reinforces that the LSS data plays a crucial role in the neutrino mass measurement.

\subsection{$\rm \Lambda$CDM+$\Neff$}
\label{sec:nuLCDM2}

\begin{figure}
    \begin{center}
        \includegraphics[width=1.0\columnwidth]{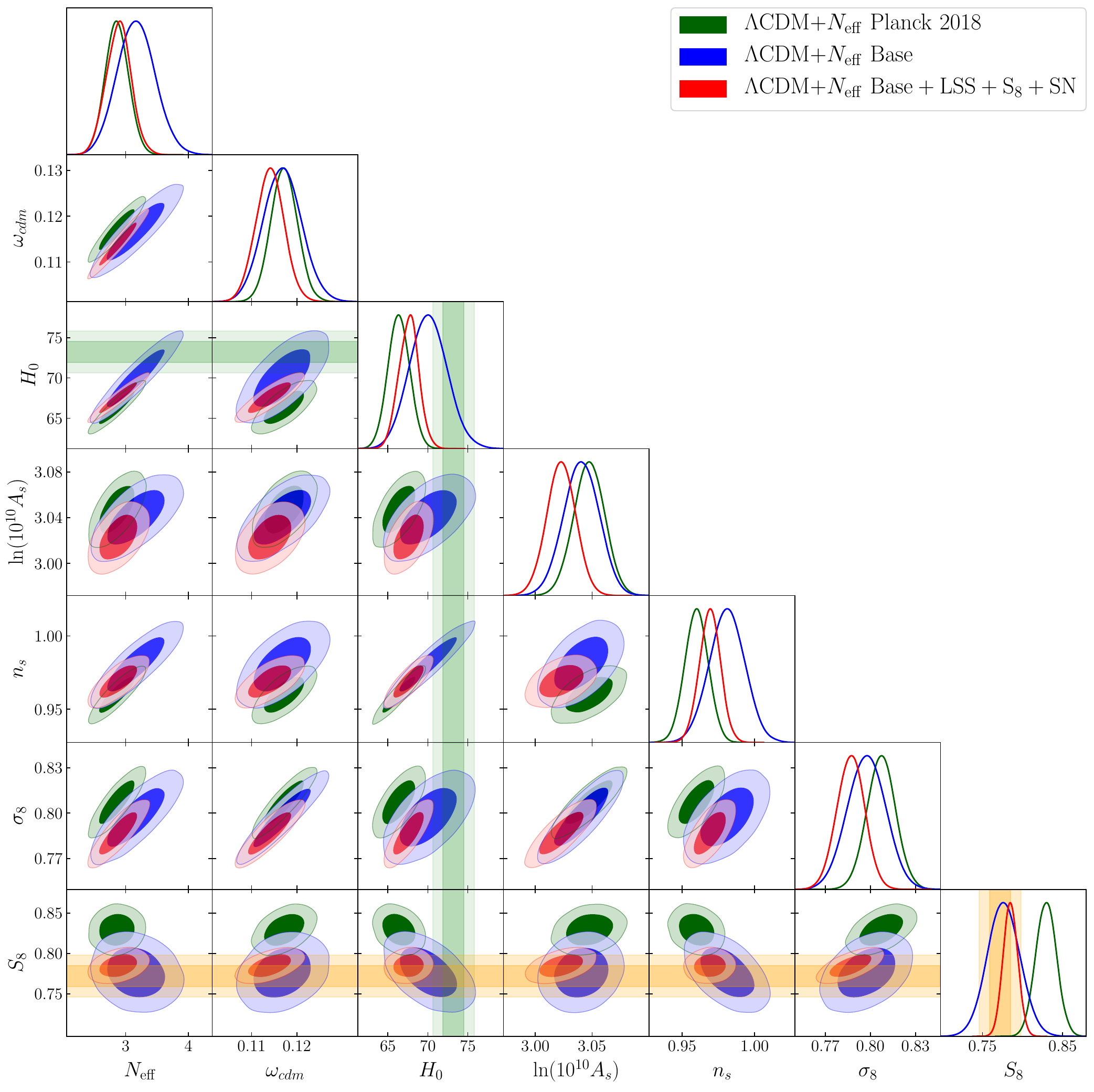}
        \caption {Marginalized 2d posterior distributions of the cosmological parameters in the $\rm \Lambda$CDM+$\Neff$ model for Planck 2018 (green), Base (blue) and $\rm Base+LSS+S_8+SN$ (red) data sets. The Gaussian prior on $\tau$ \eqref{tau} is set. The yellow bands represent $1\sigma$ and $2\sigma$ constraints on $S_8$ \eqref{S8} coming from the photometric surveys (DES-Y3, KiDS, HSC), whereas the green bands refer to the $H_0$ \eqref{H0} measurement reported by the SH0ES collaboration.}
        \label{fig:4}
    \end{center}
\end{figure}

We proceed with the $\rm \Lambda$CDM+$\Neff$ scenario. 
Tab.\,\ref{tab:4} presents the 1d marginalized constraints on cosmological parameters in the $\rm \Lambda$CDM+$\Neff$ model.
\begin{table}
	\renewcommand{\arraystretch}{1.2}
    \small
	\centering
	\begin{tabular} {|c||c|c|c|c|c|}
		\hline
		& \multicolumn{5}{c|}{$\rm \Lambda$CDM+$\Neff$}\\
		\hline
		\hline
		\multirow{2}{*}{\!\!\! Parameter\!} & \multirow{2}{*}{$\rm Planck\,2018$} & \multirow{2}{*}{$\rm Base$} & \multirow{2}{*}{$\rm Base\!+\!LSS$} & $\rm \!Base\!+\!LSS\!$
		& $\rm \!Base\!+\!LSS\!$\\
		& & & & $\rm +S_8$ & $\rm +S_8\!+\!SN$ \\
		\hline
		$100\,\omega_b$ & $2.227\pm0.021$ &
		$2.263\pm0.029$ &
		$2.236\pm0.021$ &
		$2.237\pm0.021$ & $2.244\pm0.021$ \\ 
		$10\,\omega_{cdm}$ & $1.172\pm0.029$ &
		$1.168\pm0.042$ &
		$1.161\pm0.037$ &
		$1.138\pm0.033$ & $1.150\pm0.030$ \\ 
		$H_0$ & $66.38\pm1.35$ &
		$70.00\pm2.37$ &
		$67.52\pm1.36$ &
		$67.47\pm1.36$ & $68.02^{+0.94}_{-1.08}$ \\ 
		$\tau$ & $0.059\pm0.005$ &
		$0.058\pm0.005$ &
		$0.055\pm0.005$ &
		$0.054\pm0.005$ & $0.053\pm0.005$ \\ 
		$\!{\rm ln}(10^{10} A_s)\!$ & $3.048\pm0.014$ &
		$3.040\pm0.016$ &
		$3.030\pm0.014$ &
		$3.022\pm0.013$ & $3.025\pm0.013$ \\ 
		$n_s$ & $0.960\pm0.008$ &
		$0.981\pm0.012$ &
		$0.969\pm0.008$ &
		$0.969\pm0.008$ & $0.971\pm0.007$ \\ 
		$\Neff$ & $2.86\pm0.19$ &
		$3.16\pm0.30$ &
		$2.95\pm0.22$ &
		$2.87\pm0.21$ & $2.95\pm0.19$ \\ 
		\hline   
		$\rdrag$ & $148.87\pm1.89$ &
		$147.06\pm2.78$ &
		$148.60\pm2.25$ &
		$149.66\pm2.17$ & $149.17\pm1.56$ \\    
		$\Omega_m$ & $0.318\pm0.009$ &
		$0.287\pm0.014$ &
		$0.305\pm0.007$ &
		$0.301\pm0.007$ &	$0.298\pm0.006$ \\ 
		$\sigma_8$ & $0.807\pm0.010$ &
		$0.797\pm0.013$ &
		$0.795\pm0.011$ &
		$0.786\pm0.010$ &	$0.789\pm0.009$  \\ 
		$S_8$ & $0.831\pm0.013$ &
		$0.779\pm0.021$ &
		$0.802\pm0.013$ &
		$0.787\pm0.009$ &	$0.787\pm0.009$  \\
		\hline
	\end{tabular}
	\caption {Parameter constraints in the $\rm \Lambda$CDM+$\Neff$
		model with $1\sigma$ errors. 
  Recall, the Base data set includes $\rm \!PlanckTT\text{-}low\ell\!+\!SPT$-$\rm 3G\!+\!Lens$.
  }
	\label{tab:4}
\end{table}
Fig.\,\ref{fig:4} shows the 2d posterior distributions for the various analyses.

Let us highlight the key differences between our approach and the full Planck analysis. As previously, the Base data suggests a significantly lower $S_8$ compared to the Planck 2018 analysis that makes our approach entirely consistent with the low-redshift probes \eqref{S8}. 
For the effective number of relativistic degrees of freedom we found $\Neff=3.16\pm0.30$. While our estimate agrees with the Planck 2018 result~\cite{Planck:2018vyg}, it allows for considerably larger values of $\Neff$ which leads to a moderately higher $H_0$.

It is interesting to compare our constraint with the result of the full Planck data analysis after marginalizing over the lensing information contained in the CMB spectra.
Allowing arbitrary gravitational lensing in the Planck TT,TE,EE maps one gets $H_0=68.2\pm1.6\kms$~\cite{Motloch:2019gux}. This estimate agrees well with both the Base and Planck 2018 data analyses. Unlike the $\Mnu$ limit, the error bar on $H_0$ only moderately increases compared to the Planck 2018 result. 
This effect can be attributed to the fact that the $H_0$ constraint is mainly determined from the position of the first acoustic peak which is barely affected by the CMB gravitational lensing.

Let us explore the cosmological constraints inferred from the $\rm Base+LSS$ data.
Adding the LSS information significantly improves the cosmological measurements. It also provides a $1\sigma$ lower value of $H_0$ being consistent with the Planck 2018 constraint.

Next, we add the $S_8$ data. 
Adding the $S_8$ data significantly improves only the accuracy of $S_8$ measurement, while the other parameter constraints remain largely unchanged. 

We eventually consider the supernova measurements. Adding the SN data shrinks the error bars on $H_0$ and $\rdrag$ parameters. The reason is that the supernova sample fixes the background cosmology at low redshifts which helps to lift the degeneracies between the $\Neff$ and the $\Lambda$CDM cosmological parameters. 
The $\rm Base+LSS+S_8+SN$ analysis brings
\be
\Neff=2.95\pm0.19\,.
\ee
This measurement is consistent with the Planck 2018 result. 
We conclude that the enhanced smoothing of acoustic peaks in the Planck data does not affect the $\Neff$ constraint.  Our results are in good agreement with the Planck data analysis based on the “unlensed” CMB power spectra~\cite{Motloch:2019gux}.
We also found a $1\sigma$ higher value of the Hubble parameter, $H_0=68.02^{+0.94}_{-1.08}\kms$, which moderately alleviates the Hubble tension down to the $3.2\sigma$ level, cf. with \eqref{H0}.

To assess the preference for the $\Lambda$CDM+$\Neff$ model over $\Lambda$CDM, we report the $\Delta\chi^2_{\rm min}$ and $\Delta{\rm AIC}$ values in Tab. \ref{tab:imp2}.
\begin{table}[!t]
    \renewcommand{\arraystretch}{1.1}
    \centering
    \begin{tabular} {| c || c |c | c | c |}
        \hline
        \multirow{2}{*}{Parameter}  & \multirow{2}{*}{$\rm Base$} &
		\multirow{2}{*}{$\rm Base\!+\!LSS$} &
		$\rm \!Base\!+\!LSS\!$ &
		$\rm \!Base\!+\!LSS\!$ \\
        & & & $\rm +S_8$ & $\rm +S_8\!+\!SN$\\
        \hline
        \hline
        $\Delta\chi^2_{\rm min}$ & $-1.45$  &
        $+0.1$ &
        $-1.26$ &
        $-1.68$  \\ 
        $\Delta{\rm AIC}$ & $+0.55$ &
        $+2.1$ &
        $+0.74$ &
        $+0.32$ \\ 
        \hline
    \end{tabular}
    \caption {The $\Delta\chi^2_{\rm min}$ and $\Delta{\rm AIC}$ values between the best-fit $\Lambda$CDM+$\Neff$ and $\Lambda$CDM models to different data sets.}
    \label{tab:imp2}
\end{table}
In most scenarios, the $\rm \Lambda$CDM+$\Neff$ model yields a slightly better fit to the data than $\Lambda$CDM. According to the AIC, the $\Lambda$CDM model is always preferred against $\rm \Lambda$CDM+$\Neff$.

\section{Phantom Dark Energy}
\label{sec:PDE}

In this section we explore a dark energy scenario with phantom crossing.

\subsection{Model description}
\label{sec:PDEmod}

We assume that the dark energy equation of state crosses the phantom divide, $w_{\DE} =-1$, during the course of its evolution.
According to the energy conservation equation for the dark energy fluid, $\frac{\displaystyle d\rho_{\DE}}{\displaystyle dt}=-3a^{-1}(1+w_{\DE})\rho_{\DE}$, the dark energy density should pass through an extremum at some time where $\frac{\displaystyle d\rho_{\DE}}{\displaystyle dt}$ changes its sign. Following~\cite{DiValentino:2020naf}, we expand the dark energy density around its extremum at $a=a_m$,
\begin{equation}
\label{rhoPDE}
 \rho_{\DE}(a) = \rho_0[1 + \alpha(a-a_m)^2 + \beta(a-a_m)^3].
\end{equation}
where $\rho_0$ normalizes the dark energy density, $a_m$ defines the moment when the dark energy density passes through the extremum and $\alpha,\,\beta$ describe the course of phantom crossing. Here we choose the present scale factor to be $a_0=1$. We also restrict ourselves up to the third order in the Taylor expansion because higher order terms can not be tightly measured with the present data~\cite{DiValentino:2020naf}.

Inserting \eqref{rhoPDE} into the Friedman equation for the flat space 
\begin{equation}
 H^2 = \frac{8\pi G}{3}[\rho_m + \rho_{rad} + \rho_{\DE}],
\end{equation}
we get the following evolution for the Hubble parameter,
\begin{equation}
\label{HubblePDE}
 \frac{H^2(a)}{H_0^2} = \frac{\Omega_{m}}{a^3} + \frac{\Omega_{rad}}{a^4} +
 \left(1-\Omega_{m}-\Omega_{rad}\right)
 \frac{1 + \alpha(a-a_m)^2 + \beta(a-a_m)^3}
 {1+ \alpha(1-a_m)^2 + \beta(1-a_m)^3},
\end{equation}
and for the dark energy equation of state,
\begin{equation}
\label{wPDE}
 w_{\DE}(a) = -1 - \frac{a[2\alpha(a-a_m) + 3\beta(a-a_m)^2]}{3[1+\alpha(a-a_m)^2 + \beta(a-a_m)^3]}.
\end{equation}
At early times ($a\to0$), the equation of state approaches $w_{\DE}=-1$ showing the cosmological constant behaviour. It demonstrates that the dark energy equation of state is well defined at very early times.

The PDE model is parameterised with the set of three parameters, $(a_m,\alpha,\beta)$. The PDE scenario reduces to the $\Lambda$CDM one when $\alpha=\beta=0$. Note that the parameterization \eqref{rhoPDE} allows for a negative dark energy density $\rho_{\DE}$ that introduces greater flexibility to fit the data (see e.g.~\cite{Wang:2018fng,BOSS:2014hwf,Poulin:2018zxs,Dutta:2018vmq,Bernardo:2021cxi}).

We implement the background evolution of the PDE through \eqref{HubblePDE} and \eqref{wPDE} while assuming no extra sources of clustering except for matter. We vary 9 cosmological parameters: the three PDE $(\alpha,\beta,a_m)$ and the six standard $\Lambda$CDM ($\omega_{cdm}$, $\omega_b$, $H_0$, $\ln(10^{10}A_s)$, $n_s$, $\tau$). We impose the same flat uniform priors on PDE parameters as in Ref.~\cite{DiValentino:2020naf},
 \begin{equation}
 \label{PDEprior}
 \begin{aligned}
     a_m\in[0,1],\qquad \alpha\in[0,30],\qquad \beta\in[0,30].
 \end{aligned}
 \end{equation}

\subsection{Parameter constraints}
\label{sec:PDEcon}

\begin{figure}
    \begin{center}
        \includegraphics[width=1.0\columnwidth]{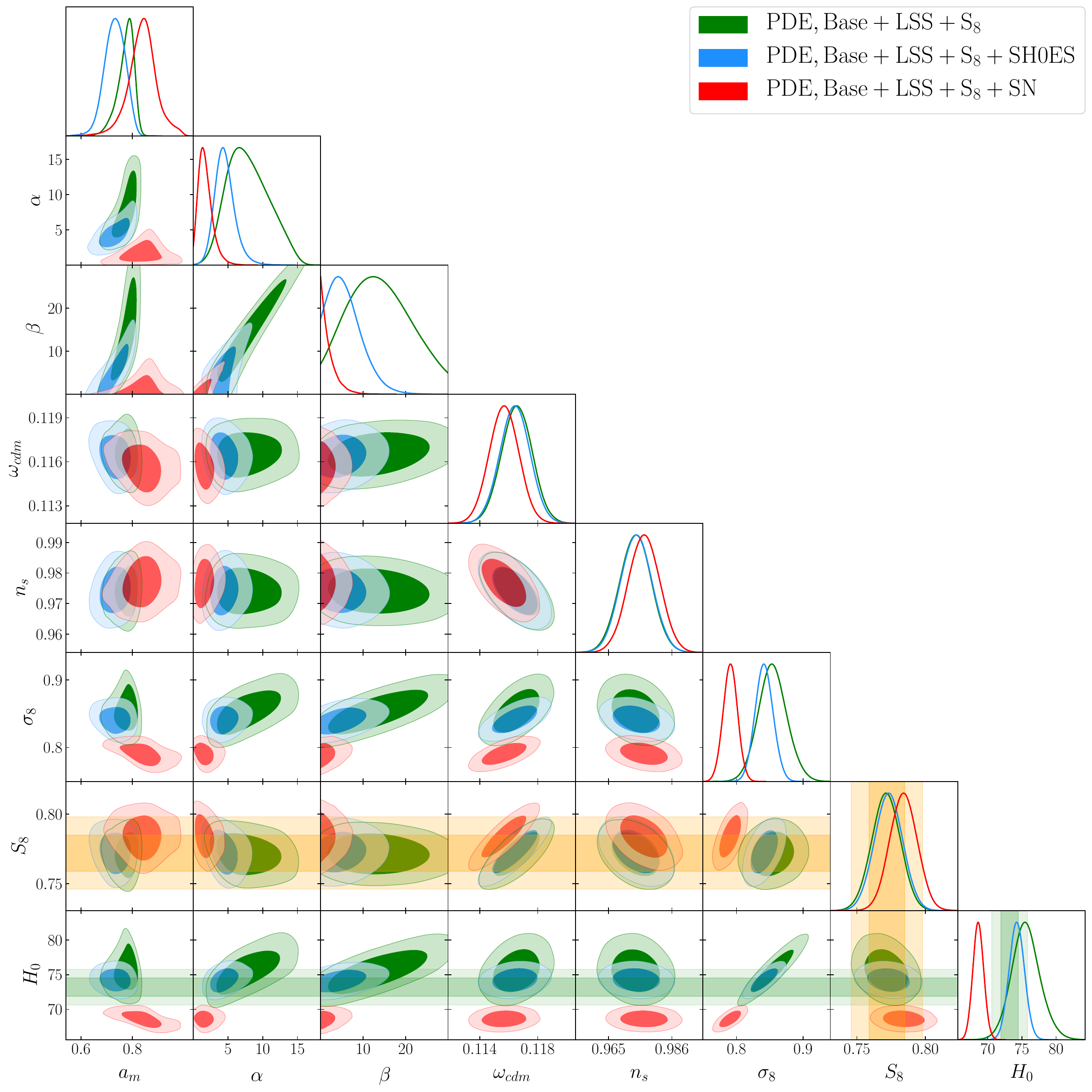}
        \caption {Marginalized 2d posterior distributions of the cosmological parameters in the PDE model for the $\rm Base+LSS+S_8$ (green), $\rm Base+LSS+S_8+SH0ES$ (blue) and $\rm Base+LSS+S_8+SN$ (red) data sets. The Gaussian prior on $\tau$ \eqref{tau} is used. The yellow bands represent $1\sigma$ and $2\sigma$ constraints on $S_8$, see Eq.\,\eqref{S8}, it comes from the photometric surveys (DES-Y3, KiDS, HSC), whereas the green bands refer to the Hubble constant $H_0$ measurement \eqref{H0} reported by the SH0ES collaboration.}
        \label{fig:5}
    \end{center}
\end{figure}

Tab.\,\ref{tab:5} presents the 1d marginalized constraints on cosmological parameters for different data set combinations in the PDE model.
\begin{table}
	\renewcommand{\arraystretch}{1.2}
    \small
	\centering
	\begin{tabular} {|c|| c |c | c |c|}
		\hline
		& \multicolumn{4}{c|}{Phantom-crossing Dark Energy (PDE)} \\
		\hline
		\hline
		\multirow{2}{*}{\!\!\! Parameter\!} & \multirow{2}{*}{$\rm Base\!+\!LSS$} & $\rm \!Base\!+\!LSS\!$
		& $\rm \!Base\!+\!LSS\!$ & $\rm \!Base\!+\!LSS\!$\\
		&  & $\rm +S_8$ & $\rm \!+S_8\!+\!SH0ES\!$ & $\rm +S_8\!+\!SN$ \\
		\hline
		$a_m$ & $0.774(0.757)^{+0.037}_{-0.020}$ &
		$0.774(0.772)^{+0.038}_{-0.020}$ &
		$0.735(0.778)^{+0.044}_{-0.036}$ &
		$0.839(0.822)^{+0.048}_{-0.049}$\\ 
		$\alpha$ & $8.1(6.6)^{+2.6}_{-3.7}$ &
		$8.0(7.6)^{+2.5}_{-3.6}$ &
		$4.7(6.3)^{+1.1}_{-1.6}$ &
		$1.8(1.3)^{+0.6}_{-1.2}$\\ 
		$\beta$ & $14.2(11.0)^{+6.7}_{-8.7}$ &
		$14.1(11.7)^{+6.8}_{-8.4}$  &
		$6.2(11.2)^{+2.2}_{-5.4}$ &
		$<2.3(0.0)$ 
        \\ 
		$100\,\omega_b$ & $2.246\pm0.019$ &
		$2.245\pm0.018$ &
		$2.247\pm0.018$ &
		$2.252\pm0.018$\\ 
		$10\,\omega_{cdm}$ & $1.165\pm0.015$ & 
		$1.166\pm0.011$ &
		$1.164\pm0.010$ &
		$1.157\pm0.010$\\ 
		$H_0$ & $75.70(75.52)^{+2.05}_{-2.32}$ & 
		$75.60(75.36)^{+1.93}_{-2.12}$ &
		$74.26(74.97)^{+1.11}_{-1.12}$ &
		$68.61(68.24)^{+0.78}_{-0.78}$\\ 
		$\tau$ & $0.057\pm0.005$ & 
		$0.057\pm0.005$ &
		$0.057\pm0.005$ &
		$0.056\pm0.005$\\ 
		$\!{\rm ln}(10^{10} A_s)\!$ & $3.038\pm0.012$ & 
		$3.038\pm0.011$ &
		$3.037\pm0.011$ &
		$3.033\pm0.011$\\ 
		$n_s$ & $0.974\pm0.006$ & 
		$0.974\pm0.005$ &
		$0.974\pm0.005$ &
		$0.977\pm0.005$\\ 
		\hline   
		$\rdrag$ & $147.93\pm0.38$ &
		$147.91\pm0.31$ &
		$147.93\pm0.31$ &
		$148.08\pm0.30$\\   
		$\Omega_m$ & $0.244\pm0.015$ & 
		$0.245\pm0.013$ &
		$0.253\pm0.008$ &
		$0.295\pm0.007$\\ 
		$\sigma_8$ & $0.854\pm0.022$ & 
		$0.855\pm0.021$ &
		$0.842\pm0.014$ &
		$0.791\pm0.011$\\ 
		$S_8$ & $0.770\pm0.017$ & 
		$0.771\pm0.010$ &
		$0.773\pm0.010$ &
		$0.784\pm0.010$\\
		\hline 
	\end{tabular}
	\caption {Parameter estimates (mean value with $1\sigma$ error bars and best fit value in the parentheses) in the phantom-crossing dark energy model. 
    The upper limits are given at $95\%$ CL.
    }
	\label{tab:5}
\end{table}
Fig.\,\ref{fig:5} shows the final 2d posterior distributions.
We do not show the Base parameter constraints because the CMB data alone can not break degeneracies present in the PDE model, hence the dark energy parameters become largely unconstrained~\cite{DiValentino:2020naf}.

We start with the Base+LSS analysis. For the dark energy parameter we found $a_m = 0.774^{+0.037}_{-0.020}$, $\alpha=8.1^{+2.6}_{-3.7}$ and $\beta=14.2^{+6.7}_{-8.7}$. This shows an indication at more than $3\sigma$ for phantom crossing in the dark energy sector.~\footnote{Note that the posterior distributions of $\alpha$ and $\beta$ parameters are highly non-gaussian.} 
We obtain $H_0=75.70^{+2.05}_{-2.32}\kms$. This constraint is now perfectly consistent within one standard deviation with the SH0ES measurement and deviates from the Planck value~\cite{Planck:2018vyg} by $3.7\sigma$. The increase of the $H_0$ parameter is due to its positive correlation with $\alpha$, $\beta$ as shown in Fig.\,\ref{fig:5}.~\footnote{Fig.\,\ref{fig:5} illustrates the posterior distributions for the data which includes the $S_8$ set. However, as we will see shortly, the $S_8$ data only minimally affects the parameter constraints.} Importantly, the Base+LSS analysis predicts a substantially lower matter density parameter compared to $\Lambda$CDM, namely $\Omega_m=0.244\pm0.015$. This result can be attributed to the phantom period of the dark energy evolution within which the $\rho_{\DE}$ increases over time resulting in a lower $\Omega_m$~\cite{Capozziello:2018jya}. 
We also found a somewhat higher value of $\sigma_8$ that can be explained by an anti-correlation between $\sigma_8$ and $H_0$ parameters shown in Fig. \ref{fig:5}.
{\it We emphasize that our analysis does not include any priors on late-time parameters but its result is fully consistent with the direct measurements of $S_8$ \eqref{S8} and $H_0$ \eqref{H0} in the late Universe.}

Next, we proceed with the $S_8$ data. 
Adding the $S_8$ information hardly impacts the posterior distributions of the PDE parameters. At the same time, it significantly improves the accuracy of the $S_8$ and $\omega_{cdm}$ measurements.
Remarkably, the mean value of $S_8$ remains virtually unchanged that illustrates an excellent agreement between the $\rm Base+LSS$ and $\rm S_8$ data sets.

Since the $\rm Base+LSS+S_8$ and SH0ES data are in agreement now, we can combine them safely together. 
We apply the entire distance ladder approach which closely reproduces the SH0ES analysis~\cite{Riess:2020fzl} instead of simply imposing a Gaussian constraint on $H_0$ (for detail see Appendix \ref{app:Prior}).
Our joint data analysis demonstrates a decisive evidence for phantom crossing in the dark energy sector, $a_m=0.735^{+0.044}_{-0.036}$, $\alpha=4.7^{+1.1}_{-1.6}$ and $\beta=6.2^{+2.2}_{-5.4}$. As shown in Fig.\,\ref{fig:5}, the SH0ES likelihood efficiently breaks the degeneracy between the PDE and standard cosmological parameters that results in significantly tighter constraints on the dark energy parameters.
We found $H_0=74.26^{+1.11}_{-1.12}\kms$ which is two-times more accurate than the $\rm Base+LSS+S_8$ constraint (without SH0ES). This result can be explained by the positive correlations between $\alpha$, $\beta$ and $H_0$ parameters. The $S_8$ constraint only barely changes being in an excellent agreement with the direct measurements.

We also explore the Pantheon sample.
The supernova absolute magnitude \eqref{Msn} that is used to derive the local $H_0$ constraint is not compatible with $M_B$ that is necessary to fit CMB, BAO and SN data (see e.g.~\cite{Camarena:2021jlr,Efstathiou:2021ocp}). Given this reason, we combine $\rm Base+LSS+S_8$ and SN data (without SH0ES). We found that the $\rm Base+LSS+S_8+SN$ data reduces the preference for phantom crossing in the dark energy sector leading to $a_m=0.839^{+0.048}_{-0.049}$, $\alpha=1.8^{+0.6}_{-1.2}$ and $\beta<2.3$ (at $95\%$ CL). But it still suggests a mild evidence for a transition in the dark energy density. 
The matter density parameter is shifted higher upon adding the SN information, namely $\Omega_m=0.295\pm0.007$, which happens to be more compatible with the Planck value~\cite{Planck:2018vyg}. Our final constraints on $S_8$ and $H_0$ parameters in the PDE scenario are 
\be
\label{H0pde}
S_8=0.784\pm0.010   \qquad\qquad   H_0=68.61\pm0.78\kms
\ee
The $S_8$ constraint is in good agreement with the low-redshifts probes. However, the $H_0$ value is significantly lower which exhibits a $3.1\sigma$ tension with the SH0ES constraint. This is because the data put strong enough constraints on background evolution which does not allow higher $H_0$ values. Our result agrees with the previous studies~\cite{Lemos:2018smw,Poulin:2018zxs,Roy:2022fif,Dinda:2021ffa,Keeley:2022ojz}, which show through the late Universe reconstruction that CMB, BAO and SN data do not allow for a higher expansion rate at low redshifts.
This conclusion has been recently reaffirmed in the context of the late Universe scenarios with a sudden transition in dark energy sector~\cite{Benevento:2020fev,Efstathiou:2021ocp,Camarena:2021jlr}.

\subsection{Discussion}
\label{sec:PDEdis}

In our analysis we combine $\rm Base+LSS+S_8$ either with the SN catalog or with the SH0ES measurement.
The basic reason is that the supernova absolute magnitude that is necessary to fit CMB, BAO and SN data is in a strong disagreement with the local astrophysical calibration via Cepheids. For instance, the $\rm Base+LSS+S_8+SN$ data predicts the following absolute magnitude of supernova
\be
\label{MsnCMB}
M_B=-19.414\pm0.018\,.
\ee
Our constraint agrees with the results from the standard inverse-distance ladder analysis~\cite{DES:2018rjw,Feeney:2018mkj} as well as a novel non-parametric approach~\cite{Camarena:2019rmj}, however it is in a $4.5\sigma$ tension with the Cepheid-based measurement \eqref{Msn}. This robustly shows that 
the SN calibration produced by CMB and BAO is not compatible with the SH0ES calibration. 
Thus, one can not combine the $\rm Base+LSS+S_8+SN$ and SH0ES data together until the source of ‘supernova absolute magnitude tension’ had not been elucidated (see e.g.~\cite{Camarena:2021jlr,Efstathiou:2021ocp}). 
In what follows, we discuss the potential origins of this tension.

The ‘supernova absolute magnitude tension’ may be caused by astrophysical systematic effects present in the distance ladder measurement. 
For instance, average standardized magnitudes of the supernova in Cepheid hosts and those in the Hubble flow sample may differ due to host-galaxy environmental effects.
The recent analyses~\cite{NearbySupernovafactory:2013qtg,Rigault:2014kaa} demonstrate that local age tracers are strongly correlated with the standardized supernova magnitude.
Using the classification based on the specific star formation rate, the study of the supernova Factory sample~\cite{NearbySupernovaFactory:2018qkd} shows that the supernova in predominantly younger environments are fainter than those in predominantly older environments by $\Delta M_B=+0.163\pm0.029$.~\footnote{The amplitude of this effect is two-time larger than the global host-stellar mass correction currently used in cosmological analyses~\cite{Pan-STARRS1:2017jku,Riess:2020fzl}. When fitting for the specific star formation rate and global stellar mass biases simultaneously, the environment-age offset remains very significant $\Delta M_B=0.129\pm0.032$, for detail see~\cite{NearbySupernovaFactory:2018qkd}.}
Importantly, the supernova from the Cepheid calibrator sample favors young stellar populations whereas those in the Hubble flow sample do not~\cite{Rigault:2014kaa}. 
It implies that the Cepheid-based calibration \eqref{Msn} may be overestimated by a certain amount that could potentially explain at least part of the ‘supernova absolute magnitude tension’~\cite{NearbySupernovaFactory:2018qkd,Briday:2021rrm}.
The importance of local supernova environmental studies remains highly debated, however (see e.g. Refs.~\cite{Jones:2015uaa,FSS:2018cey}), specifically the impact of such an astrophysical bias on the $H_0$ measurements~\cite{Riess:2016jrr,Riess:2019cxk}.

Another possible source of astrophysical systematics is related to the Cepheid calibration. 
The Ref.~\cite{Perivolaropoulos:2021bds} finds a $3\sigma$ evidence for a transition in either the color-luminosity relation or the Cepheid absolute magnitude, at a distance in the range between 10 and 20 Mpc. 
The models where these parameters are fitted by two universal values (one for low galactic distances and one for high galactic distances) are strongly favoured over the baseline analysis where no variation is allowed for the Cepheid empirical parameters. 
A transition in the color-luminosity relation may be attributed to a variation of dust properties in individual galaxies~\cite{Mortsell:2021nzg,Mortsell:2021tcx}, whereas the shift of the Cepheid absolute magnitude could be induced by an abrupt change of fundamental physic~\cite{Perivolaropoulos:2021bds}. 
These results have interesting implications in the context of the $H_0$ measurements. 
Allowing for the Cepheid color-luminosity relation to vary between galaxies, the $H_0$ constraints inferred from individual anchors ranges from $H_0=68.1\pm3.5 \kms$ to $H_0=76.7\pm2.0\kms$~\cite{Mortsell:2021nzg}. 
Next, the Ref.~\cite{Mortsell:2021tcx} investigates the sensitivity of the $H_0$ constraint to color excess cuts in the Cepheid data. By removing the reddest Cepheids in order to minimize the impact of dust extinction, they obtain $H_0=68.1\pm2.6\kms$.

The ‘supernova absolute magnitude tension’ may eventually hint at a possible failure in the standard cosmological scenario and the necessity for new physics. 
Since the two measurements, \eqref{Msn} and \eqref{MsnCMB}, are performed at different redshift ranges,~\footnote{The local astrophysical measurement \eqref{Msn} was calibrated by Cepheids at $z<0.01$ whereas the estimate \eqref{MsnCMB} was obtained at $z>0.01$ using the sound horizon at last scattering as a standard ruler.} this mismatch may indicate a transition in the absolute magnitude with amplitude $\Delta M_B\simeq-0.2$ at $z\lesssim0.01$.
Such transition can be achieved by a sudden change of the value of the effective gravitational constant which modifies the supernova intrinsic luminosity, for detail see~\cite{Marra:2021fvf,Alestas:2020zol}.~\footnote{By effective gravitational constant we refer to the strength of gravitational interactions rather than the Planck mass which determines the expansion rate of the Universe.} Ref.~\cite{Marra:2021fvf} shows that a reduction of the effective gravitational constant at $z>0.01$ by about $10\%$ will bring the Cepheid-based absolute magnitude of supernova \eqref{Msn} into agreement with the CMB calibration \eqref{MsnCMB}.
This scenario also addresses the $S_8$ tension due to the lower value of the gravitational constant at early times.
The required amplitude of the $M_B$ transition can be smaller if the transition in gravity sector is accompanied by a rapid change in the dark energy equation of state, for detail see~\cite{Alestas:2020zol}.

We conclude that the ‘supernova absolute magnitude tension’ may be affected by astrophysical systematics and/or new physics in gravity sector. The purpose of this paper is not to explore the astrophysical effects or modifications of gravity. Therefore, we adopt an agnostic approach for a possible value of the supernova standardized magnitude. To do so, we analyze the $\rm Base+LSS+S_8+SN$ and $\rm Base+LSS+S_8+SH0ES$ data separately. This makes our analysis blind to the actual $M_B$ value in the Hubble flow sample. We emphasize that the models which modify only the late Universe expansion is not capable of solving this tension~\cite{Alestas:2020zol}.

\begin{figure}[!t]
   \begin{center}
        \includegraphics[width=0.7\columnwidth]{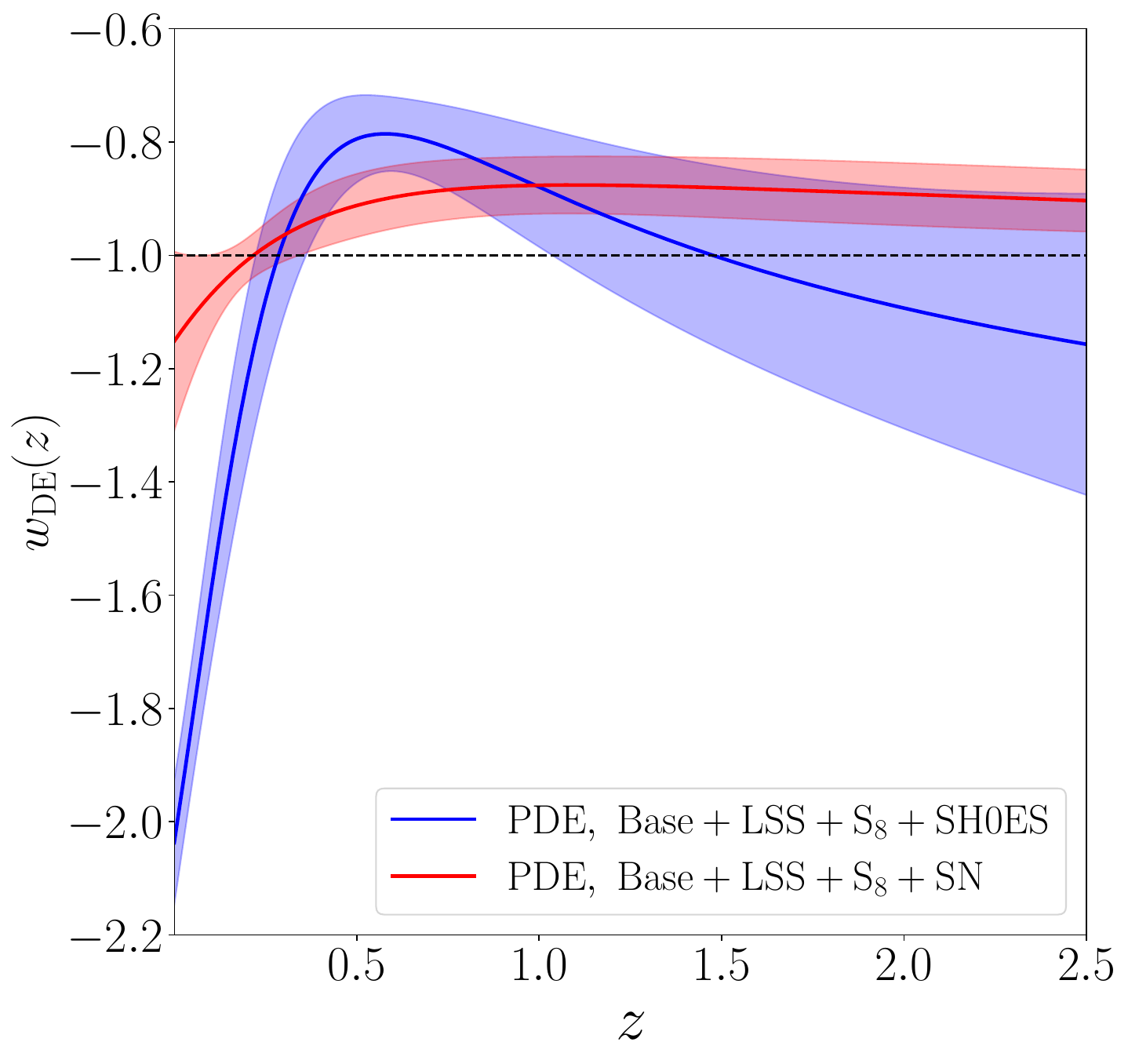} 
        \caption {Behaviour of the dark energy equation of state computed for the PDE best-fit models to the $\rm Base+LSS+S_8+SH0ES$ (blue) and $\rm Base+LSS+S_8+SN$ (red) data sets. The shaded regions represent the $1\sigma$ error band computed under the Gaussian approximation for the best-fit model. The dashed line corresponds to the cosmological constant behaviour $w_{\DE}=-1$.}
        \label{fig:7b}
    \end{center}
\end{figure}

Fig. \ref{fig:7b} shows the $w_{\DE}(z)$ evolution for the different best-fit models.
The $\rm Base+LSS+S_8+SH0ES$ analysis suggests a strong preference for phantom crossing in the dark energy sector. Interestingly, the $w_{\DE}(z)$ crosses the phantom divide multiple times. Recall that the $\rm Base+LSS+S_8+SH0ES$ data predicts the significantly lower value of $\Omega_m$ (see Tab. \ref{tab:5}). Our results thus agree with the model-independent analysis~\cite{Capozziello:2018jya} showing that multiple phantom crossings are expected for lower values of $\Omega_m$.
In contrast, $\rm Base+LSS+S_8+SN$ data shows only a slight ($\sim1\sigma$) indication in favour of the only phantom crossing.

Another important aspect of our study relates to the BAO measurements.
Tab.~\ref{tab:5} indicates that the PDE constraint on the comoving sound horizon at the end of the baryon drag epoch, $\rdrag$, remains essentially the same as in $\Lambda$CDM.~\footnote{This happens because the late-time cosmological scenarios do not alter the Universe evolution at early times.}
But in this case, one may worry that a different late-time Universe evolution might affect the relations $D_A(z)/\rdrag$ and $\rdrag H(z)$ which are precisely measured by the BAO data. Indeed, for monotonic evolution of the dark energy density the radial BAO scale can be translated to the present-day parameter combination $\rdrag H_0$~\cite{Bernal:2016gxb}, so at constant $\rdrag$ a shift in $H_0$ would spoil the fit to the BAO measurements. However, if the behaviour of $\rho_{\DE}(z)$ is not-monotonic (akin to PDE), the final result strongly depends on a particular dynamics in the dark energy sector. 
It suggests that the model with a phantom crossing is capable of fitting the BAO distances whatever the $H_0$ value is.

\begin{figure}[!t]
    \begin{center}
        \includegraphics[width=1.0\columnwidth]{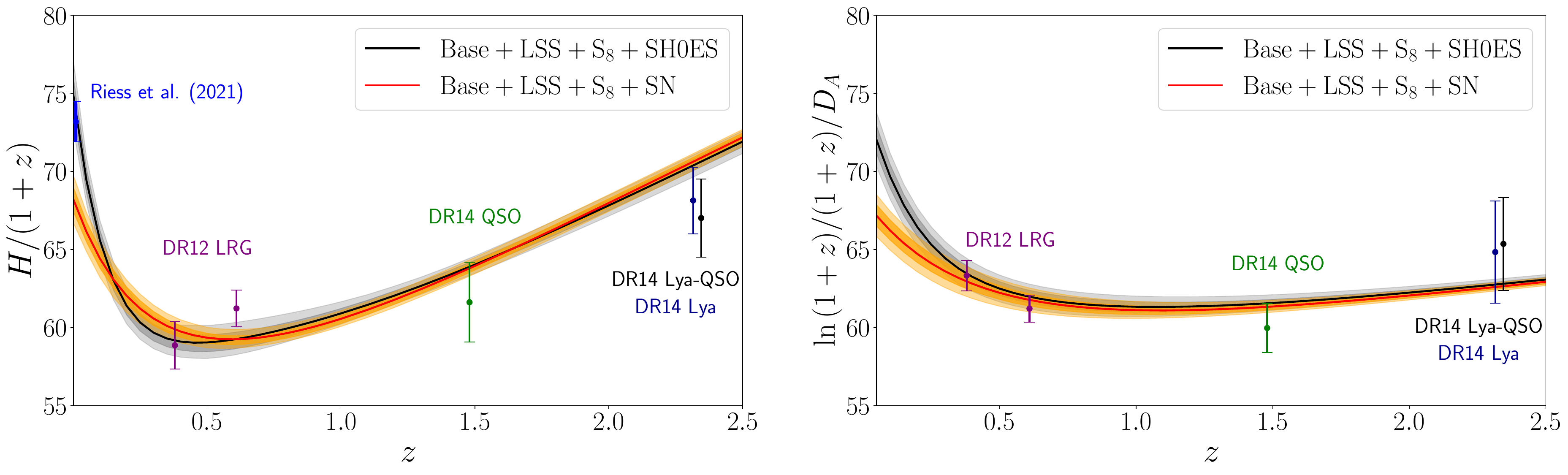}
        \caption {Behaviour of $H(z)/(1+z)$ ({\it left panel}) and $\ln(1+z)/(1+z)/D_A$ ({\it right panel}) computed for the PDE best-fit models to the $\rm Base+LSS+S_8+SH0ES$ (black line) and $\rm Base+LSS+S_8+SN$ (red line) data. The dark and light regions represent the $1\sigma$ and $2\sigma$ confidence ranges.     
        Both quantities are measured in units of km\,s$^{-1}$\,Mpc$^{-1}$. The absolute scale for the BAO measurements is set by the best-fit value of the sound horizon optimized to the Base likelihood $\rdrag=148.04$\,Mpc.}
        \label{fig:7}
    \end{center}
\end{figure}

To demonstrate the agreement with the BAO measurement, in Fig. \ref{fig:7} we show the evolution of the Hubble parameter and the inverse BAO distance for the different data combinations.
We found that the both PDE analyses agree well with the BAO measurements. Importantly, the $\rm Base+LSS+S_8+SH0ES$ model predicts $H_0\sim74\kms$ while being entirely consistent with the BAO data. It reinforces that the PDE solution is fully consistent with the BAO distances calibrated to the inferred from CMB value of $\rdrag$~\cite{DiValentino:2020naf}.

Let us compare now the dark energy evolution in Fig. \ref{fig:7b} with the results of previous analyses.
The $w_{\DE}(z)$ behaviour derived from the $\rm Base+LSS+S_8+SH0ES$ data agrees well with the model-independent analysis~\cite{Teng:2021cvy} which employs the CMB angular scale, BAO and SH0ES measurements. The multiple phantom crossings are further confirmed by the $H(z)$ reconstruction based on the Pade approximation~\cite{Capozziello:2018jya}. In turn, the $w_{\DE}(z)$ evolution predicted by the $\rm Base+LSS+S_8+SN$ data is compatible with the non-parametric Bayesian reconstraction of the dark energy evolution history~\cite{Zhao:2012aw}, which uses CMB, BAO and uncalibrated supernova sample. This is also broadly consistent with the result based on the model-independent $H(z)$ reconstruction~\cite{Roy:2022fif}. 
The mild difference can be explained by the SH0ES prior which is used by Ref.~\cite{Roy:2022fif} but absent in our analysis.

It is also interesting to compare our results with the analysis~\cite{DiValentino:2020naf} based on the same PDE framework. Using the Planck TT,TE,EE, CMB lensing, BAO, SN and SH0ES prior on $H_0$ the authors report $H_0=70.25\pm0.78\kms$.
This $H_0$ estimate is considerably higher than our $\rm Base+LSS+S_8+SN$ constraint, $H_0=68.61 \pm 0.78\kms$.
There are two main contributors to this difference. First, the Ref.~\cite{DiValentino:2020naf} includes the SH0ES prior on $H_0$ which pulls $H_0$ to higher values. Second, our analysis features the full-shape BOSS measurements and $S_8$ data which have not been considered in Ref.~\cite{DiValentino:2020naf}. We perform the direct comparison between our analysis and the full Planck approach in Appendix \ref{app:PDEcom}.

To asses the preference for the PDE model over $\Lambda$CDM, we report the $\Delta\chi^2_{\min}$ and $\Delta {\rm AIC}$ statistics to different data combinations in Tab. \ref{tab:imp3}.
\begin{table}[!t]
    \renewcommand{\arraystretch}{1.1}
    \centering
    \begin{tabular} {| c || c |c | c | c |}
        \hline
        \multirow{2}{*}{Parameter}  & 
		\multirow{2}{*}{$\rm Base\!+\!LSS$} &
		$\rm \!Base\!+\!LSS\!$ &
		$\rm \!Base\!+\!LSS\!$ &
		$\rm \!Base\!+\!LSS\!$\\
        & & $\rm +S_8$ & $\rm +S_8\!+\!SH0ES$ & $\rm +S_8\!+\!SN$\\
        \hline
        \hline
        $\Delta\chi^2_{\rm min}$ & $-10.98$ &
        $-12.24$ &
        $-21.86$ & 
        $-0.73$  \\ 
        $\Delta{\rm AIC}$ & $-4.98$ &
        $-6.24$ &
        $-17.92$ &
        $+5.27$ \\  
        \hline
        $\ln B$ & $-4.66$ &
        $-2.65$ &
        $+6.90$ &
        $-5.48$ \\ 
        \hline
    \end{tabular}
    \caption {The $\Delta\chi^2_{\rm min}$ and $\Delta{\rm AIC}$ values between the best-fit PDE and $\Lambda$CDM models to different data sets. We also show the Bayesian factors $\ln B$ calculated for the PDE model with respect
to the $\Lambda$CDM scenario. Note that the negative value of $\Delta{\rm AIC}$ indicates a preference for the PDE model, while the negative $\ln B$ shows a preference for $\Lambda$CDM.}
    \label{tab:imp3}
\end{table}
The $\rm Base+LSS$ and $\rm Base+LSS+S_8$ data show a moderate ($\gtrsim2.5\sigma$) evidence for the PDE scenario over $\Lambda$CDM. This preference is mainly driven by an improvement of the fit to the full-shape BOSS DR12 data: $\Delta\chi^2_{\rm LSS,\,full\text{-}shape}=-15.33$ and $-14.07$ for $\rm Base+LSS$ and $\rm Base+LSS+S_8$ data, respectively. 
Adding the SH0ES measurement increases the overall preference for the PDE scenario to the $4.2\sigma$ level.
In contrast, the PDE model does not significantly improve the fit to $\rm Base+LSS+S_8+SN$ compared to $\Lambda$CDM.
According to the AIC, the $\rm Base+LSS+S_8+SH0ES$ data strongly favours the dark energy with phantom crossing, whereas the $\rm Base+LSS+S_8+SN$ combination prefers the $\Lambda$CDM model.

To reliably predict the preference for the PDE scenario over $\Lambda$CDM we perform the Bayesian evidence analysis. 
Unlike the AIC, the Bayesian model selection approach penalizes models with a large volume of unconstrained parameter space. 
This method ought to be preferred in model comparison since it addresses the volume in multi-dimensional parameter space which directly controls the lack of predictivity of more complicated models~\cite{Trotta:2008qt}.~\footnote{Unlike the Bayesian evidence analysis, the AIC penalizes extra parameters regardless of whether they are constrained by the data or not.}

We compute the Bayesian evidence with the publicly available cosmological code \texttt{MCEvidence} \footnote{\href{github.com/yabebalFantaye/MCEvidence}{github.com/yabebalFantaye/MCEvidence}.}~\cite{Heavens:2017afc}.
Then, we calculate the Bayes factor defined as $\ln B\equiv\ln \mathcal{Z}_{\rm PDE}-\ln \mathcal{Z}_{\rm \Lambda CDM}$ where $\mathcal{Z}$ is the Bayesian evidence for a given model, and show the result in Tab. \ref{tab:imp3}.
A negative (positive) value of the Bayes factor $\ln B$ shows that the $\Lambda$CDM (PDE) model is preferred. 
According to the revised Jeffreys scale by Kass and Raftery~\cite{Kass:1995loi}, we will have for $0\leq |\ln B|<1$ a weak preference, for $1\leq |\ln B|<3$ a positive preference, for $3\leq |\ln B|<5$ a strong preference and for $|\ln B|\geq5$ a very strong preference.
We found that if the SH0ES data is not included the $\Lambda$CDM is always the preferred model. 
This is because the PDE model introduces new parameter degeneracy directions which are poorly constrained by the data.
In turn, the $\rm Base+LSS+S_8+SH0ES$ combination suggests a very strong preference for the PDE scenario over $\Lambda$CDM.
This happens because the available parameter space in the PDE sector significantly shrinks upon adding the SH0ES information as shown in Fig. \ref{fig:5}.

\section{Transitional Dark Energy}
\label{sec:TDE}

In this section we examine a late-time scenario with a rapid transition in the dark energy equation of state.

\subsection{Model description}
\label{sec:TDEmod}

We aim to describe a rapid transition in the dark energy sector in a more general way. To that end, we use a model-independent 4-parameter parameterization for the dark energy evolution, suggested by~\cite{Keeley:2019esp},
\bseq
\label{TDE12}
\begin{align}
\label{TDE1}
    &\rho_{\DE}(z) = \rho_{\DE,0}(1+z)^{3(1+w_{\DE}^{\rm eff}(z))},\\
\label{TDE2}
    &w_{\DE}^{\rm eff} = \frac{1}{2}\,\left((w_0 + w_1) + (w_1 - w_0)\tanh{\l\frac{z-z_{\tr}}{\Delta_{\tr}}\r}\right),
\end{align}
\eseq
where the $w_{\DE}^{\rm eff}(z)$ is an effective equation of state (see e.g.~\cite{Jassal:2006gf}) being related to the physical dark energy equation of state $w_{\DE}$ through
\begin{equation}
    w_{\DE}^{\rm eff}(z) = \frac{1}{\ln{(1+z)}}\int_{0}^{z}w_{\DE}(z')\frac{dz'}{1+z'}
\end{equation}
The $w_{\DE}^{\rm eff}(z)$ reproduces the physical equation of state $w_{\DE}(z)$ only in the regime where $w_{\DE}(z)$ is constant.
The $w_0$ and $w_1$ are two model parameters which describe the asymptotic behaviour of the TDE equation of state in the distant future ($a\to \infty$) and the distant past ($a\to 0$), respectively. 
The $z_{\tr}$ refers to the moment of the transition, whereas the $\Delta_{\tr}$ parameterize the steepness of the transition.

In the limit of instantaneous transition ($\Delta_{\tr}\to0$), the $w_{\DE}^{\rm eff}$ takes the following form
\be
\lim_{\Delta_{\tr}\to0} w_{\DE}^{\rm eff}(z)=w_0+(w_1-w_0)\times\Theta(z-z_{\tr})\,,
\ee
where the $\Theta$ denotes the Heaviside function. In this regime, the $w_0$ and $w_1$ approach the present and the early values of the physical dark energy equation of state.

The TDE model is fully parameterized by the set of four parameters ($w_0$, $w_1$, $z_{\tr}$, $\Delta_{\tr}$). Unlike the PDE parameterization, the TDE dark energy density is constrained to be positive. This generally biases towards smoother evolution of the dark energy density (see e.g.~\cite{Wang:2018fng}).

We implement the TDE background evolution through \eqref{TDE12} while assuming no perturbations in the dark energy sector. We vary all four TDE parameters ($w_0$, $w_1$, $z_{\tr}$, $\Delta_{\tr}$) along with the six $\Lambda$CDM ones ($\omega_{cdm}$, $\omega_b$, $H_0$, $\ln(10^{10}A_s)$, $n_s$, $\tau$). We impose the following uniform priors on TDE parameters: 
\begin{equation}
\label{TDEprior}
\begin{aligned}
    &w_0\in[-\infty,+\infty],\qquad  w_1\in[-4,0],\\
    &z_{\tr}\in[0,10],\qquad\qquad \Delta_{\tr}\in[0,10]
\end{aligned}
\end{equation}
Our goal is to explore the dark energy dynamics at late times, so we imply $z_{\tr}<10$ and $\Delta_{\tr}<10$.
We examine the sensitivity of the parameter constraints to the choice of the TDE priors in Appendix. \ref{app:Prior}.

\subsection{Parameter constraints}
\label{sec:TDEcon}

\begin{figure}[!htb]
    \begin{center}
        \includegraphics[width=1.0\columnwidth]{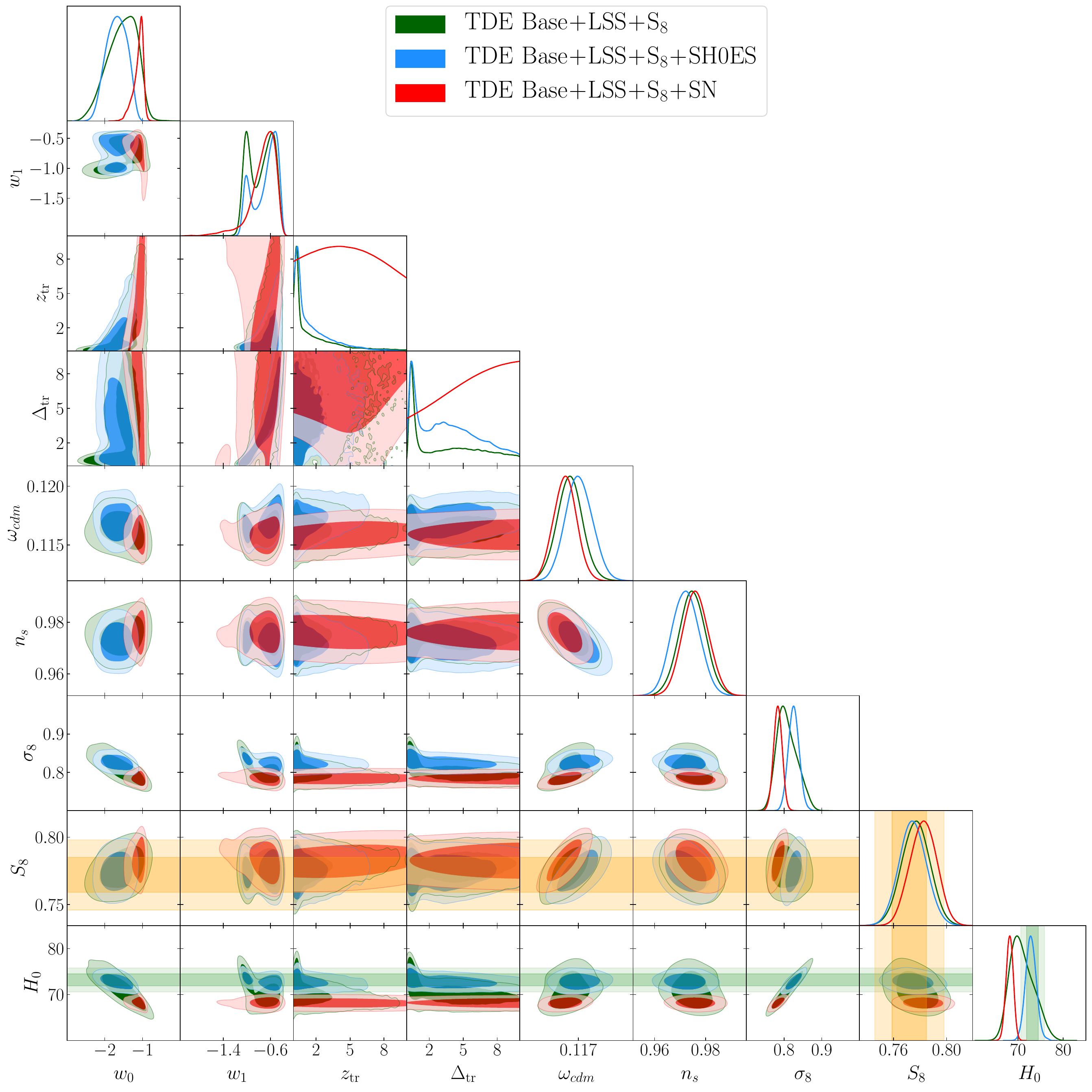}
        \caption {Marginalized 2d posterior distributions of the cosmological parameters in the TDE model for the $\rm Base+LSS+S_8$ (green), $\rm Base+LSS+S_8+SH0ES$ (blue) and $\rm Base+LSS+S_8+SN$ (red)  data sets. The Gaussian prior on $\tau$ \eqref{tau} is always adopted. The yellow bands represent $1\sigma$ and $2\sigma$ constraints on $S_8$ \eqref{S8} coming from the photometric surveys (DES-Y3, KiDS, HSC), whereas the green bands refer to the $H_0$ measurement \eqref{H0} reported by the SH0ES collaboration.}
        \label{fig:8}
    \end{center}
\end{figure}

Tab.\,\ref{tab:8} presents the 1d marginalized constraints on cosmological parameters in the TDE model. 
\begin{table}
	\renewcommand{\arraystretch}{1.2}
    \small
	\centering
	\begin{tabular} {|c|| c |c | c |c|c|}
		\hline
		& \multicolumn{4}{c|}{Transitional Dark Energy (TDE)} \\
		\hline
		\hline
		\multirow{2}{*}{$\rm \!\! Parameter\!\!$} & \multirow{2}{*}{Base\,+\,LSS} & 
		Base\,+\,LSS & Base\,+\,LSS & Base\,+\,LSS
		\\
		&& +\,S$_8$& +\,S$_8$\,+\,SH0ES & +\,S$_8$\,+\,SN
		\\
		\hline
		$w_0$ & $\!-1.46(-2.09)^{+0.46}_{-0.32}\!$  
		& $\!-1.55(-1.94)^{+0.44}_{-0.32}\!$ 
        & $\!-1.68(-1.75)^{+0.30}_{-0.26}\!$ 
        & $\!-1.11(-1.19)^{+0.16}_{-0.07}\!$
		\\ 
		$w_1$ & $\!-0.79(-1.05)^{+0.30}_{-0.30}\!$ 
		& $\!-0.78(-1.03)^{+0.30}_{-0.30}\!$ 
        & $\!-0.68(-1.04)^{+0.26}_{-0.35}\!$ 
        & $\!-0.72(-0.51)^{+0.25}_{-0.09}\!$  
		\\ 
		$z_{\rm tr}$ &  unconstrained   
		& $<6.43\,(0.34)$ 
        & $<5.26\,(0.39)$ 
        & unconstrained 
		\\ 
		$\Delta_{\rm tr}$ & unconstrained 
		& $<9.02\,(0.32)$  
        & $<8.75\,(0.28)$
        & unconstrained  
		\\ 
		$100\,\omega_b $ & $2.242\pm0.019$  
        & $2.245\pm0.019$ 
        & $2.243\pm0.019$ 
        & $2.249\pm0.018$ 
		\\ 
		$10\,\omega_{cdm}$ & $1.169\pm0.015$ 
        & $1.163\pm0.011$ 
        & $1.171\pm0.011$ 
        & $1.159\pm0.010$ 
		\\ 
		$H_0$ & $70.46(75.69)^{+1.81}_{-3.10}$  
        & $71.05(74.80)^{+2.30}_{-3.08}$ 
        & $72.83(74.36)^{+1.16}_{-1.16} $ 
        & $68.17(68.33)^{+0.82}_{-0.74}$ 
		\\ 
		$\tau$ & $0.057\pm0.005$  
        & $0.057\pm0.006$ 
        & $0.057\pm0.005$ 
        & $0.056\pm0.005$ 
		\\ 
		$\!{\rm ln}(10^{10} A_s)\!$ & $3.038\pm0.012$    
        & $3.037\pm0.011$ 
        & $3.039\pm0.011 $ 
        & $3.035\pm0.011$ 
		\\ 
		$n_s$ & $0.974\pm0.006$   
        & $0.975\pm0.005$ 
        & $0.972\pm0.005 $ 
        & $0.976\pm0.005$ 
		\\ 
		\hline   
		$\rdrag$ & $147.77\pm0.37$   
        & $147.87\pm0.32$ 
        & $147.80\pm0.31$ 
        & $148.06\pm0.30$ 
		\\    
		$\Omega_m$ & $0.283\pm0.020$   
        & $0.277\pm0.020$ 
        & $0.264\pm0.009$ 
        & $0.299\pm0.007$ 
		\\ 
		$\sigma_8$ & $0.811\pm0.027$  
        & $0.810\pm0.027$ 
        & $0.826\pm0.014 $ 
        & $0.784\pm0.011$ 
		\\ 
		$S_8$ & $0.786\pm0.018$   
        & $0.777\pm0.010$ 
        & $0.775\pm0.010$ 
        & $0.783\pm0.010$ 
		\\
		\hline
	\end{tabular}
	\caption {Parameter estimates (mean value with $1\sigma$ error bars and best fit value in the parentheses) in the transitional dark energy model. 
    }
	\label{tab:8}
\end{table}
Fig.\,\ref{fig:8} shows the resulting 2d posterior distributions.
Here we do not show the results of the Base data analysis because the CMB data alone can not break the degeneracies present in the TDE sector.

Let us start with the $\rm Base+LSS$ analysis. 
We find no evidence for a transition in the TDE equation of state, however the posteriors are consistent with this scenario. 
The posteriors of $z_{\tr}$ and $\Delta_{\tr}$ are prior-dominated, so we do not report the constraints on these parameters.
The $\rm Base+LSS$ analysis predicts $H_0=70.46^{+1.81}_{-3.10}\kms$. This $H_0$ estimate is consistent with both the Planck~\cite{Planck:2018vyg} and SH0ES~\eqref{H0} results~\cite{Planck:2018vyg}.
We found that the $\sigma_8$ constraint is consistent with the $\Lambda$CDM expectation (see Tab. \ref{tab:2}) but has a four-times larger error bar. It happens because the TDE scenario introduces extra degrees of freedom that makes low-redshift quantities more uncertain relative to the $\Lambda$CDM predictions.
The $S_8$ measurement is entirely consistent with the direct probes \eqref{S8} in the late Universe.

Next, we include the $S_8$ data. 
The data mildly prefers a transition in the TDE equation of state from phantom dark energy ($w_0<-1$) to non-phantom dark energy ($w_1>-1$).
Importantly, our analysis detects the upper limits on the TDE transition parameters: $z_{\rm tr}<6.43$ and $\Delta_{\rm tr}<9.02$ at $95\%$ CL. 
The $H_0$ constraint is now consistent within one standard deviation with the direct measurement \eqref{H0} which opens an avenue towards combining the $\rm Base+LSS+S_8$ and SH0ES data sets.
Noteworthy, the mean value of $\sigma_8$ increases compared to the $\Lambda$CDM prediction, cf. with Tab. \ref{tab:2}. 
Indeed, the data favours a phantom dark energy equation of state at present that implies a large growth rate of cosmological matter perturbations compared to $\Lambda$CDM~\cite{Alestas:2021xes}.~\footnote{It is important here that the dark energy is non-clustering. A clustering phantom dark energy predicts less growth of perturbations than $\Lambda$CDM~\cite{Alestas:2021xes}.}

We further examine the cosmological inference from the $\rm Base+LSS+S_8+SH0ES$ data. 
The preference for a transition from a phantom dark energy to quintessence increases while adding the SH0ES information.
In particular, the data prefers the $w_0<-1$ at the $3\sigma$ level.
The upper bounds on the TDE transition parameters also strengthen: $z_{\rm tr}<5.26$ and $\Delta_{\rm tr}<8.75$ at $95\%$ CL.
Our analysis predicts $H_0=72.83\pm1.16\kms$ which is in an excellent agreement with the SH0ES constraint.
Interestingly, the mean value of $\sigma_8$ increases with respect to the $\rm Base+LSS+S_8$ (without SH0ES) result. 
This effect can be attributed to a smaller $w_0$ that further decreases the growth of matter perturbations~\cite{Alestas:2021xes}.
Despite this fact, the $S_8$ constraint is entirely consistent with the direct probes \eqref{S8} due to a lower value of $\Omega_m$.

Again, the supernova absolute magnitude that is necessary to fit the CMB, BAO and SN data is not compatible with the local astrophysical calibration \eqref{Msn}, for details see Sec. \ref{sec:PDEdis}. For this reason, we combine the $\rm Base+LSS+S_8$ and SN data (without SH0ES).
Our results demonstrate no evidence for a transition in the TDE equation of state. 
The $z_{\tr}$ and $\Delta_{\tr}$ parameters become largely unconstrained, and the dynamic in the dark energy sector approaches the cosmological constant.
Therefore, the $\sigma_8$ value is now consistent with the $\Lambda$CDM prediction.
As expected, the matter density parameter is shifted towards the Planck value while adding SN data, namely $\Omega_m=0.299\pm0.007$.
Our final constraints on $S_8$ and $H_0$ in the TDE scenario read
\be
\label{H0tde}
S_8=0.783 \pm 0.010   \qquad\qquad   H_0=68.17^{+0.82}_{-0.74}\kms
\ee
The $S_8$ constraint agrees well with the direct probes of this parameter in the late Universe. In turn, the $H_0$ estimate is in a $3.3\sigma$ tension with the SH0ES measurement. 
Our results reinforce that the TDE model can not resolve the Hubble tension in agreement with the previous studies of late-time Universe modifications (see e.g.~\cite{Lemos:2018smw,Poulin:2018zxs,Roy:2022fif,Dinda:2021ffa,Keeley:2022ojz}).

\subsection{Discussion}
\label{sec:TDEdis}

For the sake of completeness, we present the constraint on the supernova absolute magnitude in the TDE scenario inferred from $\rm Base+LSS+S_8+SN$ data,
\be
\label{MsnCMB2}
M_B=-19.411\pm0.016\,.
\ee
This estimate is in perfect agreement with the PDE inference \eqref{MsnCMB} and inverse-distance ladder measurements~\cite{DES:2018rjw,Feeney:2018mkj,Camarena:2019rmj} while being in a significant $4.5\sigma$ tension with the local astrophysical calibration via Cepheids \eqref{Msn}.
It clearly shows that the $\rm Base+LSS+S_8+SN$ and SH0ES data are not compatible and, therefore, can not be combined into one data set. In Sec. \ref{sec:PDEdis} we discuss possible approaches that may alleviate the ‘supernova absolute magnitude tension’.

\begin{figure}[!t]
   \begin{center}
        \includegraphics[width=0.7\columnwidth]{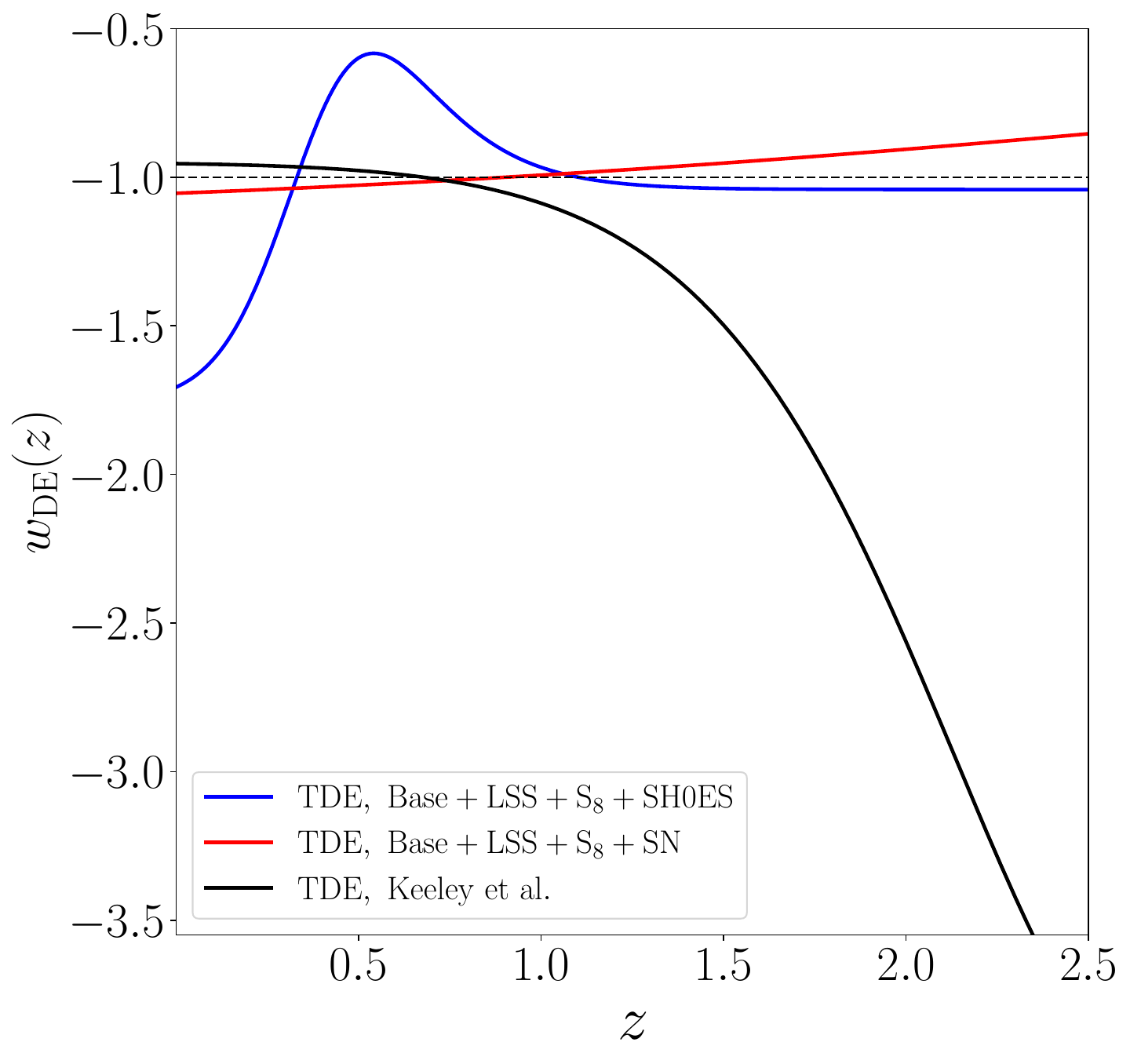} 
        \caption {Behaviour of the dark energy regular equation of state for the TDE best-fit models to $\rm Base+LSS+S_8+SH0ES$ (blue) and $\rm Base+LSS+S_8+SN$ (red) data sets, as well as the result of the Gaussian Process inference from~\cite{Keeley:2019esp} (solid black). The dashed line corresponds to $w_{\DE}= -1$.}
        \label{fig:9b}
    \end{center}
\end{figure}

In Fig. \ref{fig:9b} we show the $w_{\DE}(z)$ evolution for the different scenarios.
The $\rm Base+LSS+S_8+SH0ES$ data predicts a relatively sharp transition in the dark energy equation of state with $z_{\tr}^\bestf=0.39$ and $\Delta_{\tr}^\bestf=0.28$. 
In contrast, the $\rm Base+LSS+S_8+SN$ analysis suggests a very broad transition in the dark energy sector, namely $z_{\tr}^\bestf=6.48$ and $\Delta_{\tr}^\bestf=9.43$. 
We emphasize that the $w_{\DE}(z)$ evolution in both scenarios is consistent with the cosmological constant at $68\%$ CL, so the transitions are not statistically significant. We do not show the error band for $w_{\DE}(z)$ in order to increase readability.

\begin{figure}[t]
    \begin{center}
        \includegraphics[width=1.0\columnwidth]{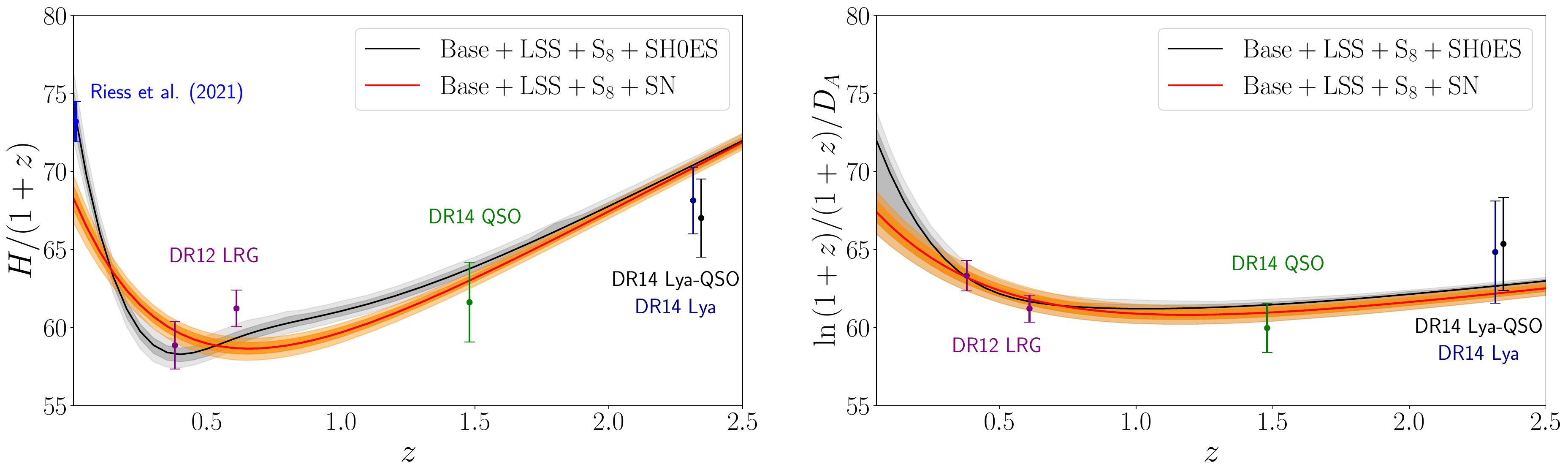} 
        \caption {Behaviour of $H(z)/(1+z)$ and $\ln(1+z)/(1+z)/D_A$ in the TDE model. Both quantities are measured in units of km\,s$^{-1}$\,Mpc$^{-1}$. The absolute scale for the BAO measurements is set by the best-fit value of the sound horizon optimized to the Base likelihood  $\rdrag=148.04$\,Mpc.}
        \label{fig:9}
    \end{center}
\end{figure}

To demonstrate consistency with the BAO measurements, in Fig. \ref{fig:9} we show
the behaviour of the Hubble parameter and the inverse BAO distance for the different data sets.
The $\rm Base+LSS+S_8+SH0ES$ analysis agrees well with the BAO distances whilst providing a higher value of $H_0$ consistent with SH0ES. The $\rm Base+LSS+S_8+SN$ prediction is also consistent with the BAO data, although it demonstrates a slightly worse agreement with the radial BAO signal inferred from the BOSS DR12 LRG sample at $z=0.57$.

It is important to compare our results with the previous analysis~\cite{Keeley:2019esp} based on the same TDE parameterization. 
Basically, the CMB, BAO, SN and the SH0ES-like $1\%$ prior on $H_0$ combined prefers a rapid transition in the dark energy equation of state from $w_{\DE}>-1$ at present to values much less than $-1$ by $z\simeq 2$.
In Fig. \ref{fig:9b} we show the median result of the Gaussian Process inference fitted in the TDE framework ($w_0 = -0.95$, $w_1 = -1.95$, $z_{\rm tr} = 2.5$, $\Delta_{\rm tr} = 0.9$)~\cite{Keeley:2019esp}.
In contrast, our $\rm Base+LSS+S_8+SH0ES$ analysis suggests a sharper transition from a phantom dark energy at present to nearly the cosmological constant at $z>1$. 
The difference in the parameter inference can be attributed to the fact that the Ref.~\cite{Keeley:2019esp} utilizes the SH0ES-like measurement and SN data which are not compatible when combining with CMB+BAO (see e.g.~\cite{Lemos:2018smw,Poulin:2018zxs,Roy:2022fif,Dinda:2021ffa,Keeley:2022ojz}).~\footnote{Specifically, the BAO distances calibrated by the CMB-inferred value of $\rdrag$ disagree with the supernova distances calibrated by the SH0ES-like prior on $H_0$ as shown in Figs. 1 and 10 of Ref.~\cite{Keeley:2019esp}. Note that our analysis is consistent with the BAO measurements as shown in Fig. \ref{fig:9}.}
Accordingly, the TDE analysis~\cite{Keeley:2019esp} predicts a slower growth rate of matter perturbations today than $\Lambda$CDM, whereas our analysis features a faster growth of cosmic structure compared to the concordance model.~\footnote{The difference in the growth history can be attributed to the $w_{\DE}(z)$ behaviour: the TDE analysis~\cite{Keeley:2019esp} favours $w_{\DE}>-1$ at present leading to a slower growth of perturbations, whereas our analysis predicts $w_{\DE}<-1$ today and, hence, an enhanced growth of cosmic structures~\cite{Alestas:2021xes}.}

Ref.~\cite{Sharma:2022ifr} presented a model-independent analysis of evolving dark energy with massive neutrinos. Specifically, the authors use a four parameter model for the physical dark energy equation of state $w_{\DE}(z)$ which is different from our parameterization of $w_{\DE}^{\rm eff}(z)$.
This analysis also features a neutrino mass as free parameter whereas we assume $\sum m_\nu=0.06\eV$. 
When all data are put together, the authors report the $w_{\DE}(z)$ evolution being broadly consistent with the cosmological constant.
We can not directly compare the parameter constraints provided the differences in our approaches, however our results are nominally consistent with those of~\cite{Sharma:2022ifr}.

Finally, we evaluate the performance of the TDE and $\Lambda$CDM models. Tab. \ref{tab:imp4} presents the $\Delta\chi^2_{\rm min}$ and $\Delta{\rm AIC}$ values for different data sets.
\begin{table}[!h]
    \renewcommand{\arraystretch}{1.1}
    \centering
    \begin{tabular} {| c || c |c | c | c |}
        \hline
        \multirow{2}{*}{Parameter}  & 
		\multirow{2}{*}{$\rm Base\!+\!LSS$} &
		$\rm \!Base\!+\!LSS\!$ &
		$\rm \!Base\!+\!LSS\!$ &
		$\rm \!Base\!+\!LSS\!$\\
        & & $\rm +S_8$ & $\rm +S_8\!+\!SH0ES$ & $\rm +S_8\!+\!SN$\\
        \hline
        \hline
        $\Delta\chi^2_{\rm min}$ & $-10.94$ &
        $-12.99$ &
        $-22.68$ & 
        $-0.61$  \\ 
        $\Delta{\rm AIC}$ & $-2.94$ &
        $-4.99$ &
        $-14.68$ &
        $+7.39$ \\  
        \hline
        $\ln \, B$ &
        $-13.91$ &
        $-7.48$ &
        $+1.19$ &
        $-6.51$ \\ \hline
    \end{tabular}
    \caption {The $\Delta\chi^2_{\rm min}$ and $\Delta{\rm AIC}$ values between the best-fit TDE and $\Lambda$CDM models to different data sets. We also show the Bayesian factors $\ln B$ calculated for the TDE model with respect
to the $\Lambda$CDM scenario. Note that the negative value of $\Delta{\rm AIC}$ indicates a preference for the TDE scenario, while the negative $\ln B$ shows a preference for $\Lambda$CDM.}
    \label{tab:imp4}
\end{table}
The $\rm Base+LSS+S_8+SH0ES$ data suggests a $3.8\sigma$ preference for the TDE model. This effect is attributed to the higher value of $H_0$ in the TDE scenario which significantly improves the fit to the SH0ES data, namely $\Delta\chi^2_{\rm SH0ES}=-12.47$. In turn, the $\rm Base+LSS+S_8+SN$ data shows a marginal evidence for the TDE scenario over $\Lambda$CDM. According to the AIC, the $\rm Base+LSS+S_8+SH0ES$ data strongly favours the TDE scenario against $\Lambda$CDM, whereas the $\rm Base+LSS+S_8+SN$ combination prefers the base $\Lambda$CDM model.

We also compute the Bayes factor $\ln B$ for different data sets and show the results in Tab. \ref{tab:imp4}.
The $\rm Base+LSS+S_8+SH0ES$ data suggests only a weak preference for the TDE model over $\Lambda$CDM. This result can be explained by largely unconstrained parameter space in the TDE sector that harshly penalized this model (see Fig. \ref{fig:8}).

\section{Summary and Conclusions}
\label{sec:concl}

In this work we have presented new limits on the cosmological parameters in $\Lambda$CDM and its several motivated extensions using the alternative CMB measurements along with large-scale structure and supernova data. As the primary CMB data we consider the SPT-3G polarization, SPTpol gravitational lensing and Planck temperature CMB measurements.

Our analysis leads to systematically lower values of $S_8$ being entirely consistent with the low-redshift cosmological probes.
In contrast, the baseline Planck analysis exhibits the $S_8$ tension at the $3.3\sigma$ significance level.
Our results suggests that the current discordance between the Planck results and the weak lensing and photometric galaxy clustering data is largely driven by the extra smoothing of the Planck TT power spectrum peaks and troughs that pulls the late-time amplitude to higher values.
Combining the primary CMB data (Base) with the large-scale structure ($\rm LSS+S_8$) and uncalibrated supernova (SN) measurements within $\Lambda$CDM we found $H_0=68.47 \pm 0.38\kms$ and $S_8= 0.790 \pm 0.009$.

Then we explore the cosmological inference in the $\rm \Lambda$CDM+$\sum m_\nu$ model. The $\rm Base+LSS+S_8+SN$ data suggests a $3.9\sigma$ preference for nonzero neutrino masses, $\sum m_\nu=0.22\pm0.06\eV$. 
We found that breaking the CMB degeneracies between $\Mnu$ and the cosmological parameters by the LSS data is a major contribution to our neutrino mass measurements.
The Planck lensing-like anomaly strengthens the constraints on neutrino masses making such higher values of $\sum m_\nu$ implausible~\cite{Planck:2018vyg}.
We conclude that the future CMB data must be considered before safely ruling out the region $\sum m_\nu\gtrsim 0.2\eV$.
The Simons Observatory~\cite{SimonsObservatory:2018koc} and CMB-S4 experiment~\cite{CMB-S4:2016ple} will provide the exquisite CMB measurements that help to elucidate the source behind the Planck lensing-like anomaly.

In addition, we revisit the parameter constraints in the $\rm \Lambda$CDM+$\Neff$ scenario. We found the cosmological measurements compatible with the Planck baseline analysis.

Finally, we investigate the possibility of dynamical dark energy with two model-independent approaches. 
The PDE scenario is based on reconstruction of the dark energy density whereas the TDE approach utilizes a general four parameter model for the effective dark energy equation of state.
Since the supernova calibrations provided by CMB+BAO and Cepheids disagree, we consider the measurements of uncalibrated supernovae luminosity distance (SN) and local distance ladder (SH0ES) separately.
In the both models, the $\rm Base+LSS+S_8+SN$ data suggests $H_0\simeq68\kms$ being in moderate $(\sim3\sigma)$ tension with the SH0ES constraint.
However, when the local Type Ia supernovae are calibrated by Cepheids the dynamical dark energy approaches predict significantly higher values of $H_0$ consistent with SH0ES.  
Using the Bayesian evidence ratio, the $\rm Base+LSS+S_8+SH0ES$ combination strongly favours the PDE scenario, whereas the TDE model is only weakly preferred compared to $\Lambda$CDM.

Our work underlines the importance of the calibration of the supernova absolute magnitude to reconcile the $H_0$ tension.
Using the sound horizon at last scattering as a standard ruler and the local astophysical calibration by Cepheids lead to values of the supernova absolute magnitude $M_B$ that are inconsistent at a level more than $4\sigma$.
This inconsistency is the basis of the $H_0$ tension because the difference in $M_B$ is easily translated to a difference in the values of the Hubble constant given the degeneracy between $M_B$ and $H_0$.
The supernova absolute magnitude tension may be caused by astrophysical systematics in the distance ladder and/or new physics in gravity sector.
We reinforce that the models which modify only the late Universe expansion is not capable of solving this tension.

Our work can be extended in multiple ways. A natural extension of our analysis would be to include the recent SPT-3G measurements of TT power spectrum~\cite{SPT-3G:2022hvq}. 
In addition, it would be interesting to consider the alternative ACT-DR4 CMB measurements at small angular scales~\cite{ACT:2020gnv}.
Finally, our analysis can be improved by including the full-shape analysis of the eBOSS quasar sample~\cite{Simon:2022csv,Chudaykin:2022nru} and the galaxy bispectrum multipoles~\cite{Ivanov:2023qzb} which can potentially yield a significant information gain in extended cosmological scenarios.
We leave these tasks to future work.

\vspace{1cm}
\section*{Acknowledgments}

We thank to Mikhail M. Ivanov for fruitful discussions. 
We are grateful to Neal Dalal and Jessie Muir for important suggestions that improved our analysis.
The work on neutrino masses is supported by the RSF grant 22-12-00271. The work on dark energy models is carried out within the framework of the scientific program of the National Center for Physics and
Mathematics, the project “Particle Physics and Cosmology”. 
All numerical calculations have been performed with the HybriLIT heterogeneous computing platform (LIT, JINR) (\href{http://hlit.jinr.ru}{http://hlit.jinr.ru}) and the Helios cluster at the Institute for Advanced Study, Princeton.


\appendix 

\section{$\chi^2_{\min}$ per experiment}
\label{app:chi2}

In this appendix we provide the best-fit $\chi^2_{\min}$ values per experiment. Tab. \ref{tab:chi2_4} presents the results for the $\Lambda$CDM, $\rm \Lambda$CDM+$\sum m_\nu$ and $\rm \Lambda$CDM+$\Neff$ models, whereas Tab. \ref{tab:chi2_5} shows the results in the PDE and TDE scenarios.

\begin{table}
	\renewcommand{\arraystretch}{1.0}
	\centering
    \small
	\begin{tabular} { c || c |c |c|c}
		\hline
        \hline
        \multirow{1}{*}{$\Lambda$CDM} & 
		\multirow{1}{*}{$\rm Base$} &
		\multirow{1}{*}{$\rm Base\!+\!LSS$} &
		\multirow{1}{*}{$\rm Base\!+\!LSS\!+\!S_8$} &
		\multirow{1}{*}{$\rm Base\!+\!LSS\!+\!S_8\!+\!SN$}  \\
		\hline
		$\rm SPT$-$\rm 3G$ & 
        $522.31$ & 
        $523.48$ &
		$523.87$ &
		$523.55$\\
		$\rm Planck\, TT, \ell<30$ & 
		$21.15$ & 
		$21.76$ &
		$21.86$ &
		$21.69$\\
		$\rm Planck\, TT, 30\leq\ell<1000$ & 
		$406.05$ & 
		$406.57$ &
		$405.77$ &
		$406.27$\\
		$\rm Lens$ & 
		$5.57$ & 
		$5.57$ &
		$5.40$ &
		$5.43$   \\
		$\tau$-prior & 
		$0.01$ &
		$0.11$ &
		$1.49$ &
		$0.54$\\
		LSS, full-shape & 
		$-$ &
		$1074.95$ &
		$1074.17$ &
		$1073.69$\\
		LSS, BAO & 
		$-$ &
		$7.28$ &
		$7.57$ &
		$7.40$\\
		$\rm S_8$ &
		$-$ &
		$-$ &
		$1.59$ &
		$3.40$\\
		$\rm SN$ & 
		$-$ &
		$-$ &
		$-$ &
		$1027.81$\\
		\hline
		Total $\chi^2_{\min}$ & 
        $955.09$ & 
        $2039.71$ &
		$2041.72$ &
		$3069.78$\\
		\hline
        \hline
        \multirow{1}{*}{$\Lambda$CDM+$\sum m_\nu$} & 
		\multirow{1}{*}{$\rm Base$} &
		\multirow{1}{*}{$\rm Base\!+\!LSS$} &
		\multirow{1}{*}{$\rm Base\!+\!LSS\!+\!S_8$} &
		\multirow{1}{*}{$\rm Base\!+\!LSS\!+\!S_8\!+\!SN$}  \\
		\hline
		$\rm SPT$-$\rm 3G$ & 
        $520.65$ &
		$523.79$ &
		$523.77$ &
		$523.38$\\
		$\rm Planck\, TT, \ell<30$ & 
		$20.92$ &
		$20.73$ &
		$20.81$ &
		$20.80$\\
		$\rm Planck\, TT, 30\leq\ell<1000$ & 
		$406.66$ &
		$406.53$ &
		$406.43$ &
		$407.41$\\
		$\rm Lens$ & 
		$5.64$ &
		$6.45$ &
		$6.38$ &
		$5.90$   \\
		$\tau$-prior & 
		$0.0$ &
		$0.01$ &
		$0.0$ &
		$0.86$\\
		LSS, full-shape & 
		$-$ &
		$1070.59$ &
		$1070.66$ &
		$1071.08$\\
		LSS, BAO & 
		$-$ &
		$7.24$ &
		$7.23$ &
		$7.30$\\
		$\rm S_8$ & 
		$-$ &
		$-$ &
		$0.22$ &
		$0.01$\\
		$\rm SN$ & 
		$-$ &
		$-$ &
		$-$ &
		$1027.13$\\
		\hline
		Total $\chi^2_{\min}$ & 
        $953.87$ &
		$2035.34$ &
		$2035.50$ &
		$3063.87$\\
		  $\Delta\chi^2_{\min}$ & 
		  $-1.22$ &
		  $-4.37$ &
		  $-6.22$ &
		  $-5.91$\\
        \hline
        \hline
        \multirow{1}{*}{$\Lambda$CDM+$\Neff$} & 
		\multirow{1}{*}{$\rm Base$} &
		\multirow{1}{*}{$\rm Base\!+\!LSS$} &
		\multirow{1}{*}{$\rm Base\!+\!LSS\!+\!S_8$} &
		\multirow{1}{*}{$\rm Base\!+\!LSS\!+\!S_8\!+\!SN$}  \\
		\hline
		$\rm SPT$-$\rm 3G$ & 
        $521.80$ &
		$523.11$ &
		$521.75$ &
		$522.0$\\
		$\rm Planck\, TT, \ell<30$ & 
		$21.00$ &
		$22.59$ &
		$21.60$ &
		$21.61$\\
		$\rm Planck\, TT, 30\leq\ell<1000$ & 
		$405.28$ &
		$406.52$ &
		$406.75$ &
		$407.71$\\
		$\rm Lens$ & 
		$5.55$ &
		$5.59$ &
		$5.53$ &
		$5.54$   \\
		$\tau$-prior & 
		$0.01$ &
		$2.05$ &
		$0.46$ &
		$0.91$\\
		LSS, full-shape & 
		$-$ &
		$1072.65$ &
		$1073.57$ &
		$1074.33$\\
		LSS, BAO & 
		$-$ &
		$7.30$ &
		$7.40$ &
		$7.50$\\
		$\rm S_8$ & 
		$-$ &
		$-$ &
		$3.40$ &
		$1.51$\\
		$\rm SN$ & 
		$-$ &
		$-$ &
		$-$ &
		$1026.99$\\
		\hline
		Total $\chi^2_{\min}$ & 
        $953.64$ &
		$2039.81$ &
		$2040.46$ &
		$3068.10$\\
		  $\Delta\chi^2_{\min}$ & 
		  $-1.45$ &
		  $+0.1$ &
		  $-1.26$ &
		  $-1.68$\\
		\hline
	\end{tabular}
	  \caption {$\chi^2_{\min}$ values for the best-fit $\Lambda$CDM, $\Lambda$CDM+$\sum m_\nu$ and $\Lambda$CDM+$\Neff$ models to the Base, $\rm Base+LSS$, $\rm Base+LSS+S_8$ and $\rm Base+LSS+S_8+SN$ data sets.}
	\label{tab:chi2_4}
\end{table}

\begin{table}
	\renewcommand{\arraystretch}{1.0}
	\centering
    \small
	\begin{tabular} { c || c |c |c|c}
		\hline
        \hline
		\multirow{2}{*}{PDE} & 
		\multirow{2}{*}{$\rm Base\!+\!LSS$} &
		$\rm \!Base\!+\!LSS\!$ &
		$\rm \!Base\!+\!LSS\!$ &
		$\rm \!Base\!+\!LSS\!$\\
		& & $\rm +S_8$ & $\rm \!+S_8\!+\!SH0ES\!$ & $\rm +S_8\!+\!SN$\\
		\hline
		$\rm SPT$-$\rm 3G$ & 
        $522.64$ &
		$523.11$ &
		$522.84$ &
		$521.36$\\
		$\rm Planck\, TT, \ell<30$ & 
		$22.03$ &
		$21.62$ &
		$21.39$ &
		$20.99$\\
		$\rm Planck\, TT, 30\leq\ell<1000$ & 
		$407.80$ &
		$406.74$ &
		$408.95$ &
		$407.54$\\
		$\rm Lens$ &
		$6.53$ &
		$6.88$ &
		$7.11$ &
		$6.56$   \\
		$\tau$-prior & 
		$0.24$ &
		$1.06$ &
		$0.36$ &
		$0.13$\\
		LSS, full-shape &
		$1059.62$ &
		$1060.10$ &
		$1060.47$ &
		$1075.28$\\
		LSS, BAO &
		$9.87$ &
		$9.97$ &
		$9.51$ &
		$7.79$\\
		$\rm S_8$ & 
		$-$ &
		$0.0$ &
		$0.18$ &
		$0.56$\\
		$\rm SH0ES$ & 
		$-$ &
		$-$ &
		$214.12$ &
		$-$\\
		$\rm SN$ & 
		$-$ &
		$-$ &
		$-$ &
		$1028.84$\\
		\hline
		Total $\chi^2_{\min}$ & 
        $2028.73$ &
		$2029.48$ &
		$2244.92$ &
		$3069.05$\\
		  $\Delta\chi^2_{\min}$ & 
		$-10.98$ &
		$-12.24$ &
		$-21.86$ & 
		$-0.73$\\
		\hline
        \hline
		\multirow{2}{*}{TDE} & 
		\multirow{2}{*}{$\rm Base\!+\!LSS$} &
		$\rm \!Base\!+\!LSS\!$ &
		$\rm \!Base\!+\!LSS\!$ &
		$\rm \!Base\!+\!LSS\!$\\
		& & $\rm +S_8$ & $\rm +S_8\!+\!SH0ES\!$ & $\rm +S_8\!+\!SN $\\
		\hline
		$\rm SPT$-$\rm 3G$ & 
        $522.09$ & 
		$523.31$ &
		$523.88$ & 
		$522.14$\\
		$\rm Planck\, TT, \ell<30$ & 
		$21.10$ & 
		$20.98$ & 
		$21.15$ & 
		$21.56$\\
		$\rm Planck\, TT, 30\leq\ell<1000$ & 
		$407.66$ & 
		$406.92$ & 
		$408.97$ & 
		$407.54$\\
		$\rm Lens$ &
		$7.02$ & 
		$7.05$ & 
		$7.42$ & 
		$6.73$\\
		$\tau$-prior & 
		$0.00$ &
		$0.00$ & 
		$0.36$ & 
		$0.67$\\
		LSS, full-shape &
		$1060.87$ & 
		$1059.64$ & 
		$1061.72$ & 
		$1075.63$\\
		LSS, BAO &
		$10.01$ & 
		$10.63$ & 
		$10.28$ & 
		$7.69$\\
		$\rm S_8$ & 
		$-$ &
		$0.20$ & 
		$0.09$ & 
		$0.61$\\
		$\rm SH0ES$ & 
		$-$ &
		$-$ &
		$210.07$ & 
		$-$\\
		$\rm SN$ & 
		$-$ &
		$-$ &
		$-$ &
		$1026.62$\\
		\hline
		Total $\chi^2_{\min}$ & 
        $2028.77$ &
		$2028.73$ &
		$2243.94$ & 
		$3069.17$\\
		$\Delta\chi^2_{\min}$ & 
		$-10.94$ &
		$-12.99 $ &
		$-22.68$ & 
		$-0.61$\\
		\hline
	\end{tabular}
	\caption {$\chi^2_{\min}$ values for the best-fit PDE and TDE models to the $\rm Base+LSS$, $\rm Base+LSS+S_8$, $\rm Base+LSS+S_8+SH0ES$ and $\rm Base+LSS+S_8+SN$ data sets.}
	\label{tab:chi2_5}
\end{table}

\section{Choice of multipole cutoff}
\label{app:split}

Here, we quantify the consistency between the Base data set and the Planck TT $\ell^{\rm TT}>1000$ spectrum.~\footnote{For clarity, in this section we will refer to multipoles in the TT power spectra as $\ell^{\rm TT}$ in order to discriminate between temperature and polarization multipole ranges.} We also assess the impact of adding the Planck TT high-$\ell$ data gradually to the Base data set combination.

\begin{figure}[!t]
    \begin{center}
        \includegraphics[width=1.\columnwidth]{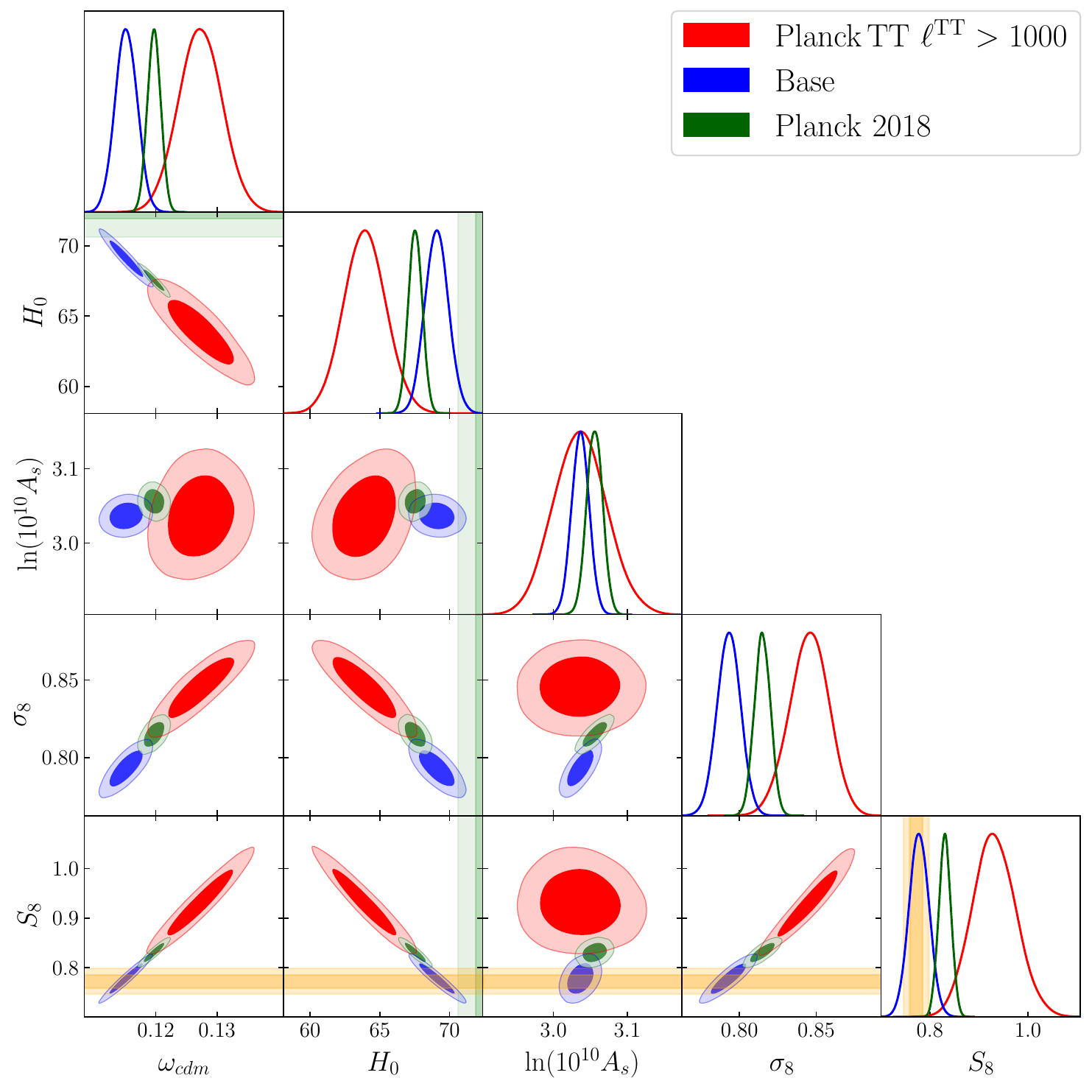}
        \caption{2d posterior distributions of the cosmological parameters inferred for the Planck TT $\ell^{\rm TT}>1000$ (red), Base (blue) and Planck 2018 (green) data sets. The Gaussian prior on $\tau$ \eqref{tau} is adopted.}
        \label{fig:10a}
    \end{center}
\end{figure}

First, we explore the consistency between the Base and Planck TT $\ell^{\rm TT}>1000$ data at the level of posterior distributions. 
Fig.~\ref{fig:10a} shows the two-dimensional parameter constraints inferred from these data sets together with the Planck 2018 results.  
The corresponding 1d marginalized parameter constraints are listed in Tab. \ref{tab:highl}.
\begin{table}[ht]
	\renewcommand{\arraystretch}{1.2}
    \small
	\centering
	\begin{tabular} {|c|| c |c | c |}
		\hline
		& \multicolumn{3}{c|}{$\Lambda$CDM}\\
		\hline
		\hline
		\multirow{1}{*}{\!\!\! Parameter\!} & \multirow{1}{*}{$\rm Planck\,2018$} & \multirow{1}{*}{$\rm Base$} & \multirow{1}{*}{Planck TT $\ell^{\rm TT}>1000$} \\
		\hline
		$100\,\omega_b$ & $2.241\pm0.015$ &
		$2.255\pm0.020$ &
		$2.115\pm0.094$ \\ 
		$10\,\omega_{cdm}$ & $1.197\pm0.011$ &
		$1.151\pm0.018$ &
		$1.273\pm0.037$ \\ 
		$H_0$ & $67.53\pm0.50$ &
		$69.09\pm0.84$ &
		$63.93\pm1.56$ \\ 
		$\tau$ & $0.060\pm0.005$ &
		$0.058\pm0.005$ &
		$0.058\pm0.005$  \\ 
		$\ln(10^{10} A_s)\!\!$ & $3.055\pm0.011$ &
		$3.036\pm0.012$ &
		$3.036\pm0.037$   \\ 
		$n_s$ & $0.967\pm0.004$ &
		$0.977\pm0.006$ &
		$1.000\pm0.048$  \\ 
		\hline   
		$\Omega_m$ & $0.313\pm0.007$ &
		$0.290\pm0.010$ &
		$0.366\pm0.025$ \\ 
		$\sigma_8$ & $0.815\pm0.005$ &
		$0.793\pm0.008$ &
		$0.845\pm0.013$  \\ 
		$S_8$ & $0.833\pm0.013$ &
		$0.780\pm0.020$ &
		$0.933\pm0.044$ \\ 
        \hline
	\end{tabular}
	\caption {Parameter constraints for different data sets with $1\sigma$ errors in the $\Lambda$CDM model.}
	\label{tab:highl}
\end{table}

Our results show that the Base combination and the Planck TT $\ell^{\rm TT}>1000$ data lead to significantly different parameter constraints. The Planck TT high-$\ell$ measurements predict a $3.4\sigma$ higher value of $\sigma_8$ compared to the Base data analysis. Combined with a moderately higher $\Omega_m$, it results in $S_8=0.933 \pm 0.044$, which exhibits a $3.5\sigma$ tension with the low-redshift cosmological probes \eqref{S8}. The Planck TT $\ell^{\rm TT}>1000$ data also predicts a considerably lower value of the Hubble parameter, $H_0=63.93\pm1.56\kms$, which is in a $4.6\sigma$ tension with the SH0ES measurement. It also deviates from the Planck 2018 result by $2.2\sigma$.

Even though the posterior distributions give insight into the parameter discrepancy, it is important to assess the significance of the corresponding tension in the full $\Lambda$CDM parameter space.
To quantify the overall consistency between disjoint data sets
we consider the metric
\be
\label{chi2d}
\chi^2=(\textbf{p}_1-\textbf{p}_2)^{\rm T} (C_1+C_2)^{-1}(\textbf{p}_1-\textbf{p}_2)\,,
\ee
where $\textbf{p}_i$ is the vector of parameter means and $C_i$ is the posterior covariance, both for a given experiment $i$. We carry out the comparison in the 5-parameter space, namely ($\omega_{cdm}$, $\omega_b$, $H_0$, $n_s$, $\ln(10^{10} A_s)$). 
We ignore $\tau$ because the $\tau$ information went into both sets of estimated parameters through the Gaussian prior \eqref{tau}.~\footnote{We checked that the comparison in the parameter space ($\omega_{cdm}$, $\omega_b$, $H_0$, $n_s$, $A_se^{-2\tau}$) gives identical results.}

Then, we compute the probability to exceed $\chi^2$ (for a $\chi^2$ distribution with degrees of freedom equal to the number of free parameters) and convert it into the equivalent number of $\sigma$ using the standard Gaussian interpretation. We also scan for $\max(|\Delta \textbf{p}/\sigma_p|)$ ($\sigma_p$ is the posterior error given by the square root of a diagonal element of $C_1+C_2$) and report the most deviant parameter(s). We cite the corresponding difference in units of $\sigma_p$.

Our results are summarized in Tab. \ref{tab:chi2_2}
\begin{table}[h]
	\renewcommand{\arraystretch}{1.1}
	\centering
	\begin{tabular} {| c  c |c c|}
		\hline
        \hline
        Data set 1: &  Data set 2: & \multicolumn{2}{c|}{Test} \\
		$\rm SPT$-$\rm 3G\!+\!Lens\!+\!Planck\,TT$ & Planck TT & $\chi^2$ & max-param \\
		\hline
		\hline
		$\ell^{\rm TT}<1000$ (Base) 
        & $\ell^{\rm TT}>1000$
        & $2.4\sigma$  
        & $3.0\sigma$ ($\omega_{cdm}$, $H_0$) \\
        $\ell^{\rm TT}<800$ 
        & $\ell^{\rm TT}>800$
        & $2.3\sigma$  
        & $3.1\sigma$ ($\omega_{cdm}$) \\
        \hline
	\end{tabular}
	\caption {Consistency of different data sets (first and second columns) as determined from the metric \eqref{chi2d} (third column) and the shift in the most deviant parameter(s) (fourth column).}
	\label{tab:chi2_2}
\end{table}
with a comparison of the Base and Planck TT $\ell^{\rm TT}>1000$ data given in the first raw. We identified a $2.4\sigma$ tension between these data sets in the 5-dimensional parameter space. Note that individual cosmological parameters, like $\omega_{cdm}$ and $H_0$, deviate by $3\sigma$. These parameters are of the most interest because they relate to the low-redshift cosmological probes. Indeed, the $H_0$ measurement is currently the center of great attention due to the Hubble tension, whereas $\omega_{cdm}$ determines the broadband shape of the galaxy power spectrum and controls the growth rate of cosmological matter perturbations.
We also assess consistency between the Base combination and Planck TT $\ell^{\rm TT}>1000$ data in the parameter space ($\omega_{cdm}$, $\omega_b$, $H_0$, $n_s$, $\sigma_8$), where we consider the late-time $\sigma_8$ instead of $\ln(10^{10} A_s)$.
In this case, the significance of the overall tension between the data sets increases by $0.2\sigma$ compared to that in Tab. \ref{tab:chi2_2}.

To test the robustness of our findings we assess the effect of splitting the Planck TT spectrum at $\ell^{\rm TT}=800$. This choice roughly corresponds to an even division of the Planck TT constraining power on $\Lambda$CDM parameters coming from the $\ell^{\rm TT}<800$ and $\ell^{\rm TT}>800$ multipole ranges which has been extensively discussed in~\cite{Aghanim:2016sns}.
Specifically, we perform a comparison of the $\rm SPT$-$\rm 3G\!+\!Lens\!+\!PlanckTT$ $(\ell^{\rm TT}<800)$ and Planck TT $\ell^{\rm TT}>800$ data sets and show the results in the second raw of Tab. \ref{tab:chi2_2}.
Our findings are insensitive to the particular choice of the multipole cutoff in the TT power spectrum.

One comment is in order here. The measure \eqref{chi2d} gives a reasonable estimate of the parameter discrepancy only in the limit of multivariate Gaussian posterior distribution. As shown in Fig. \ref{fig:10a}, the parameter posteriors inferred from the Base and Planck TT $\ell^{\rm TT}>1000$ data sets are reasonably Gaussian, so the metrics defined above gives a good measure of consistency in the full parameter space.

\begin{figure}[!htb]
    \begin{center}
        \includegraphics[width=1.\columnwidth]{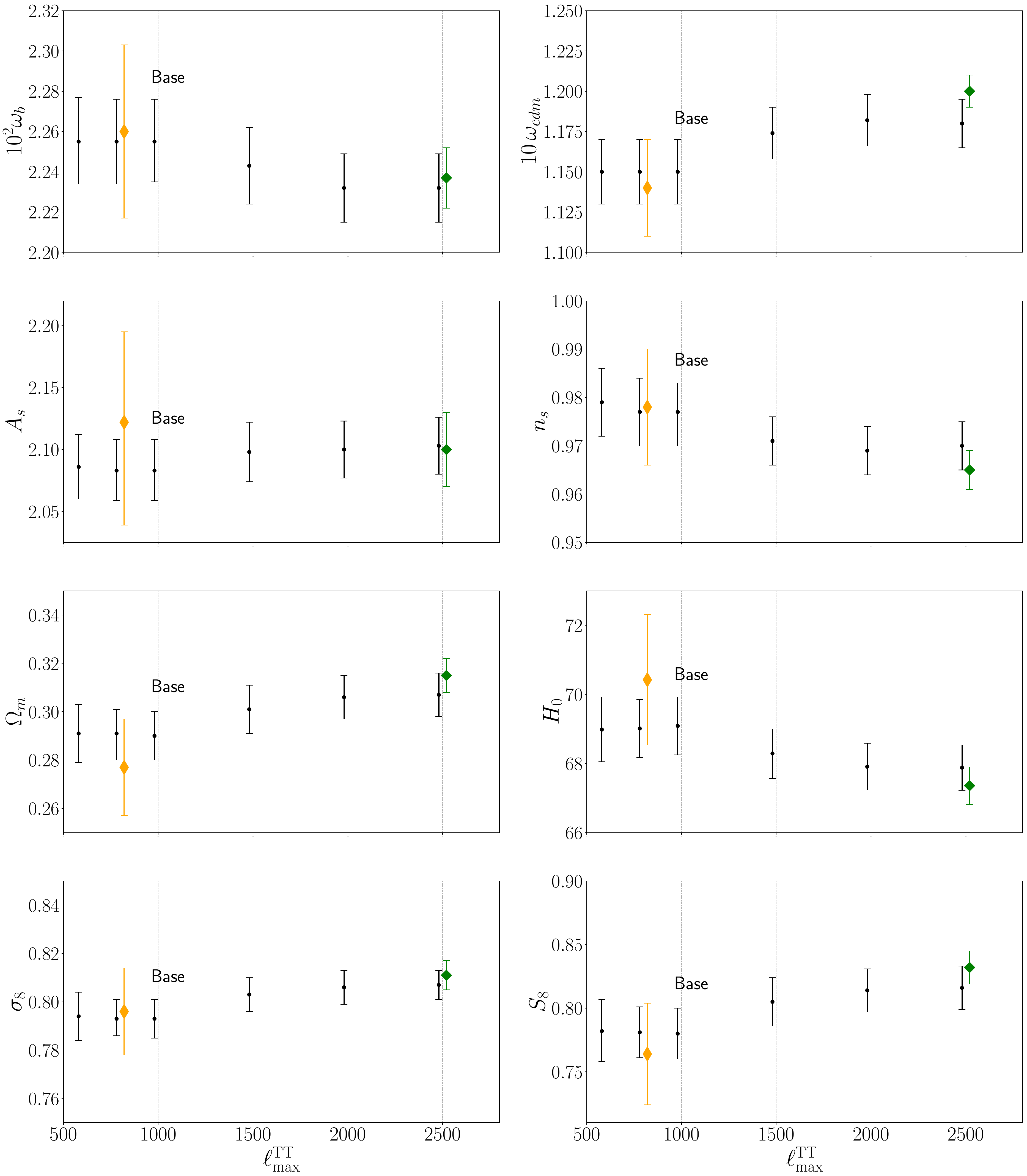}
        \caption {Marginalized parameter estimates (1$\sigma$ error bars) inferred from the $\rm \!SPT$-$\rm 3G\!+\!SPTlens\!+\!Planck\,TT$ data with the Planck TT spectrum analyzed up to a certain cutoff point $\ell^{\rm TT}<\lmax^{\rm TT}$.
        We also display the results of the official Planck TT $\ell^{\rm TT}<800$ analysis~\cite{Aghanim:2016sns} (yellow diamonds) as well as the Planck legacy release constraints~\cite{Planck:2018vyg} (green diamonds).}
        \label{fig:10}
    \end{center}
\end{figure}

We also explore the sensitivity of our CMB-based parameter constraints to the choice of a Planck TT data cutoff. To that end, we perform an analysis of the $\rm SPT$-$\rm 3G\!+\!SPTlens\!+\!Planck\,TT$ data with the Planck TT spectrum taken at $\ell^{\rm TT}<\lmax^{\rm TT}$.
In Fig. \ref{fig:10} we show the resulting parameter constraints for the following multipole cuts $\lmax^{\rm TT}=800$, $1000$, $1500$, $2000$ and $2500$.~\footnote{To avoid unnecessary details, we write $\lmax^{\rm TT}$ of $800$, $1000$, $1500$, $2000$, and $2500$, even though the true values of $\lmax^{\rm TT}$ are $801$, $999$, $1503$, $1996$ and $2508$ (where the nearest data bin falls). We also assume the minimal multipole $\ell_{\min}^{\rm TT}$ always to be $2$, e.g. $\ell^{\rm TT}<1000$ means $2\leq\ell^{\rm TT}<1000$.}
Note that the choice $\lmax^{\rm TT}=1000$ corresponds to the Base combination whereas the $\lmax^{\rm TT}=2500$ refers to the entire Planck TT power spectrum.

We found that the parameter measurements are stable across $\lmax^{\rm TT}\in[600,1000]$. Remarkably, the combined data approach leads to significantly tighter constraints on all cosmological parameters compared to the Planck TT $\ell<800$ analysis from Ref.~\cite{Aghanim:2016sns} (shown by yellow diamonds).~\footnote{Note that the mean of $A_s$ reported in~\cite{Aghanim:2016sns} is systematically higher due to a larger value of optical depth, $\tau=0.07\pm0.02$, adopted in this analysis. For clarity, we decided to show the original results from Ref.~\cite{Aghanim:2016sns}.} This effect is attributed to the SPT polarization and gravitational lensing measurements which sharp the parameter constraints by a factor of $2$. For $\lmax^{\rm TT}>1000$ the means of cosmological parameters drift away from the values found in our baseline analysis (labeled as Base). As far as the entire Planck TT data is included, we found $\leq1.4\sigma$ shifts in the parameter posteriors from the Base results. This difference originates from the Planck high-$\ell$ TT spectrum which, as we showed before, is in a $2.4\sigma$ tension with the Base data combination. Although the deviation is not very significant, we choose not to combine the Base and the Planck TT $\ell^{\rm TT}>1000$ measurements into one data set.

Our baseline choice $\lmax^{\rm TT}=1000$ roughly corresponds to the maximum multipole accessible to WMAP~\cite{Planck:2015bpv}. 
This data cut was extensively discussed in Ref.~\cite{Addison:2015wyg}. 
Note that the exact choice of $\lmax^{\rm TT}$ is arbitrary since the final parameter constraints are stable across $\lmax^{\rm TT}\in[600,1000]$ as shown in Fig. \ref{fig:10}.

\section{PDE in full Planck approach}
\label{app:PDEcom}

In this Appendix we explore the difference in parameter inference between our approach and the full Planck data analysis inside the PDE framework.

\begin{figure}
    \begin{center}
        \includegraphics[width=1.0\columnwidth]{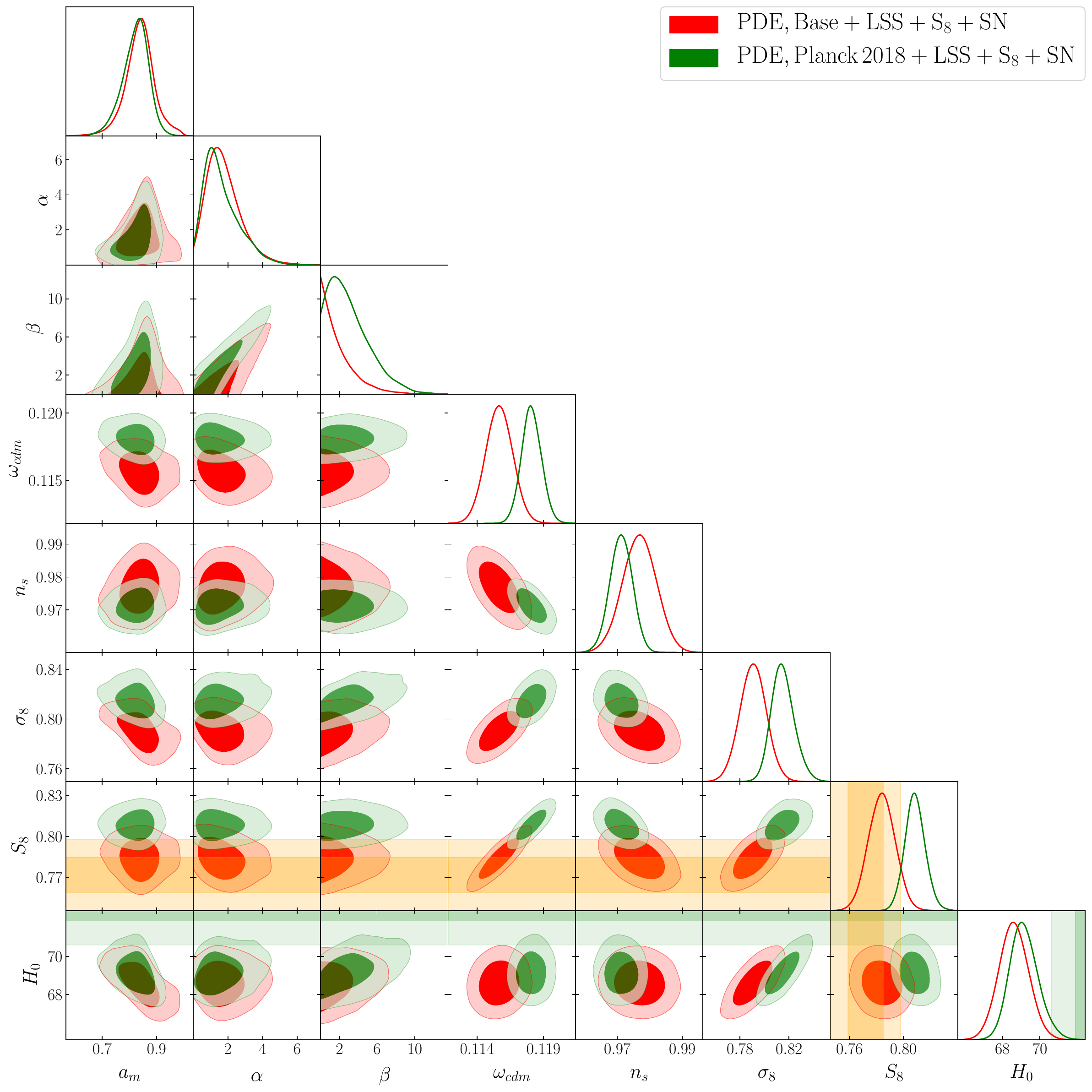}
        \caption {Marginalized 2d posterior distributions of the cosmological parameters in the PDE model for the $\rm Base+LSS+S_8+SN$ (red) and the $\rm Planck\,2018+LSS+S_8+SN$ (green) data sets. The Gaussian prior on $\tau$ \eqref{tau} is set. The yellow bands represent $1\sigma$ and $2\sigma$ constraints on $S_8$, see Eq.\,\eqref{S8}, it comes from the photometric surveys (DES-Y3, KiDS, HSC), whereas the green bands refer to the Hubble constant $H_0$ measurement \eqref{H0} reported by the SH0ES collaboration.}
        \label{fig:51}
    \end{center}
\end{figure}

We analyze the complete $\rm Planck\,2018+LSS+S_8+SN$ data.
The marginalized parameter constraints are tabulated in Tab.\,\ref{tab:51}.
\begin{table}
	\renewcommand{\arraystretch}{1.2}
    \small
	\centering
	\begin{tabular} {|c|| c |c|}
		\hline
		 & \multicolumn{2}{c|}{PDE} \\
		\hline
		\hline
        \multirow{1}{*}{\!\!\! Parameter\!} & $\rm \!Base\!+\!LSS\!+\!S_8\!+\!SN$ & $\rm \!Planck\,2018\!+\!LSS\!+S_8\!+\!SN$\\
		\hline
		$a_m$ & $0.839(0.822)^{+0.048}_{-0.049}$ &
		$0.822(0.817)^{+0.053}_{-0.039}$\\ 
		$\alpha$ & $1.8(1.3)^{+0.6}_{-1.2}$ &
		$1.7(1.3)^{+0.5}_{-1.3}$\\ 
		$\beta$ & $(0.0)<2.3$ &
		$3.1(2.0)^{+1.0}_{-0.9}$\\ 
		$100\,\omega_b$ & $2.252\pm0.018$ &
		$2.253\pm0.012$\\ 
		$10\,\omega_{cdm}$ & $1.157\pm0.010$ & 
		$1.181\pm0.007$\\ 
		$H_0$ & $68.61(68.24)\pm0.78$ & 
		$69.16(68.97)\pm0.76$\\ 
		$\tau$ & $0.056\pm0.005$ & 
		$0.058\pm0.005$\\ 
		${\rm ln}(10^{10} A_s)\!\!$ & $3.033\pm0.011$ & 
		$3.047\pm0.010$\\ 
		$n_s$ & $0.977\pm0.005$ & 
		$0.971\pm0.003$\\ 
		\hline   
		$\rdrag$ & $148.08\pm0.30$ &
		$147.42\pm0.21$\\   
		$\Omega_m$ & $0.295\pm0.007$ & 
		$0.295\pm0.007$\\ 
		$\sigma_8$ & $0.791\pm0.011$ & 
		$0.815\pm0.009$\\ 
		$S_8$ & $0.784\pm0.010$ & 
		$0.809\pm0.008$\\
		\hline 
	\end{tabular}
	\caption {Parameter constraints (mean value with $1\sigma$ error bars and best fit value in the parentheses) in the PDE model.}
	\label{tab:51}
\end{table}
The 2d posterior distributions of cosmological parameters are shown in Fig. \ref{fig:51}.
For comparison we also show our baseline results based on the $\rm Base+LSS+S_8+SN$ data.

We found that the constraints on the dark energy parameters in the both analyses agree, although the $\rm Planck\,2018+LSS+S_8+SN$ data favours considerably larger $\beta$.
Importantly, the analysis based on the full Planck likelihood predicts a $2.2\sigma$ higher value of the late-time fluctuation amplitude, $\sigma_8=0.815 \pm 0.009$.
This leads to $S_8=0.809 \pm 0.008$ which is in a $2.4\sigma$ tension with the direct probes \eqref{S8}.
This difference can be explained by the enhanced smoothing of acoustic peaks in the Planck data which pulls the late-time amplitude to higher values. 
Our analysis is free from this feature and entirely consistent with the direct measurements of $S_8$ in the late Universe. 
Interestingly, the $\rm Planck\,2018+LSS+S_8+SN$ combination predicts a slightly higher value of the Hubble constant, $H_0=69.16 \pm 0.76\kms$. 
This effect can be attributed to the observed degeneracy direction $\sigma_8 h^{-1.2}$ that pulls $H_0$ to higher values.

In essence, the PDE scenario with the entire Planck data slightly alleviates the $S_8$ tension, whereas the $\rm Base+LSS+S_8+SN$ analysis is entirely consistent with the direct measurements of $S_8$ \eqref{S8}. The $H_0$ constraint only barely changes.

\section{Distanceladder vs. Gaussian prior on $H_0$}
\label{app:H0}

Here we illustrate the difference between the entire distance ladder approach embedded in the package \texttt{distanceladder}~\cite{Greene:2021shv} and the traditional Gaussian prior on $H_0$ within the PDE model. 

In many studies the distance ladder measurements are reduced to a simple Gaussian constraint on $H_0$. In cosmological scenarios which are phenomenologically close to $\Lambda$CDM at late time (including those which only modify the early Universe), this approximation is accurate. However, when analyzing models which deviate significantly from $\Lambda$CDM at $z\lesssim 1$ using the traditional Gaussian prior on $H_0$ can bias results and even lead to the spurious detection of new physics~\cite{Benevento:2020fev,Greene:2021shv}.
The reason is that the local distance ladder measures distances to supernova in the Hubble flow at $z\gtrsim0.02$ rather than simply constrain $H_0$.
Thus, the entire distance ladder approach is required for any model which modifies the Universe expansion in this redshift range.

\begin{figure}[!t]
    \begin{center}
        \includegraphics[width=1.0\columnwidth]{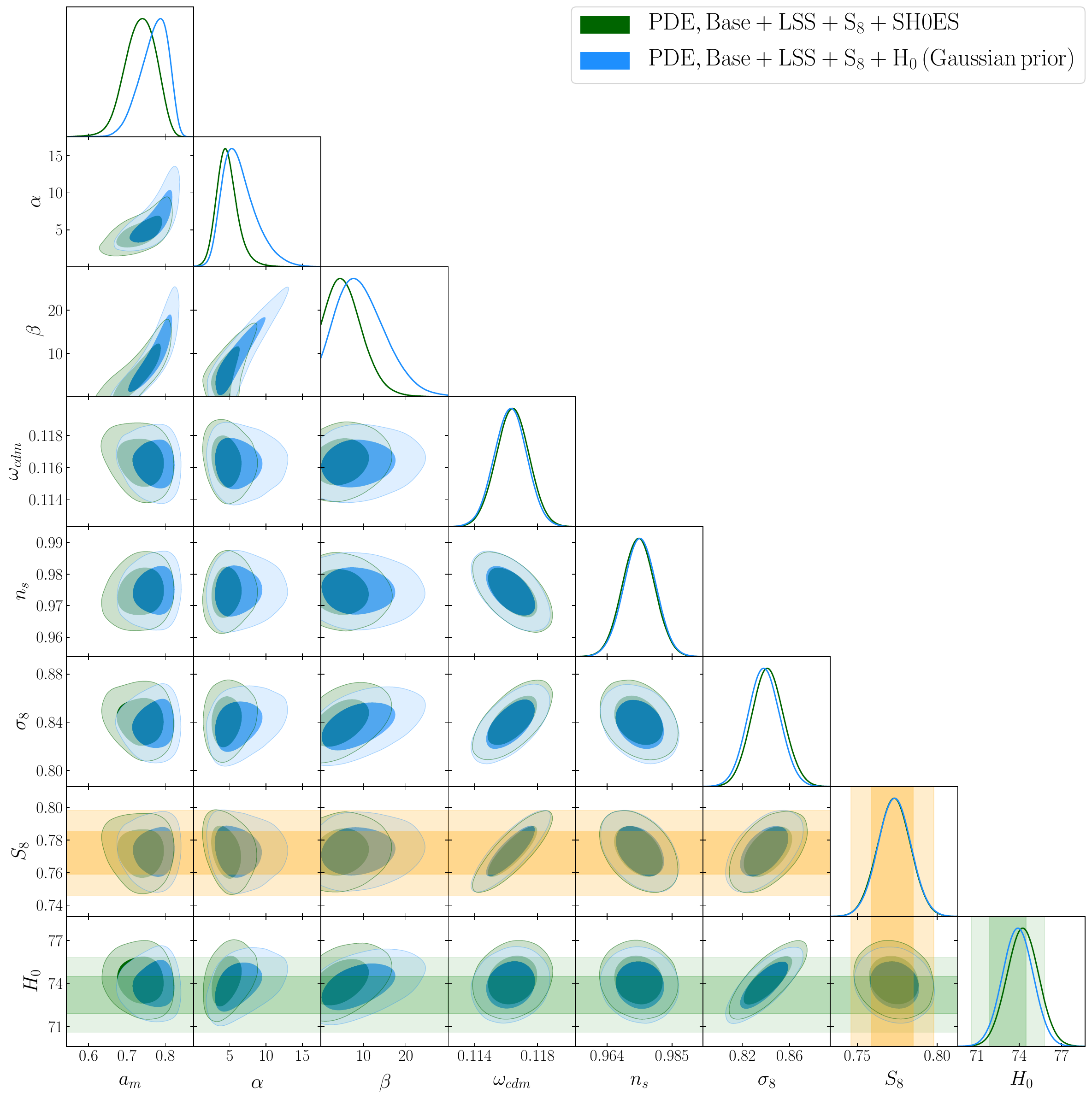} 
        \caption {2d posterior distributions of the cosmological parameters in the PDE model inferred from the $\rm Base+LSS+S_8+SH0ES$ data using the entire distance ladder (green) and the $\rm Base+LSS+S_8+H_0$ combination utilizing the Gaussian constraint on $H_0$ \eqref{H0} (blue).}
        \label{fig:12}
    \end{center}
\end{figure}

To showcase the difference in the parameter inference between these two approaches, we explore the parameter constraints in the PDE scenario using the Gaussian constraint on $H_0$ \eqref{H0} (dubbed $\rm H_0$). We analyze the $\rm Base+LSS+S_8+H_0$ data and show the resulting posterior distributions of cosmological parameters in Fig. \ref{fig:12}. 

We found that the results of using the entire distance ladder and the Gaussian prior on $H_0$ are in good agreement. The actual distance measurements have impact on the distributions of the PDE parameters while the constraints on the $\Lambda$CDM cosmological parameters remain virtually intact. This result can be explained by a smooth background evolution in the PDE model (see e.g. Fig. \ref{fig:7b}). Note that a sudden low-redshift discontinuity in the Hubble rate breaks down the standard cosmographic expansion of the luminosity distance to supernova that will make the traditional Gaussian constraint on $H_0$ inadequate~\cite{Greene:2021shv}.

\section{TDE prior dependence}
\label{app:Prior}

In the baseline analysis we have followed the previous work~\cite{Keeley:2019esp} and assumed the uniform priors on $z_{\tr}$ and $\Delta_{\tr}$ defined in \eqref{TDEprior}. Here, we examine the sensitivity of parameter constraints to the choice of the TDE priors.

\begin{figure}[!t]
   \begin{center}
        \includegraphics[width=1.0\columnwidth]{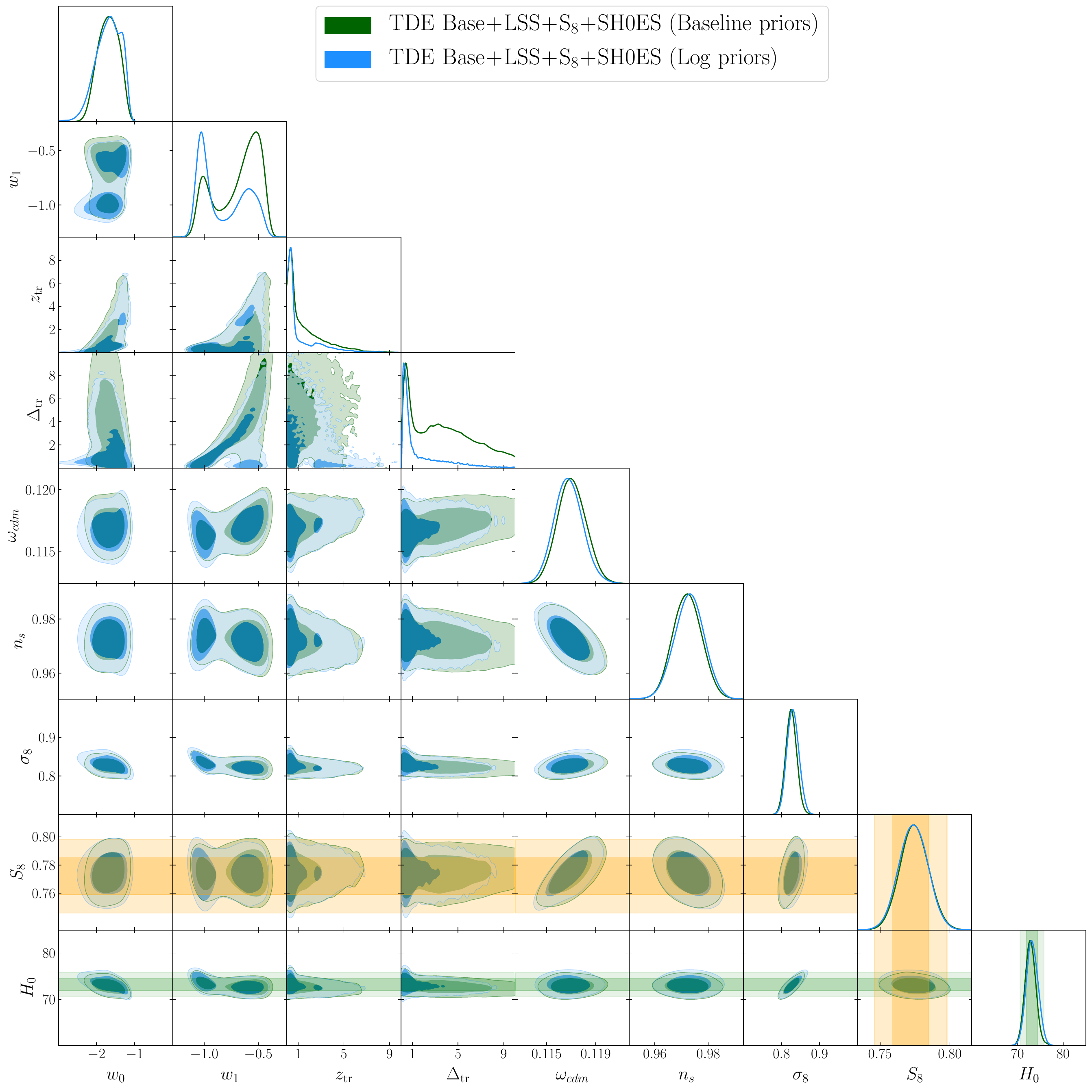} 
        \caption {Marginalized 2d posterior distributions of the cosmological parameters in the TDE scenario inferred from the $\rm Base+LSS+S_8+SH0ES$ data  with the baseline priors \eqref{TDEprior} (green) and new priors \eqref{TDEprior2} (blue).}
        \label{fig:13}
    \end{center}
\end{figure}

To elucidate the prior dependence, we repeat a MCMC analysis with uniform priors imposed on $\log_{10}(1+z_{\tr})$ and $\log_{10}\Delta_{\tr}$, namely
\be
\label{TDEprior2}
\begin{gathered}
    \log_{10}(1+z_{\tr})\in[0,1.041],\qquad \log_{10}\Delta_{\tr}\in[-1,1]
\end{gathered}
\ee
Note that the bounds on $\log_{10}(1+z_{\tr})$ and the upper limit on $\log_{10}\Delta_{\tr}$ are chosen to match \eqref{TDEprior}. 
We keep the flat priors on $w_0$ and $w_1$ as in \eqref{TDEprior}.
To showcase the impact of new priors, we examine the cosmological inference from the $\rm Base+LSS+S_8+SH0ES$ data
which suggests the most prominent transition in the dark energy equation of state (see e.g. Fig. \ref{fig:9b}). 
The parameter constraints are tabulated in Tab. \ref{tab:coTDE}.
\begin{table}
	\renewcommand{\arraystretch}{1.2}
    \small
	\centering
	\begin{tabular} {|c|| c |c | }
		\hline 
		& \multicolumn{2}{c|}{TDE} \\
		\hline
		\hline
		Parameter & 
		Baseline priors & Log priors
		\\
		\hline
		$w_0 $ & $-1.68(-1.75)^{+0.30}_{-0.26}$ & $-1.67(-1.69)^{+0.42}_{-0.22}$ 
		\\ 
		$w_1$ & $-0.68(-1.04)^{+0.26}_{-0.35}$ & $-0.81(-1.07)^{+0.31}_{-0.28}$ 
		\\ 
		$z_{tr}$ & $<5.26 (0.39) $ & $<4.89 (0.40)$ 
		\\ 
        $\Delta_{tr}$ & $<8.75 (0.28)$ & $1.79(0.24)^{+0.20}_{-1.73}$ 
		\\ 
        $100\,\omega_b $ & 
		$2.243\pm0.019 $ & $2.243\pm0.019$ 
		\\ 
		$10\,\omega_{cdm}$ & 
		$1.171\pm0.011 $ & $1.168\pm0.012$
		\\ 
		$H_0$ & 
		$72.83(74.36)\pm1.16 $ & $73.17(74.95)^{+1.30}_{-1.27}$ 
		\\ 
		$\tau$ & 
		$0.057\pm0.005 $ & $0.057\pm0.005$ 
		\\ 
		$\ln(10^{10} A_s)\!\!$ &
		$3.039\pm0.011 $ & $3.038\pm0.012$
		\\ 
		$n_s$ & 
		$0.972\pm0.005 $ & $0.973\pm0.005$
		\\ 
        \hline   
		$\rdrag$ & $147.80\pm0.31$ &
		$147.87\pm0.32$\\    
		$\Omega_m$ & $0.264\pm0.009$ & 
		$0.262\pm0.010$\\ 
		$\sigma_8$ & $0.826\pm0.014$ & 
		$0.829\pm0.016$\\ 
		$S_8$ & $0.775\pm0.010$ & 
		$0.774\pm0.011$\\
		\hline 
	\end{tabular}
	\caption {Parameter estimates (mean value with $1\sigma$ error bars and best fit value in the parentheses) inferred for the $\rm Base+LSS+S_8+SH0ES$ data with the baseline priors \eqref{TDEprior} (Baseline priors) and new priors \eqref{TDEprior2} (Log priors).}
	\label{tab:coTDE}
\end{table}
The corresponding posterior distributions are shown in  Fig. \ref{fig:13}.

We found that uniform priors on $\log_{10}(1+z_{\tr})$ and $\log_{10}\Delta_{\tr}$ impose a stronger preference for {\it small} values of $z_{\tr}$ and $\Delta_{\tr}$. In particular, the distribution of $\Delta_{\tr}$ is highly peaked at zero, and the second maximum at larger values of this parameter disappears. This effect is not surprising since the logarithmic priors imply strong weight toward small $z_{\tr}$ and $\Delta_{\tr}$ values. 
The best-fit parameter values given in Tab. \ref{tab:coTDE} indicate that the TDE dynamics remains essentially unchanged.
We found the difference in the best-fit $\chi^2$ statistics between these two analyses to be not significant, namely $\Delta\chi^2_{\min}=\chi^2_{\rm \min}({\rm Log\,priors})-\chi^2_{\rm \min}({\rm Baseline priors})=-0.5$.

Our findings demonstrate the modest impact of the TDE priors on the dark energy parameters. The $\Lambda$CDM parameter constraints are robust against the choice of the priors.

\bibliographystyle{JHEP}
\bibliography{short.bib}

\providecommand{\href}[2]{#2}\begingroup\raggedright\begin{thebibliography}{100}

\bibitem{Abdalla:2022yfr}
E.~Abdalla et~al., \emph{{Cosmology intertwined: A review of the particle
  physics, astrophysics, and cosmology associated with the cosmological
  tensions and anomalies}},
  \href{https://doi.org/10.1016/j.jheap.2022.04.002}{\emph{JHEAp} {\bfseries
  34} (2022) 49} [\href{https://arxiv.org/abs/2203.06142}{{\ttfamily
  2203.06142}}].

\bibitem{Riess:2020fzl}
A.~G. Riess, S.~Casertano, W.~Yuan, J.~B. Bowers, L.~Macri, J.~C. Zinn et~al.,
  \emph{{Cosmic Distances Calibrated to 1\% Precision with Gaia EDR3 Parallaxes
  and Hubble Space Telescope Photometry of 75 Milky Way Cepheids Confirm
  Tension with $\Lambda$CDM}},
  \href{https://doi.org/10.3847/2041-8213/abdbaf}{\emph{Astrophys. J. Lett.}
  {\bfseries 908} (2021) L6}
  [\href{https://arxiv.org/abs/2012.08534}{{\ttfamily 2012.08534}}].

\bibitem{Planck:2018vyg}
{\scshape Planck} collaboration, N.~Aghanim et~al., \emph{{Planck 2018 results.
  VI. Cosmological parameters}},
  \href{https://doi.org/10.1051/0004-6361/201833910}{\emph{Astron. Astrophys.}
  {\bfseries 641} (2020) A6}
  [\href{https://arxiv.org/abs/1807.06209}{{\ttfamily 1807.06209}}].

\bibitem{Riess:2021jrx}
A.~G. Riess et~al., \emph{{A Comprehensive Measurement of the Local Value of
  the Hubble Constant with 1 km/s/Mpc Uncertainty from the Hubble Space
  Telescope and the SH0ES Team}},
  \href{https://doi.org/10.3847/2041-8213/ac5c5b}{\emph{Astrophys. J. Lett.}
  {\bfseries 934} (2022) L7}
  [\href{https://arxiv.org/abs/2112.04510}{{\ttfamily 2112.04510}}].

\bibitem{Freedman:2020dne}
W.~L. Freedman, B.~F. Madore, T.~Hoyt, I.~S. Jang, R.~Beaton, M.~G. Lee et~al.,
  \emph{{Calibration of the Tip of the Red Giant Branch (TRGB)}},
  \href{https://arxiv.org/abs/2002.01550}{{\ttfamily 2002.01550}}.

\bibitem{Wong:2019kwg}
K.~C. Wong et~al., \emph{{H0LiCOW \textendash{} XIII. A 2.4 per cent
  measurement of H0 from lensed quasars: 5.3\ensuremath{\sigma} tension between
  early- and late-Universe probes}},
  \href{https://doi.org/10.1093/mnras/stz3094}{\emph{Mon. Not. Roy. Astron.
  Soc.} {\bfseries 498} (2020) 1420}
  [\href{https://arxiv.org/abs/1907.04869}{{\ttfamily 1907.04869}}].

\bibitem{Birrer:2020tax}
S.~Birrer et~al., \emph{{TDCOSMO - IV. Hierarchical time-delay cosmography
  \textendash{} joint inference of the Hubble constant and galaxy density
  profiles}}, \href{https://doi.org/10.1051/0004-6361/202038861}{\emph{Astron.
  Astrophys.} {\bfseries 643} (2020) A165}
  [\href{https://arxiv.org/abs/2007.02941}{{\ttfamily 2007.02941}}].

\bibitem{DiValentino:2020vvd}
E.~Di~Valentino et~al., \emph{{Cosmology Intertwined III: $f \sigma_8$ and
  $S_8$}},  \href{https://arxiv.org/abs/2008.11285}{{\ttfamily 2008.11285}}.

\bibitem{DES:2021wwk}
{\scshape DES} collaboration, T.~M.~C. Abbott et~al., \emph{{Dark Energy Survey
  Year 3 Results: Cosmological Constraints from Galaxy Clustering and Weak
  Lensing}},  \href{https://arxiv.org/abs/2105.13549}{{\ttfamily 2105.13549}}.

\bibitem{KiDS:2020suj}
{\scshape KiDS} collaboration, M.~Asgari et~al., \emph{{KiDS-1000 Cosmology:
  Cosmic shear constraints and comparison between two point statistics}},
  \href{https://doi.org/10.1051/0004-6361/202039070}{\emph{Astron. Astrophys.}
  {\bfseries 645} (2021) A104}
  [\href{https://arxiv.org/abs/2007.15633}{{\ttfamily 2007.15633}}].

\bibitem{Dalal:2023olq}
R.~Dalal et~al., \emph{{Hyper Suprime-Cam Year 3 Results: Cosmology from Cosmic
  Shear Power Spectra}},  \href{https://arxiv.org/abs/2304.00701}{{\ttfamily
  2304.00701}}.

\bibitem{Philcox:2021kcw}
O.~H.~E. Philcox and M.~M. Ivanov, \emph{{The BOSS DR12 Full-Shape Cosmology:
  $\Lambda$CDM Constraints from the Large-Scale Galaxy Power Spectrum and
  Bispectrum Monopole}},  \href{https://arxiv.org/abs/2112.04515}{{\ttfamily
  2112.04515}}.

\bibitem{Nunes:2021ipq}
R.~C. Nunes and S.~Vagnozzi, \emph{{Arbitrating the S8 discrepancy with growth
  rate measurements from redshift-space distortions}},
  \href{https://doi.org/10.1093/mnras/stab1613}{\emph{Mon. Not. Roy. Astron.
  Soc.} {\bfseries 505} (2021) 5427}
  [\href{https://arxiv.org/abs/2106.01208}{{\ttfamily 2106.01208}}].

\bibitem{Aghanim:2016sns}
{\scshape Planck} collaboration, N.~Aghanim et~al., \emph{{Planck intermediate
  results. LI. Features in the cosmic microwave background temperature power
  spectrum and shifts in cosmological parameters}},
  \href{https://doi.org/10.1051/0004-6361/201629504}{\emph{Astron. Astrophys.}
  {\bfseries 607} (2017) A95}
  [\href{https://arxiv.org/abs/1608.02487}{{\ttfamily 1608.02487}}].

\bibitem{Motloch:2019gux}
P.~Motloch and W.~Hu, \emph{{Lensinglike tensions in the $Planck$ legacy
  release}}, \href{https://doi.org/10.1103/PhysRevD.101.083515}{\emph{Phys.
  Rev.} {\bfseries D101} (2020) 083515}
  [\href{https://arxiv.org/abs/1912.06601}{{\ttfamily 1912.06601}}].

\bibitem{SPT-3G:2021eoc}
{\scshape SPT-3G} collaboration, D.~Dutcher et~al., \emph{{Measurements of the
  E-Mode Polarization and Temperature-E-Mode Correlation of the CMB from SPT-3G
  2018 Data}},  \href{https://arxiv.org/abs/2101.01684}{{\ttfamily
  2101.01684}}.

\bibitem{ACT:2020gnv}
{\scshape ACT} collaboration, S.~Aiola et~al., \emph{{The Atacama Cosmology
  Telescope: DR4 Maps and Cosmological Parameters}},
  \href{https://doi.org/10.1088/1475-7516/2020/12/047}{\emph{JCAP} {\bfseries
  12} (2020) 047} [\href{https://arxiv.org/abs/2007.07288}{{\ttfamily
  2007.07288}}].

\bibitem{Chudaykin:2020acu}
A.~Chudaykin, D.~Gorbunov and N.~Nedelko, \emph{{Combined analysis of Planck
  and SPTPol data favors the early dark energy models}},
  \href{https://doi.org/10.1088/1475-7516/2020/08/013}{\emph{JCAP} {\bfseries
  2008} (2020) 013} [\href{https://arxiv.org/abs/2004.13046}{{\ttfamily
  2004.13046}}].

\bibitem{Chudaykin:2020igl}
A.~Chudaykin, D.~Gorbunov and N.~Nedelko, \emph{{Exploring an early dark energy
  solution to the Hubble tension with Planck and SPTPol data}},
  \href{https://doi.org/10.1103/PhysRevD.103.043529}{\emph{Phys. Rev. D}
  {\bfseries 103} (2021) 043529}
  [\href{https://arxiv.org/abs/2011.04682}{{\ttfamily 2011.04682}}].

\bibitem{Yang:2018qmz}
W.~Yang, S.~Pan, E.~Di~Valentino, E.~N. Saridakis and S.~Chakraborty,
  \emph{{Observational constraints on one-parameter dynamical dark-energy
  parametrizations and the $H_0$ tension}},
  \href{https://doi.org/10.1103/PhysRevD.99.043543}{\emph{Phys. Rev. D}
  {\bfseries 99} (2019) 043543}
  [\href{https://arxiv.org/abs/1810.05141}{{\ttfamily 1810.05141}}].

\bibitem{DiValentino:2019dzu}
E.~Di~Valentino, A.~Melchiorri and J.~Silk, \emph{{Cosmological constraints in
  extended parameter space from the Planck 2018 Legacy release}},
  \href{https://doi.org/10.1088/1475-7516/2020/01/013}{\emph{JCAP} {\bfseries
  01} (2020) 013} [\href{https://arxiv.org/abs/1908.01391}{{\ttfamily
  1908.01391}}].

\bibitem{Vagnozzi:2019ezj}
S.~Vagnozzi, \emph{{New physics in light of the $H_0$ tension: An alternative
  view}}, \href{https://doi.org/10.1103/PhysRevD.102.023518}{\emph{Phys. Rev.
  D} {\bfseries 102} (2020) 023518}
  [\href{https://arxiv.org/abs/1907.07569}{{\ttfamily 1907.07569}}].

\bibitem{Keeley:2019esp}
R.~E. Keeley, S.~Joudaki, M.~Kaplinghat and D.~Kirkby, \emph{{Implications of a
  transition in the dark energy equation of state for the $H_0$ and $\sigma_8$
  tensions}}, \href{https://doi.org/10.1088/1475-7516/2019/12/035}{\emph{JCAP}
  {\bfseries 12} (2019) 035}
  [\href{https://arxiv.org/abs/1905.10198}{{\ttfamily 1905.10198}}].

\bibitem{Yang:2021flj}
W.~Yang, E.~Di~Valentino, S.~Pan, Y.~Wu and J.~Lu, \emph{{Dynamical dark energy
  after Planck CMB final release and $H_0$ tension}},
  \href{https://doi.org/10.1093/mnras/staa3914}{\emph{Mon. Not. Roy. Astron.
  Soc.} {\bfseries 501} (2021) 5845}
  [\href{https://arxiv.org/abs/2101.02168}{{\ttfamily 2101.02168}}].

\bibitem{Roy:2022fif}
N.~Roy, S.~Goswami and S.~Das, \emph{{Quintessence or phantom: Study of scalar
  field dark energy models through a general parametrization of the Hubble
  parameter}}, \href{https://doi.org/10.1016/j.dark.2022.101037}{\emph{Phys.
  Dark Univ.} {\bfseries 36} (2022) 101037}
  [\href{https://arxiv.org/abs/2201.09306}{{\ttfamily 2201.09306}}].

\bibitem{Sharma:2022ifr}
R.~K. Sharma, K.~L. Pandey and S.~Das, \emph{{Implications of an Extended Dark
  Energy Model with Massive Neutrinos}},
  \href{https://doi.org/10.3847/1538-4357/ac7a33}{\emph{Astrophys. J.}
  {\bfseries 934} (2022) 113}
  [\href{https://arxiv.org/abs/2202.01749}{{\ttfamily 2202.01749}}].

\bibitem{Zhao:2012aw}
G.-B. Zhao, R.~G. Crittenden, L.~Pogosian and X.~Zhang, \emph{{Examining the
  evidence for dynamical dark energy}},
  \href{https://doi.org/10.1103/PhysRevLett.109.171301}{\emph{Phys. Rev. Lett.}
  {\bfseries 109} (2012) 171301}
  [\href{https://arxiv.org/abs/1207.3804}{{\ttfamily 1207.3804}}].

\bibitem{Zhao:2017cud}
G.-B. Zhao et~al., \emph{{Dynamical dark energy in light of the latest
  observations}}, \href{https://doi.org/10.1038/s41550-017-0216-z}{\emph{Nature
  Astron.} {\bfseries 1} (2017) 627}
  [\href{https://arxiv.org/abs/1701.08165}{{\ttfamily 1701.08165}}].

\bibitem{Wang:2018fng}
Y.~Wang, L.~Pogosian, G.-B. Zhao and A.~Zucca, \emph{{Evolution of dark energy
  reconstructed from the latest observations}},
  \href{https://doi.org/10.3847/2041-8213/aaf238}{\emph{Astrophys. J. Lett.}
  {\bfseries 869} (2018) L8}
  [\href{https://arxiv.org/abs/1807.03772}{{\ttfamily 1807.03772}}].

\bibitem{Dutta:2018vmq}
K.~Dutta, Ruchika, A.~Roy, A.~A. Sen and M.~M. Sheikh-Jabbari, \emph{{Beyond
  $\Lambda $CDM with low and high redshift data: implications for dark
  energy}}, \href{https://doi.org/10.1007/s10714-020-2665-4}{\emph{Gen. Rel.
  Grav.} {\bfseries 52} (2020) 15}
  [\href{https://arxiv.org/abs/1808.06623}{{\ttfamily 1808.06623}}].

\bibitem{Capozziello:2018jya}
S.~Capozziello, Ruchika and A.~A. Sen, \emph{{Model independent constraints on
  dark energy evolution from low-redshift observations}},
  \href{https://doi.org/10.1093/mnras/stz176}{\emph{Mon. Not. Roy. Astron.
  Soc.} {\bfseries 484} (2019) 4484}
  [\href{https://arxiv.org/abs/1806.03943}{{\ttfamily 1806.03943}}].

\bibitem{Heisenberg:2022lob}
L.~Heisenberg, H.~Villarrubia-Rojo and J.~Zosso, \emph{{Simultaneously solving
  the $H_0$ and $\sigma_8$ tensions with late dark energy}},
  \href{https://arxiv.org/abs/2201.11623}{{\ttfamily 2201.11623}}.

\bibitem{Alestas:2021xes}
G.~Alestas and L.~Perivolaropoulos, \emph{{Late-time approaches to the Hubble
  tension deforming H(z), worsen the growth tension}},
  \href{https://doi.org/10.1093/mnras/stab1070}{\emph{Mon. Not. Roy. Astron.
  Soc.} {\bfseries 504} (2021) 3956}
  [\href{https://arxiv.org/abs/2103.04045}{{\ttfamily 2103.04045}}].

\bibitem{DiValentino:2020naf}
E.~Di~Valentino, A.~Mukherjee and A.~A. Sen, \emph{{Dark Energy with Phantom
  Crossing and the $H_0$ Tension}},
  \href{https://doi.org/10.3390/e23040404}{\emph{Entropy} {\bfseries 23} (2021)
  404} [\href{https://arxiv.org/abs/2005.12587}{{\ttfamily 2005.12587}}].

\bibitem{Jassal:2006gf}
H.~K. Jassal, J.~S. Bagla and T.~Padmanabhan, \emph{{Understanding the origin
  of CMB constraints on Dark Energy}},
  \href{https://doi.org/10.1111/j.1365-2966.2010.16647.x}{\emph{Mon. Not. Roy.
  Astron. Soc.} {\bfseries 405} (2010) 2639}
  [\href{https://arxiv.org/abs/astro-ph/0601389}{{\ttfamily
  astro-ph/0601389}}].

\bibitem{Henning:2017nuy}
{\scshape SPT} collaboration, J.~Henning et~al., \emph{{Measurements of the
  Temperature and E-Mode Polarization of the CMB from 500 Square Degrees of
  SPTpol Data}},
  \href{https://doi.org/10.3847/1538-4357/aa9ff4}{\emph{Astrophys. J.}
  {\bfseries 852} (2018) 97}
  [\href{https://arxiv.org/abs/1707.09353}{{\ttfamily 1707.09353}}].

\bibitem{Ivanov:2019pdj}
M.~M. Ivanov, M.~Simonovi\'c and M.~Zaldarriaga, \emph{{Cosmological Parameters
  from the BOSS Galaxy Power Spectrum}},
  \href{https://arxiv.org/abs/1909.05277}{{\ttfamily 1909.05277}}.

\bibitem{Ivanov:2021haa}
M.~M. Ivanov, O.~H.~E. Philcox, M.~Simonovi\'c, M.~Zaldarriaga, T.~Nishimichi
  and M.~Takada, \emph{{Cosmological constraints without fingers of God}},
  \href{https://arxiv.org/abs/2110.00006}{{\ttfamily 2110.00006}}.

\bibitem{Philcox:2020vvt}
O.~H.~E. Philcox, M.~M. Ivanov, M.~Simonovi\'c and M.~Zaldarriaga,
  \emph{{Combining Full-Shape and BAO Analyses of Galaxy Power Spectra: A
  1.6\textbackslash{}\% CMB-independent constraint on H$_0$}},
  \href{https://doi.org/10.1088/1475-7516/2020/05/032}{\emph{JCAP} {\bfseries
  05} (2020) 032} [\href{https://arxiv.org/abs/2002.04035}{{\ttfamily
  2002.04035}}].

\bibitem{Chudaykin:2020aoj}
A.~Chudaykin, M.~M. Ivanov, O.~H.~E. Philcox and M.~Simonović,
  \emph{{Non-linear perturbation theory extension of the Boltzmann code
  CLASS}}, \href{https://doi.org/10.1103/PhysRevD.102.063533}{\emph{Phys. Rev.}
  {\bfseries D102} (2020) 063533}
  [\href{https://arxiv.org/abs/2004.10607}{{\ttfamily 2004.10607}}].

\bibitem{Audren:2012wb}
B.~Audren, J.~Lesgourgues, K.~Benabed and S.~Prunet, \emph{{Conservative
  Constraints on Early Cosmology: an illustration of the Monte Python
  cosmological parameter inference code}},
  \href{https://doi.org/10.1088/1475-7516/2013/02/001}{\emph{JCAP} {\bfseries
  1302} (2013) 001} [\href{https://arxiv.org/abs/1210.7183}{{\ttfamily
  1210.7183}}].

\bibitem{Brinckmann:2018cvx}
T.~Brinckmann and J.~Lesgourgues, \emph{{MontePython 3: boosted MCMC sampler
  and other features}},
  \href{https://doi.org/10.1016/j.dark.2018.100260}{\emph{Phys. Dark Univ.}
  {\bfseries 24} (2019) 100260}
  [\href{https://arxiv.org/abs/1804.07261}{{\ttfamily 1804.07261}}].

\bibitem{Lewis:2002ah}
A.~Lewis and S.~Bridle, \emph{{Cosmological parameters from CMB and other data:
  A Monte Carlo approach}},
  \href{https://doi.org/10.1103/PhysRevD.66.103511}{\emph{Phys. Rev. D}
  {\bfseries 66} (2002) 103511}
  [\href{https://arxiv.org/abs/astro-ph/0205436}{{\ttfamily
  astro-ph/0205436}}].

\bibitem{Lewis:2013hha}
A.~Lewis, \emph{{Efficient sampling of fast and slow cosmological parameters}},
  \href{https://doi.org/10.1103/PhysRevD.87.103529}{\emph{Phys. Rev. D}
  {\bfseries 87} (2013) 103529}
  [\href{https://arxiv.org/abs/1304.4473}{{\ttfamily 1304.4473}}].

\bibitem{Lewis:2019xzd}
A.~Lewis, \emph{{GetDist: a Python package for analysing Monte Carlo samples}},
   \href{https://arxiv.org/abs/1910.13970}{{\ttfamily 1910.13970}}.

\bibitem{Wu:2019hek}
W.~L.~K. Wu et~al., \emph{{A Measurement of the Cosmic Microwave Background
  Lensing Potential and Power Spectrum from 500 deg$^2$ of SPTpol Temperature
  and Polarization Data}},
  \href{https://doi.org/10.3847/1538-4357/ab4186}{\emph{Astrophys. J.}
  {\bfseries 884} (2019) 70}
  [\href{https://arxiv.org/abs/1905.05777}{{\ttfamily 1905.05777}}].

\bibitem{DeBelsunce:2021xcp}
R.~De~Belsunce, S.~Gratton, W.~Coulton and G.~Efstathiou, \emph{{Inference of
  the optical depth to reionization from low multipole temperature and
  polarisation Planck data}},
  \href{https://arxiv.org/abs/2103.14378}{{\ttfamily 2103.14378}}.

\bibitem{Philcox:2020vbm}
O.~H.~E. Philcox, \emph{{Cosmology without window functions: Quadratic
  estimators for the galaxy power spectrum}},
  \href{https://doi.org/10.1103/PhysRevD.103.103504}{\emph{Phys. Rev. D}
  {\bfseries 103} (2021) 103504}
  [\href{https://arxiv.org/abs/2012.09389}{{\ttfamily 2012.09389}}].

\bibitem{Philcox:2021ukg}
O.~H.~E. Philcox, \emph{{Cosmology without window functions. II. Cubic
  estimators for the galaxy bispectrum}},
  \href{https://doi.org/10.1103/PhysRevD.104.123529}{\emph{Phys. Rev. D}
  {\bfseries 104} (2021) 123529}
  [\href{https://arxiv.org/abs/2107.06287}{{\ttfamily 2107.06287}}].

\bibitem{Beutler:2021eqq}
F.~Beutler and P.~McDonald, \emph{{Unified galaxy power spectrum measurements
  from 6dFGS, BOSS, and eBOSS}},
  \href{https://doi.org/10.1088/1475-7516/2021/11/031}{\emph{JCAP} {\bfseries
  11} (2021) 031} [\href{https://arxiv.org/abs/2106.06324}{{\ttfamily
  2106.06324}}].

\bibitem{Chudaykin:2020ghx}
A.~Chudaykin, K.~Dolgikh and M.~M. Ivanov, \emph{{Constraints on the curvature
  of the Universe and dynamical dark energy from the Full-shape and BAO data}},
  \href{https://doi.org/10.1103/PhysRevD.103.023507}{\emph{Phys. Rev. D}
  {\bfseries 103} (2021) 023507}
  [\href{https://arxiv.org/abs/2009.10106}{{\ttfamily 2009.10106}}].

\bibitem{Ross:2014qpa}
A.~J. Ross, L.~Samushia, C.~Howlett, W.~J. Percival, A.~Burden and M.~Manera,
  \emph{{The clustering of the SDSS DR7 main Galaxy sample \textendash{} I. A 4
  per cent distance measure at $z = 0.15$}},
  \href{https://doi.org/10.1093/mnras/stv154}{\emph{Mon. Not. Roy. Astron.
  Soc.} {\bfseries 449} (2015) 835}
  [\href{https://arxiv.org/abs/1409.3242}{{\ttfamily 1409.3242}}].

\bibitem{Beutler:2011hx}
F.~Beutler, C.~Blake, M.~Colless, D.~H. Jones, L.~Staveley-Smith, L.~Campbell
  et~al., \emph{{The 6dF Galaxy Survey: Baryon Acoustic Oscillations and the
  Local Hubble Constant}},
  \href{https://doi.org/10.1111/j.1365-2966.2011.19250.x}{\emph{Mon. Not. Roy.
  Astron. Soc.} {\bfseries 416} (2011) 3017}
  [\href{https://arxiv.org/abs/1106.3366}{{\ttfamily 1106.3366}}].

\bibitem{Neveux:2020voa}
R.~Neveux et~al., \emph{{The completed SDSS-IV extended Baryon Oscillation
  Spectroscopic Survey: BAO and RSD measurements from the anisotropic power
  spectrum of the quasar sample between redshift 0.8 and 2.2}},
  \href{https://doi.org/10.1093/mnras/staa2780}{\emph{Mon. Not. Roy. Astron.
  Soc.} {\bfseries 499} (2020) 210}
  [\href{https://arxiv.org/abs/2007.08999}{{\ttfamily 2007.08999}}].

\bibitem{duMasdesBourboux:2020pck}
H.~du~Mas~des Bourboux et~al., \emph{{The Completed SDSS-IV Extended Baryon
  Oscillation Spectroscopic Survey: Baryon Acoustic Oscillations with
  Ly\ensuremath{\alpha} Forests}},
  \href{https://doi.org/10.3847/1538-4357/abb085}{\emph{Astrophys. J.}
  {\bfseries 901} (2020) 153}
  [\href{https://arxiv.org/abs/2007.08995}{{\ttfamily 2007.08995}}].

\bibitem{deMattia:2020fkb}
A.~de~Mattia et~al., \emph{{The Completed SDSS-IV extended Baryon Oscillation
  Spectroscopic Survey: measurement of the BAO and growth rate of structure of
  the emission line galaxy sample from the anisotropic power spectrum between
  redshift 0.6 and 1.1}},
  \href{https://doi.org/10.1093/mnras/staa3891}{\emph{Mon. Not. Roy. Astron.
  Soc.} {\bfseries 501} (2021) 5616}
  [\href{https://arxiv.org/abs/2007.09008}{{\ttfamily 2007.09008}}].

\bibitem{Ivanov:2021zmi}
M.~M. Ivanov, \emph{{Cosmological constraints from the power spectrum of eBOSS
  emission line galaxies}},  \href{https://arxiv.org/abs/2106.12580}{{\ttfamily
  2106.12580}}.

\bibitem{HSC:2018mrq}
{\scshape HSC} collaboration, C.~Hikage et~al., \emph{{Cosmology from cosmic
  shear power spectra with Subaru Hyper Suprime-Cam first-year data}},
  \href{https://doi.org/10.1093/pasj/psz010}{\emph{Publ. Astron. Soc. Jap.}
  {\bfseries 71} (2019) 43} [\href{https://arxiv.org/abs/1809.09148}{{\ttfamily
  1809.09148}}].

\bibitem{Greene:2021shv}
K.~L. Greene and F.-Y. Cyr-Racine, \emph{{Hubble distancing: focusing on
  distance measurements in cosmology}},
  \href{https://doi.org/10.1088/1475-7516/2022/06/002}{\emph{JCAP} {\bfseries
  06} (2022) 002} [\href{https://arxiv.org/abs/2112.11567}{{\ttfamily
  2112.11567}}].

\bibitem{Camarena:2021jlr}
D.~Camarena and V.~Marra, \emph{{On the use of the local prior on the absolute
  magnitude of Type Ia supernovae in cosmological inference}},
  \href{https://doi.org/10.1093/mnras/stab1200}{\emph{Mon. Not. Roy. Astron.
  Soc.} {\bfseries 504} (2021) 5164}
  [\href{https://arxiv.org/abs/2101.08641}{{\ttfamily 2101.08641}}].

\bibitem{Pan-STARRS1:2017jku}
{\scshape Pan-STARRS1} collaboration, D.~M. Scolnic et~al., \emph{{The Complete
  Light-curve Sample of Spectroscopically Confirmed SNe Ia from Pan-STARRS1 and
  Cosmological Constraints from the Combined Pantheon Sample}},
  \href{https://doi.org/10.3847/1538-4357/aab9bb}{\emph{Astrophys. J.}
  {\bfseries 859} (2018) 101}
  [\href{https://arxiv.org/abs/1710.00845}{{\ttfamily 1710.00845}}].

\bibitem{DiValentino:2021imh}
E.~Di~Valentino and A.~Melchiorri, \emph{{Neutrino Mass Bounds in the era of
  Tension Cosmology}},  \href{https://arxiv.org/abs/2112.02993}{{\ttfamily
  2112.02993}}.

\bibitem{Lemos:2018smw}
P.~Lemos, E.~Lee, G.~Efstathiou and S.~Gratton, \emph{{Model independent $H(z)$
  reconstruction using the cosmic inverse distance ladder}},
  \href{https://doi.org/10.1093/mnras/sty3082}{\emph{Mon. Not. Roy. Astron.
  Soc.} {\bfseries 483} (2019) 4803}
  [\href{https://arxiv.org/abs/1806.06781}{{\ttfamily 1806.06781}}].

\bibitem{Poulin:2018zxs}
V.~Poulin, K.~K. Boddy, S.~Bird and M.~Kamionkowski, \emph{{Implications of an
  extended dark energy cosmology with massive neutrinos for cosmological
  tensions}}, \href{https://doi.org/10.1103/PhysRevD.97.123504}{\emph{Phys.
  Rev. D} {\bfseries 97} (2018) 123504}
  [\href{https://arxiv.org/abs/1803.02474}{{\ttfamily 1803.02474}}].

\bibitem{Dinda:2021ffa}
B.~R. Dinda, \emph{{Cosmic expansion parametrization: Implication for curvature
  and H0 tension}},
  \href{https://doi.org/10.1103/PhysRevD.105.063524}{\emph{Phys. Rev. D}
  {\bfseries 105} (2022) 063524}
  [\href{https://arxiv.org/abs/2106.02963}{{\ttfamily 2106.02963}}].

\bibitem{Keeley:2022ojz}
R.~E. Keeley and A.~Shafieloo, \emph{{Ruling Out New Physics at Low Redshift as
  a solution to the $H_0$ Tension}},
  \href{https://arxiv.org/abs/2206.08440}{{\ttfamily 2206.08440}}.

\bibitem{Schmittfull:2013uea}
M.~M. Schmittfull, A.~Challinor, D.~Hanson and A.~Lewis, \emph{{Joint analysis
  of CMB temperature and lensing-reconstruction power spectra}},
  \href{https://doi.org/10.1103/PhysRevD.88.063012}{\emph{Phys. Rev. D}
  {\bfseries 88} (2013) 063012}
  [\href{https://arxiv.org/abs/1308.0286}{{\ttfamily 1308.0286}}].

\bibitem{Peloton:2016kbw}
J.~Peloton, M.~Schmittfull, A.~Lewis, J.~Carron and O.~Zahn, \emph{{Full
  covariance of CMB and lensing reconstruction power spectra}},
  \href{https://doi.org/10.1103/PhysRevD.95.043508}{\emph{Phys. Rev. D}
  {\bfseries 95} (2017) 043508}
  [\href{https://arxiv.org/abs/1611.01446}{{\ttfamily 1611.01446}}].

\bibitem{Addison:2015wyg}
G.~Addison, Y.~Huang, D.~Watts, C.~Bennett, M.~Halpern, G.~Hinshaw et~al.,
  \emph{{Quantifying discordance in the 2015 Planck CMB spectrum}},
  \href{https://doi.org/10.3847/0004-637X/818/2/132}{\emph{Astrophys. J.}
  {\bfseries 818} (2016) 132}
  [\href{https://arxiv.org/abs/1511.00055}{{\ttfamily 1511.00055}}].

\bibitem{Planck:2015bpv}
{\scshape Planck} collaboration, N.~Aghanim et~al., \emph{{Planck 2015 results.
  XI. CMB power spectra, likelihoods, and robustness of parameters}},
  \href{https://doi.org/10.1051/0004-6361/201526926}{\emph{Astron. Astrophys.}
  {\bfseries 594} (2016) A11}
  [\href{https://arxiv.org/abs/1507.02704}{{\ttfamily 1507.02704}}].

\bibitem{Bianchini:2019vxp}
{\scshape SPT} collaboration, F.~Bianchini et~al., \emph{{Constraints on
  Cosmological Parameters from the 500 deg$^2$ SPTpol Lensing Power Spectrum}},
  \href{https://doi.org/10.3847/1538-4357/ab6082}{\emph{Astrophys. J.}
  {\bfseries 888} (2020) 119}
  [\href{https://arxiv.org/abs/1910.07157}{{\ttfamily 1910.07157}}].

\bibitem{Motloch:2018pjy}
P.~Motloch and W.~Hu, \emph{{Tensions between direct measurements of the lens
  power spectrum from Planck data}},
  \href{https://doi.org/10.1103/PhysRevD.97.103536}{\emph{Phys. Rev. D}
  {\bfseries 97} (2018) 103536}
  [\href{https://arxiv.org/abs/1803.11526}{{\ttfamily 1803.11526}}].

\bibitem{1100705}
H.~Akaike, \emph{A new look at the statistical model identification},
  \href{https://doi.org/10.1109/TAC.1974.1100705}{\emph{IEEE Transactions on
  Automatic Control} {\bfseries 19} (1974) 716}.

\bibitem{BOSS:2014hwf}
{\scshape BOSS} collaboration, T.~Delubac et~al., \emph{{Baryon acoustic
  oscillations in the Ly\ensuremath{\alpha} forest of BOSS DR11 quasars}},
  \href{https://doi.org/10.1051/0004-6361/201423969}{\emph{Astron. Astrophys.}
  {\bfseries 574} (2015) A59}
  [\href{https://arxiv.org/abs/1404.1801}{{\ttfamily 1404.1801}}].

\bibitem{Bernardo:2021cxi}
R.~C. Bernardo, D.~Grand\'on, J.~L. Said and V.~H. C\'ardenas,
  \emph{{Parametric and nonparametric methods hint dark energy evolution}},
  \href{https://arxiv.org/abs/2111.08289}{{\ttfamily 2111.08289}}.

\bibitem{Efstathiou:2021ocp}
G.~Efstathiou, \emph{{To H0 or not to H0?}},
  \href{https://doi.org/10.1093/mnras/stab1588}{\emph{Mon. Not. Roy. Astron.
  Soc.} {\bfseries 505} (2021) 3866}
  [\href{https://arxiv.org/abs/2103.08723}{{\ttfamily 2103.08723}}].

\bibitem{Benevento:2020fev}
G.~Benevento, W.~Hu and M.~Raveri, \emph{{Can Late Dark Energy Transitions
  Raise the Hubble constant?}},
  \href{https://doi.org/10.1103/PhysRevD.101.103517}{\emph{Phys. Rev. D}
  {\bfseries 101} (2020) 103517}
  [\href{https://arxiv.org/abs/2002.11707}{{\ttfamily 2002.11707}}].

\bibitem{DES:2018rjw}
{\scshape DES} collaboration, E.~Macaulay et~al., \emph{{First Cosmological
  Results using Type Ia Supernovae from the Dark Energy Survey: Measurement of
  the Hubble Constant}}, \href{https://doi.org/10.1093/mnras/stz978}{\emph{Mon.
  Not. Roy. Astron. Soc.} {\bfseries 486} (2019) 2184}
  [\href{https://arxiv.org/abs/1811.02376}{{\ttfamily 1811.02376}}].

\bibitem{Feeney:2018mkj}
S.~M. Feeney, H.~V. Peiris, A.~R. Williamson, S.~M. Nissanke, D.~J. Mortlock,
  J.~Alsing et~al., \emph{{Prospects for resolving the Hubble constant tension
  with standard sirens}},
  \href{https://doi.org/10.1103/PhysRevLett.122.061105}{\emph{Phys. Rev. Lett.}
  {\bfseries 122} (2019) 061105}
  [\href{https://arxiv.org/abs/1802.03404}{{\ttfamily 1802.03404}}].

\bibitem{Camarena:2019rmj}
D.~Camarena and V.~Marra, \emph{{A new method to build the (inverse) distance
  ladder}}, \href{https://doi.org/10.1093/mnras/staa770}{\emph{Mon. Not. Roy.
  Astron. Soc.} {\bfseries 495} (2020) 2630}
  [\href{https://arxiv.org/abs/1910.14125}{{\ttfamily 1910.14125}}].

\bibitem{NearbySupernovafactory:2013qtg}
{\scshape Nearby Supernova factory} collaboration, M.~Rigault et~al.,
  \emph{{Evidence of Environmental Dependencies of Type Ia Supernovae from the
  Nearby Supernova Factory indicated by Local H$\alpha$}},
  \href{https://doi.org/10.1051/0004-6361/201322104}{\emph{Astron. Astrophys.}
  {\bfseries 560} (2013) A66}
  [\href{https://arxiv.org/abs/1309.1182}{{\ttfamily 1309.1182}}].

\bibitem{Rigault:2014kaa}
M.~Rigault et~al., \emph{{Confirmation of a Star Formation Bias in Type Ia
  Supernova Distances and its Effect on Measurement of the Hubble Constant}},
  \href{https://doi.org/10.1088/0004-637X/802/1/20}{\emph{Astrophys. J.}
  {\bfseries 802} (2015) 20} [\href{https://arxiv.org/abs/1412.6501}{{\ttfamily
  1412.6501}}].

\bibitem{NearbySupernovaFactory:2018qkd}
{\scshape Nearby Supernova Factory} collaboration, M.~Rigault et~al.,
  \emph{{Strong Dependence of Type Ia Supernova Standardization on the Local
  Specific Star Formation Rate}},
  \href{https://doi.org/10.1051/0004-6361/201730404}{\emph{Astron. Astrophys.}
  {\bfseries 644} (2020) A176}
  [\href{https://arxiv.org/abs/1806.03849}{{\ttfamily 1806.03849}}].

\bibitem{Briday:2021rrm}
M.~Briday et~al., \emph{{Accuracy of environmental tracers and consequences for
  determining the Type Ia supernova magnitude step}},
  \href{https://doi.org/10.1051/0004-6361/202141160}{\emph{Astron. Astrophys.}
  {\bfseries 657} (2022) A22}
  [\href{https://arxiv.org/abs/2109.02456}{{\ttfamily 2109.02456}}].

\bibitem{Jones:2015uaa}
D.~O. Jones, A.~G. Riess and D.~M. Scolnic, \emph{{Reconsidering the Effects of
  Local Star Formation On Type Ia Supernova Cosmology}},
  \href{https://doi.org/10.1088/0004-637X/812/1/31}{\emph{Astrophys. J.}
  {\bfseries 812} (2015) 31}
  [\href{https://arxiv.org/abs/1506.02637}{{\ttfamily 1506.02637}}].

\bibitem{FSS:2018cey}
{\scshape FSS} collaboration, D.~O. Jones et~al., \emph{{The Foundation
  Supernova Survey: Measuring Cosmological Parameters with Supernovae from a
  Single Telescope}},
  \href{https://doi.org/10.3847/1538-4357/ab2bec}{\emph{Astrophys. J.}
  {\bfseries 881} (2019) 19}
  [\href{https://arxiv.org/abs/1811.09286}{{\ttfamily 1811.09286}}].

\bibitem{Riess:2016jrr}
A.~G. Riess et~al., \emph{{A 2.4\% Determination of the Local Value of the
  Hubble Constant}},
  \href{https://doi.org/10.3847/0004-637X/826/1/56}{\emph{Astrophys. J.}
  {\bfseries 826} (2016) 56}
  [\href{https://arxiv.org/abs/1604.01424}{{\ttfamily 1604.01424}}].

\bibitem{Riess:2019cxk}
A.~G. Riess, S.~Casertano, W.~Yuan, L.~M. Macri and D.~Scolnic, \emph{{Large
  Magellanic Cloud Cepheid Standards Provide a 1\% Foundation for the
  Determination of the Hubble Constant and Stronger Evidence for Physics beyond
  $\Lambda$CDM}},
  \href{https://doi.org/10.3847/1538-4357/ab1422}{\emph{Astrophys. J.}
  {\bfseries 876} (2019) 85}
  [\href{https://arxiv.org/abs/1903.07603}{{\ttfamily 1903.07603}}].

\bibitem{Perivolaropoulos:2021bds}
L.~Perivolaropoulos and F.~Skara, \emph{{Hubble tension or a transition of the
  Cepheid SnIa calibrator parameters?}},
  \href{https://doi.org/10.1103/PhysRevD.104.123511}{\emph{Phys. Rev. D}
  {\bfseries 104} (2021) 123511}
  [\href{https://arxiv.org/abs/2109.04406}{{\ttfamily 2109.04406}}].

\bibitem{Mortsell:2021nzg}
E.~Mortsell, A.~Goobar, J.~Johansson and S.~Dhawan, \emph{{Sensitivity of the
  Hubble Constant Determination to Cepheid Calibration}},
  \href{https://doi.org/10.3847/1538-4357/ac756e}{\emph{Astrophys. J.}
  {\bfseries 933} (2022) 212}
  [\href{https://arxiv.org/abs/2105.11461}{{\ttfamily 2105.11461}}].

\bibitem{Mortsell:2021tcx}
E.~Mortsell, A.~Goobar, J.~Johansson and S.~Dhawan, \emph{{The Hubble Tension
  Revisited: Additional Local Distance Ladder Uncertainties}},
  \href{https://doi.org/10.3847/1538-4357/ac7c19}{\emph{Astrophys. J.}
  {\bfseries 935} (2022) 58}
  [\href{https://arxiv.org/abs/2106.09400}{{\ttfamily 2106.09400}}].

\bibitem{Marra:2021fvf}
V.~Marra and L.~Perivolaropoulos, \emph{{Rapid transition of Geff at
  zt\ensuremath{\simeq}0.01 as a possible solution of the Hubble and growth
  tensions}}, \href{https://doi.org/10.1103/PhysRevD.104.L021303}{\emph{Phys.
  Rev. D} {\bfseries 104} (2021) L021303}
  [\href{https://arxiv.org/abs/2102.06012}{{\ttfamily 2102.06012}}].

\bibitem{Alestas:2020zol}
G.~Alestas, L.~Kazantzidis and L.~Perivolaropoulos, \emph{{$w-M$ phantom
  transition at $z_t$ \ensuremath{<}0.1 as a resolution of the Hubble
  tension}}, \href{https://doi.org/10.1103/PhysRevD.103.083517}{\emph{Phys.
  Rev. D} {\bfseries 103} (2021) 083517}
  [\href{https://arxiv.org/abs/2012.13932}{{\ttfamily 2012.13932}}].

\bibitem{Bernal:2016gxb}
J.~L. Bernal, L.~Verde and A.~G. Riess, \emph{{The trouble with $H_0$}},
  \href{https://doi.org/10.1088/1475-7516/2016/10/019}{\emph{JCAP} {\bfseries
  10} (2016) 019} [\href{https://arxiv.org/abs/1607.05617}{{\ttfamily
  1607.05617}}].

\bibitem{Teng:2021cvy}
Y.-P. Teng, W.~Lee and K.-W. Ng, \emph{{Constraining the dark-energy equation
  of state with cosmological data}},
  \href{https://doi.org/10.1103/PhysRevD.104.083519}{\emph{Phys. Rev. D}
  {\bfseries 104} (2021) 083519}
  [\href{https://arxiv.org/abs/2105.02667}{{\ttfamily 2105.02667}}].

\bibitem{Trotta:2008qt}
R.~Trotta, \emph{{Bayes in the sky: Bayesian inference and model selection in
  cosmology}}, \href{https://doi.org/10.1080/00107510802066753}{\emph{Contemp.
  Phys.} {\bfseries 49} (2008) 71}
  [\href{https://arxiv.org/abs/0803.4089}{{\ttfamily 0803.4089}}].

\bibitem{Heavens:2017afc}
A.~Heavens, Y.~Fantaye, A.~Mootoovaloo, H.~Eggers, Z.~Hosenie, S.~Kroon et~al.,
  \emph{{Marginal Likelihoods from Monte Carlo Markov Chains}},
  \href{https://arxiv.org/abs/1704.03472}{{\ttfamily 1704.03472}}.

\bibitem{Kass:1995loi}
R.~E. Kass and A.~E. Raftery, \emph{{Bayes Factors}},
  \href{https://doi.org/10.1080/01621459.1995.10476572}{\emph{J. Am. Statist.
  Assoc.} {\bfseries 90} (1995) 773}.

\bibitem{SimonsObservatory:2018koc}
{\scshape Simons Observatory} collaboration, P.~Ade et~al., \emph{{The Simons
  Observatory: Science goals and forecasts}},
  \href{https://doi.org/10.1088/1475-7516/2019/02/056}{\emph{JCAP} {\bfseries
  02} (2019) 056} [\href{https://arxiv.org/abs/1808.07445}{{\ttfamily
  1808.07445}}].

\bibitem{CMB-S4:2016ple}
{\scshape CMB-S4} collaboration, K.~N. Abazajian et~al., \emph{{CMB-S4 Science
  Book, First Edition}},  \href{https://arxiv.org/abs/1610.02743}{{\ttfamily
  1610.02743}}.

\bibitem{SPT-3G:2022hvq}
{\scshape SPT-3G} collaboration, L.~Balkenhol et~al., \emph{{A Measurement of
  the CMB Temperature Power Spectrum and Constraints on Cosmology from the
  SPT-3G 2018 TT/TE/EE Data Set}},
  \href{https://arxiv.org/abs/2212.05642}{{\ttfamily 2212.05642}}.

\bibitem{Simon:2022csv}
T.~Simon, P.~Zhang and V.~Poulin, \emph{{Cosmological inference from the
  EFTofLSS: the eBOSS QSO full-shape analysis}},
  \href{https://arxiv.org/abs/2210.14931}{{\ttfamily 2210.14931}}.

\bibitem{Chudaykin:2022nru}
A.~Chudaykin and M.~M. Ivanov, \emph{{Cosmological constraints from the power
  spectrum of eBOSS quasars}},
  \href{https://doi.org/10.1103/PhysRevD.107.043518}{\emph{Phys. Rev. D}
  {\bfseries 107} (2023) 043518}
  [\href{https://arxiv.org/abs/2210.17044}{{\ttfamily 2210.17044}}].

\bibitem{Ivanov:2023qzb}
M.~M. Ivanov, O.~H.~E. Philcox, G.~Cabass, T.~Nishimichi, M.~Simonovi\'c and
  M.~Zaldarriaga, \emph{{Cosmology with the Galaxy Bispectrum Multipoles:
  Optimal Estimation and Application to BOSS Data}},
  \href{https://arxiv.org/abs/2302.04414}{{\ttfamily 2302.04414}}.

\end{thebibliography}\endgroup

\end{document}